\documentclass[a4paper,11pt]{article}
\usepackage[utf8]{inputenc}
\usepackage{amsmath}
\usepackage{stackrel,amsfonts}
\usepackage{color}
\usepackage{mathabx}
\usepackage{cite,ulem}
\usepackage{slashed}
\usepackage{graphicx}
\usepackage[hidelinks]{hyperref}
\allowdisplaybreaks

\date{\today}
\def\lsim{\raise0.3ex\hbox{$\;<$\kern-0.75em\raise-1.1ex\hbox{$\sim\;$}}}
\def\gsim{\raise0.3ex\hbox{$\;>$\kern-0.75em\raise-1.1ex\hbox{$\sim\;$}}}
\title{Decays of a bino-like particle in the low-mass regime}
\author{Florian Domingo\footnote{domingo@th.physik.uni-bonn.de}, Herbi K. Dreiner\footnote{dreiner@uni-bonn.de}\\
{\small Bethe Center for Theoretical Physics \& Physikalisches Institut der 
Universit\"at Bonn,}\\ {\small Nu{\ss}allee 12, 53115 Bonn, Germany}}
\date{}

\begin{document}

\maketitle
\vspace{-6cm}\rightline{BONN-TH-2022-13}\vspace{6cm}

\begin{abstract}
We study the phenomenology associated with a light bino-like neutralino with mass $\lsim m_{\tau}$ in the context of the R-parity violating 
Minimal Supersymmetric Standard Model. This is a well-motivated example of scenarios producing potentially light and long-lived exotic 
particles, which might be testable in far-detector experiments, such as the FASER experiment at the Large Hadron Collider. A quantitative 
assessment of the discovery potential or the extraction of limits run through a detailed understanding of the interactions of the light exotic 
fermion with Standard Model matter, in particular, the hadronic sector. Here, we propose a systematic analysis of the decays of such a 
particle and proceed to a model-independent derivation of the low-energy effects, so that this formalism may be transposed to other 
UV-completions or even stand as an independent effective field theory. We then stress the diversity of the possible phenomenology and 
more specifically discuss the features associated with the R-parity violating supersymmetric framework, for example neutron-antineutron 
oscillations.
\end{abstract}

\tableofcontents

\section{Introduction}
While baryon and lepton numbers are classical symmetries of the Standard Model (SM) of particle physics, the observed baryon-antibaryon 
asymmetry of the Universe \cite{Steigman:1976ev,Canetti:2012zc} provides clear evidence of their violation \cite{Sakharov:1967dj}. In fact, 
attempts to unify the interactions of Yang-Mills type \cite{Georgi:1974sy,Raby:2006sk} generically break such symmetries in an explicit fashion, 
suggesting that their apparent conservation results from the accidental composition of the low-energy spectrum rather than from deep conceptual 
roots. On the other hand, matter stability --- see \textit{e.g.}~the reviews in Refs.\cite{Raby:2002wc,Nath:2006ut,Ellis:2019fwf} --- strongly limits the 
magnitude that baryon- and lepton-number violation can reach, pushing the corresponding dynamics to very high energy scales. This situation 
leaves open the question of whether baryon- and / or lepton-number violation can be detected in low-energy processes.

In supersymmetric (SUSY) extensions of the SM \cite{Nilles:1983ge,Martin:1997ns}, such as the Minimal SUSY SM (MSSM), conservation of 
baryon- and lepton-number usually appears as a consequence of the requirement of R-parity conservation \cite{Farrar:1978xj}. Yet, this property 
does not extend to non-renormalizable operators, as expected when taking the low-energy limit of \textit{e.g.}~a grand-unified ultraviolet (UV) completion. 
(See instead the discrete symmetry proton hexality \cite{Dreiner:2005rd,Dreiner:2007vp,Dreiner:2012ae} to cover also the non-renormalizable 
operators.) On the other hand, through the enforced stability of the lightest R-odd particle, R-parity has a deep impact on the phenomenology of the 
model, both in view of cosmological (thermal relics of the stable exotic particle) and collider observables (missing energy in SUSY decay chains). 
While this path has been copiously investigated in the course of the past forty years, its benefits with respect to matter stability may not outweigh the 
rich alternative, that of R-parity violation (RpV) \cite{Dreiner:1997uz,Barbier:2004ez,Dreiner:1991pe,Dercks:2017lfq}. In this latter case, explicit 
baryon- and / or lepton-number violating terms enter the superpotential, even though matter stability might still be protected by either of these discrete 
symmetries. For completeness, we remind here the form of the renormalizable RpV terms completing the superpotential of the MSSM 
\cite{Weinberg:1981wj}:
\begin{equation}
{W_{\text{RpV}}}=\mu_i \hat{H}_u \cdot \hat{L}_i+\frac{1}{2}\lambda_{ijk}\hat{L}_i\cdot \hat{L}_j
(\hat{E}^c)_k +\lambda'_{ijk}\hat{L}_i\cdot \hat{Q}_{j\alpha} (\hat{D}^c)_k^\alpha+\frac{1}{2}\lambda''_{ijk}
	\varepsilon_{\alpha\beta\gamma}(\hat{U}^{c})^{\alpha}_i(\hat{D}^{c})^{\beta}_j(\hat{D}^{c})^{\gamma}_k,
\label{eq:RpVSuperpotential}
\end{equation}
with $\hat{Q}$, $\hat{U}^c$, $\hat{D}^c$, $\hat{L}$, $\hat{E}^c$ denoting the usual matter chiral superfields \cite{Allanach:2003eb} and appearing in 
three generations while $\hat{H}_d$ and $\hat{H}_u$ represent the Higgs supermultiplets. Latin subscripts refer to flavor and Greek superscripts to 
color. The $SU(2)_L$ invariant product is indicated with `$\cdot$'. Without loss of generality, $\lambda_{ijk}$ is antisymmetric under exchange of its 
two first indices, and $\lambda''_{ijk}$, under exchange of its two last.

A remarkable scenario of the RpV MSSM \cite{Allanach:2003eb} involves a bino-dominated neutralino with mass in or below the GeV range 
\cite{Choudhury:1999tn,Dreiner:2003wh}. Indeed, as long as the bino mass is not correlated with other gaugino masses, such a light particle escapes 
constraints from laboratory experiments \cite{Dreiner:2009ic}, the main limit being set by the bound on the invisible $Z$-decay 
\cite{Choudhury:1999tn}, to which the bino would only contribute through subdominant higgsino components. At the LHC, searches with a missing 
energy signature \cite{Brandt:2008nq} are in general inapplicable to the RpV case \cite{Dreiner:1991pe,Allanach:2006st,Dercks:2017lfq}. Similarly, 
constraints from astrophysical \cite{Dreiner:2003wh,Kachelriess:2000dz,Dreiner:2013tja} or cosmological \cite{Cowsik:1972gh,Lee:1977ua,Dreiner:2011fp}
sources would only apply to a stable particle (or long-lived on cosmological scales). Search proposals for a light unstable or long-lived neutralino in 
heavy meson, $Z$ or tau decays have been suggested in 
\textit{e.g.}~Refs.~\cite{Choudhury:1999tn,Dedes:2001zia,Dreiner:2009er,deVries:2015mfw,Helo:2018qej,Dercks:2018wum,Wang:2019orr,Wang:2019xvx,Dreiner:2020qbi,Dey:2020juy}. 
Nevertheless, it is clear that a preliminary understanding of the phenomenology of the light fermion, especially its interplay with hadronic matter, is 
needed in any attempt at quantitatively interpreting the outcome of such searches. Common approximations, such as focusing on individual channels 
or trivializing mixing effects may be helpful for a qualitative comprehension of the phenomena at stake, but remain insufficient for actual tests and 
simulations.

In this paper, we specialize to the case of a light bino-dominated neutralino with mass below that of the $\tau$ lepton and investigate its decays in 
an unprejudiced fashion: given the considered field content, we build the most general low-energy effective field theory (EFT) mediating neutralino 
decays up to dimension $6$ terms. While we explicitly fix the short-distance physics through matching conditions with the RpV MSSM, the low-energy 
analysis should remain valid for a wider range of models predicting a light unstable exotic fermion, such as a right-handed neutrino, see for example
Refs.~\cite{Kling:2018wct,Helo:2018qej,DeVries:2020jbs}. We also resum QCD UV logarithms at leading order through the renormalization group 
equations (RGEs). In order to derive the decay widths, in particular in the hadronic and semi-leptonic channels, we call upon available `dictionaries' 
between partonic and hadronic physics. This systematic study definitely improves on earlier estimates and we collect our predictions in a Mathematica 
package, which is available from the authors upon request. 

In the following section, we construct and analyze the low-energy EFT that mediates the decays of the light neutralino, listing all possible operators of 
dimension $6$, matching them to the RpV MSSM and computing the associated RGEs. In section~3, we exploit conversion recipes between the 
partonic and hadronic languages, in order to derive realistic decay widths of the exotic fermion in terms of the Wilson coefficients of the EFT. 
Section~4 offers a brief discussion of possible production modes. In section~5, we demonstrate a few numerical applications before a short 
discussion and conclusion in Section~6. In eleven appendices we present some technical details.

\section{Low-energy Effective Field Theory for the Decays of a Light Neutralino}
We consider a neutralino state with mass $\lsim m_{\tau}$. At such energies, the active partonic degrees of freedom consist of the light quarks (in 
two-component left-handed spinor notation \cite{Dreiner:2008tw}) $q_i\in\{u,d,s\}$ and antiquarks $q^c_i\in\{u^c,d^c,s^c\}$, the light charged leptons 
$e_i\in\{e,\mu\}$ and antileptons $e^c_i\in\{e^c,\mu^c\}$, the neutrinos $\nu_i\in\{\nu_e,\nu_{\mu},\nu_{\tau}\}$, the photon $A_{\mu}$, the gluons 
$G^a_{\mu}$ and the light neutralino $\psi$. We assume that there exists no additional light `exotic' degree of freedom (\textit{e.g.}~axinos 
\cite{Dreiner:2014eda}, gravitinos \cite{Takayama:2000uz}). For a left-handed spinor $f$, we associate the right-handed conjugate spinor $\bar{f}$. 
For compactness, we will also employ the four-component 
spinor notation, with generically $F\equiv(f,\bar{f^c})$, and $Q_i\equiv(q_i,\bar{q}^c_i)^T$, and $E_i\equiv(e_i,\bar{e}^c_i)^T$ representing the Dirac 
spinors for the quarks and charged leptons, $N_i\equiv(\nu_i,\bar{\nu}_i)^T$ and $\Psi\equiv(\psi,\bar{\psi})^T$  representing the Majorana spinors for
the neutrinos and light neutralino, while $P_{L,R}$ denote the chiral projectors. The relevant symmetries are the gauged electromagnetism and 
chromodynamics $U(1)_{\text{em}}\times SU(3)_c$. The light neutralino state is (by definition) a color singlet; limits on light charged $SU(2)_L$ 
partners, as well as on invisible / exotic $Z$ decays, additionally restrict realistic choices to a dominantly electroweak singlet state. With the MSSM 
field content, $\psi$ is thus necessarily bino-like \cite{Choudhury:1999tn,Dreiner:2009ic}.

The most general renormalizable Lagrangian density at low energy reads:
\begin{equation}\label{eq:renlag}
{\cal L}_{\text{ren.}}=-\frac{1}{4}F^{\mu\nu}F_{\mu\nu}-\frac{1}{4}G^{a\,\mu\nu}G^a_{\mu\nu}+\!\!\!\sum_{F=Q_i,E_i}\!\!\!\bar{F}\left[i
\slashed{D}-m_F\right]F+\!\!\sum_{F=N_i,\Psi}\!\frac{1}{2}\bar{F}\left[i\slashed{\partial}-m_f\right]F-m_{\psi\nu_i}\bar{\Psi}N_i\,,
\end{equation}
where $F_{\mu\nu}$ and $G^a_{\mu\nu}$ represent the field-strength tensors associated with the photon and gluon fields, $D_{\mu}$ is the covariant 
derivative, and $m_f$ is the mass of the fermion species $F$. The mass term $m_{\psi\nu_i}$ induces mixing between the neutralino and neutrino 
fields. It can be absorbed within the definition of $\psi$, through diagonalization of the mass-matrix. The $\nu_i$ are hence no longer electroweak 
gauge eigenstates, but light mass eigenstates of the neutralino-neutrino sector.  So we set $m_{\psi\nu_i}\stackrel[]{!}{=}0$ 
\cite{Hall:1983id,Dreiner:2003hw}. We also systematically neglect the neutrino masses $m_{\nu_i}\ll m_{\psi},m_e$ below. Without interactions at the 
renormalizable level, the light neutralino cannot decay via this Lagrangian. The short-distance effects governing this phenomenon must be introduced 
through operators of higher dimension.


\subsection{Dimension  $6$ operators for neutralino decay}
Given the field content and the symmetries, one can build the following dimension ($5$ and) $6$ operators (together with their hermitian 
conjugates) involving one neutralino and thus possibly mediating its decays. (We use here the more compact four-component notation. The lengthier 
but for computations more convenient two-component expressions are given in Appendix~\ref{ap:opin2comp}.)
\begin{itemize}
\item Electromagnetic dipoles ($i=1,2,3$): 
\begin{equation}\label{eq:elmdipop}
{\cal E}_i\equiv\frac{e}{32\pi^2}(\bar{\Psi}\Sigma^{\mu\nu}P_L N_i)F_{\mu\nu}\,.
\end{equation}
\item Leptonic operators (the chirality indices $J,K$ may take the values $L,R$; the indices $i,j,k$ correspond to the fermion generation):
\begin{align}\label{eq:leptop}
&\widetilde{\cal N}_{ijk}\equiv(\bar{\Psi}P_LN_i)(\bar{N}^c_jP_LN_k)\,,\nonumber\\
&\null\hspace{1.7cm}(i,j,k)\in\{(1,2,2),(1,3,3),(1,2,3),(2,1,1),(2,3,3),(2,1,3),(3,1,1),(3,2,2)\}\,,\nonumber\\[1.6mm]
&{\cal N}_{ijk}\equiv(\bar{\Psi}\gamma^{\mu}P_LN_i)(\bar{N}_j\gamma_{\mu}P_LN_k)\,,\ \ \ (i\leq k)\,,\\[1.6mm]
&{\cal S}_{ijk}^{\nu e\,K}\equiv(\bar{\Psi}P_LN_i)(\bar{E}_jP_KE_k)\,,\,\qquad {\cal V}_{ijk}^{\nu e\,K}\equiv(\bar{\Psi}\gamma^{\mu}P_LN_i)(\bar{E}
_j\gamma_{\mu}P_KE_k)\,,\, \nonumber\\[1.6mm]
&{\cal T}_{ijk}^{\nu e}\equiv\frac{1}{4}(\bar{\Psi}\Sigma^{\mu\nu}P_LN_i)(\bar{E}_j\Sigma_{\mu\nu}P_LE_k).\nonumber
\end{align}
\item Semi-leptonic operators ($q=u,d$ separately for the operators involving a neutrino):
\begin{align}\label{eq:seleptop}
&{\cal S}_{ijk}^{eq\,JK}\equiv(\bar{\Psi}P_JE_i)(\bar{D}_jP_KU_k)\,,& & {\cal S}_{ijk}^{\nu q\,K}\equiv(\bar{\Psi}P_LN_i)(\bar{Q}_jP_KQ_k)\,,
\nonumber\\
&{\cal V}_{ijk}^{eq\,JK}\equiv(\bar{\Psi}\gamma^{\mu}P_JE_i)(\bar{D}_j\gamma_{\mu}P_KU_k)\,,& & {\cal V}_{ijk}^{\nu q\,K}\equiv(\bar{\Psi}
\gamma^{\mu}P_LN_i)(\bar{Q}_j\gamma_{\mu}P_KQ_k)\,,\\
&{\cal T}_{ijk}^{eq\,J}\equiv\frac{1}{4}(\bar{\Psi}\Sigma^{\mu\nu}P_JE_i)(\bar{D}_j\Sigma_{\mu\nu}P_JU_k)\,,& & {\cal T}^{\nu q}_{ijk}\equiv\frac
{1}{4}(\bar{\Psi}\Sigma^{\mu\nu}P_LN_i)(\bar{Q}_j\Sigma_{\mu\nu}P_LQ_k)\,.\nonumber
\end{align}
\item Hadronic operators (greek indices refer to color; $F^c$ is the charged-conjugate of the fermion $F$ in four-component notation):
\begin{align}\label{eq:hadop}
&{\cal H}_{ijk}^{JK}\equiv\varepsilon_{\alpha\beta\gamma}(\bar{\Psi}P_JU^{c\,\alpha}_i)(\bar{D}^{\ \beta}_jP_KD^{c\,\gamma}_k)\ \ (j<k)\,,\nonumber\\
&\widetilde{\cal H}_{ijk}^{JK}\equiv\varepsilon_{\alpha\beta\gamma}(\bar{\Psi}P_JD^{c\,\beta}_j)(\bar{U}^{\ \alpha}_iP_KD^{c\,\gamma}_k)\,.
\end{align}
The identity $\widetilde{\cal H}_{ijk}^{LL}-\widetilde{\cal H}_{ikj}^{LL}+{\cal H}_{ijk}^{LL}=0$ (and similarly for $LL\to RR$) implies some 
redundancy (which does not matter as long as the corresponding contributions are not double-counted). 
\end{itemize}

All other combinations of the light fields of dimension $6$ can be written in terms of the previous ones, through the two-component spinor 
(Fierz) identities  \cite{Dreiner:2008tw} applied to the expressions in Appendix~\ref{ap:opin2comp}, \textit{e.g.}:
\begin{align}
&(f_1f_2)(f_3f_4)+(f_1f_3)(f_2f_4)+(f_1f_4)(f_2f_3)=0\,,\nonumber\\
&(\bar{f_1}\bar{\sigma}^{\mu}f_2)(\bar{f_3}\bar{\sigma}_{\mu}f_4)=2(\bar{f_1}\bar{f_3})(f_2f_4)\,,\\
&(\bar{f_1}\bar{\sigma}^{\mu}f_2)(f_3\sigma_{\mu}\bar{f_4})=-2(\bar{f_1}\bar{f_4})(f_2f_3)\,,\nonumber\\
&(f_1\sigma^{\mu\nu}f_2)(f_3\sigma_{\mu\nu}f_4)=-2(f_1f_4)(f_2f_3)-(f_1f_2)(f_3f_4)=2(f_1f_3)(f_2f_4)+(f_1f_2)(f_3f_4)\,.\nonumber
\end{align}

Forgetting about neutralino-neutrino mixing, the electromagnetic dipole, leptonic and semi-leptonic operators violate lepton number, by three 
units for $\widetilde{\cal N}$ and one unit for the others. The hadronic operators violate baryon number by one unit. We define the Wilson 
coefficients $C[\Omega]$ associated with any of these operators $\Omega$ and add the resulting terms to the Lagrangian density of 
Eq.~(\ref{eq:renlag}), hence obtaining our effective field theory with non-renormalizable operators:
\begin{equation}
{\cal L}^{\text{EFT}}_{\text{dim 6}}={\cal L}_{\text{ren.}}+\sum_{\Omega}C[\Omega]\,\Omega\,.
\end{equation}

\subsection{Matching to the RpV MSSM\label{sec:matching}}
For almost any model leading to the considered spectrum in the GeV range, the low-energy dynamics associated with the decays of the light exotic 
fermion field $\psi$ can be described at leading order by the above EFT.\footnote{The exception is if specific counting rules suppress the dimension $6$ 
operators with respect to higher-order ones.} In this paper, however, we choose to focus on a specific type of UV completion, that associated with the 
RpV MSSM. Then, the short-distance effects contributing to the non-renormalizable operators depend on their mediation by heavy SUSY particles, typically 
sfermions, or electroweak gauge bosons. We denote the scale corresponding to this UV dynamics as $\mu_0\sim M_{\text{SUSY}}$ and assume that it 
is not very far from the electroweak scale, \textit{i.e.} in the TeV range. This scale is the high-energy threshold at which the EFT should be matched to the 
RpV MSSM through the identification of scattering amplitudes (although further intermediate thresholds can be considered). Below, we restrict ourselves to 
a matching at leading order.

Two types of topologies are involved in mediating the amplitudes that contribute to the lepton- or baryon-number-violating operators leading to neutralino 
decay. The vector topology employs the electroweak gauge bosons $W$ and $Z$ as mediators and relies exclusively on lepton-electrowikino mixing to 
generate lepton-number violating amplitudes. On the other hand, the scalar topology proceeds through the exchange of squarks or Higgs-slepton 
admixtures. Mediators of Higgs-type would convey lepton-number violation through lepton-electrowikino mixing, similarly to the vector topology, and / or 
through Higgs-slepton mixing. In the case of mediators of sfermion-type, lepton- or baryon-number violation emerges through the trilinear RpV couplings.

For commodity, we introduce the following notation for couplings between a vector $V_\mu$ or a scalar $S$ and two fermions $F_i$, $F_j$:
\begin{equation}
\label{eq:couplingdef}
{\cal L}_{\text{MSSM}}\ni V_{\mu}\,\bar{F}_i\,\gamma^{\mu}g_{L,R}^{Vf_i^cf_j}P_{L,R}\,F_j+S\,\bar{F}_i\,g_{L,R}^{Sf_i^cf_j}P_{L,R}\,F_j\,.
\end{equation}
Employing the usual algebra, one can derive the following identities:
\begin{eqnarray}
g_{L,R}^{Vf_i^cf_j}&=&\left(g_{L,R}^{V^*f_j^cf_i}\right)^*=-g_{R,L}^{Vf_jf_i^c}=-\left(g_{R,L}^{V^*f_if_j^c}\right)^*\,,\\[2mm]
g_{L,R}^{Sf_i^cf_j}&=&\left(g_{R,L}^{S^*f_j^cf_i}\right)^*=g_{L,R}^{Sf_jf_i^c}=\left(g_{R,L}^{S^*f_if_j^c}\right)^*\;.
\end{eqnarray}
These rules (for uncolored fields) suffer an exception in the case of non-trivial color products, as those generated by the $\lambda''$ couplings, see also
Ref.~\cite{Dreiner:1999qz}. We list the relevant couplings in Appendix~\ref{ap:couplingdef}. Here, we stress \cite{Allanach:2003eb} that the neutrinos $\nu_i$ 
and the light neutralino $\psi$ are just special (light) occurrences of the neutralino mass eigenstates $\chi^0_k$, while the charged leptons $e_i$ and 
$e_i^c$ also correspond more generally to light chargino mass eigenstates $\chi_k^{\mp}$. Similarly, Higgs-slepton mixing leads to a collection of real neutral 
(pseudo-)scalars $S^0_{p}$, built out of the neutral Higgs and sneutrino components, as well as to charged Higgs-slepton admixtures $S_{p}^{\pm}$. 
Squarks are denoted as $\tilde{Q}_p\in\{\tilde{U}_p,\tilde{D}_p\}$ and in general involve left-right as well as generation mixing. The mixing among fields 
of the RpV MSSM is shortly described in Appendix~\ref{ap:mixing}.

Let us first consider the decay $\psi\to\bar{\nu}_i\nu_j\bar{\nu}_k$  ($i\leq k$): the amplitude in the EFT reads (where $\breve
{\Omega}$ represents the spinor structure resulting from the operator $\Omega$):
\begin{equation}
	{\cal A}^{\text{EFT}}[\psi\to\bar{\nu}_i\nu_j\bar{\nu}_k]=i(1+\delta_{ik})\,C[{\cal N}_{ijk}](\mu_0)\,\breve{\cal N}_{ijk}\,.
\end{equation}
The corresponding amplitude in the RpV MSSM (neglecting Yukawa couplings) is mediated either by a $Z$-boson or by a neutral 
scalar, and reads (with implicit summation):
\begin{equation}
	{\cal A}^{\text{MSSM}}[\psi\to\bar{\nu}_i\nu_j\bar{\nu}_k]=i\left\{\frac{1}{2m_{S_p^0}^2} g_R^{S_p^0\nu_j^c\psi}g_L^{S_p^0\nu_i\nu_k}-\frac{1}
	{M_Z^2} \left[g_L^{Z\psi \nu_i}g_L^{Z\nu_j^c\nu_k}+g_L^{Z\psi \nu_k}g_L^{Z\nu_j^c\nu_i}\right]\right\}\breve{\cal N}_{ijk}\,.
\end{equation}
Identifying both amplitudes, we deduce:
\begin{equation}
C[{\cal N}_{ijk}](\mu_0)\stackrel[]{!}{=}\frac{1}{1+\delta_{ik}}\left\{\frac{1}{2m_{S_p^0}^2} g_R^{S_p^0\nu_j^c\psi}g_L^{S_p^0\nu_i\nu_k}-\frac{1}{M_Z^2} 
\left[g_L^{Z\psi \nu_i}g_L^{Z\nu_j^c\nu_k}+g_L^{Z\psi \nu_k}g_L^{Z\nu_j^c\nu_i}\right]\right\}\,.
\end{equation}

Performing the same operation for the decay $\psi\to\bar{\nu}_i\bar{\nu}_j\bar{\nu}_k$ (for the same combinations of flavor indices as those of the 
operators $\widetilde{\cal N}_{ijk}$ in Eq.~(\ref{eq:leptop}), and in particular $i\neq j,k$, $j\leq k$), we can determine: 
\begin{equation}
C[\widetilde{\cal N}_{ijk}](\mu_0)\stackrel[]{!}{=}\frac{1}{(1+\delta_{jk})m^2_{S_p^0}}\left[g_L^{S_p^0\psi\nu_i}g_L^{S_p^0\nu_j\nu_k}-g_L^{S_p^0\psi\nu_k}
g_L^{S_p^0\nu_j\nu_i}\right]\,.
\end{equation}
Given that the corresponding operators violate lepton-number by three units, they require a high degree of mixing and may thus be seen as typically 
suppressed.

Turning to the decay $\psi\to\bar{\nu}_ie_j\bar{e}_k$, where both vector and scalar mediators contribute, we can determine the remaining Wilson 
coefficients of leptonic type:
\begin{align}\label{eq:matchnulep}
&C[{\cal S}^{\nu e\,L}_{ijk}](\mu_0)\stackrel[]{!}{=}\tfrac{1}{m_{S_p^0}^2}g_L^{S_p^0\psi\nu_i}g_L^{S_p^0e_j^ce_k}-\tfrac{1}{2m_{S_p^{\pm}}^2}\big[g_L^{S_p^+\psi e_k}g_L^{S_p^-e_j^c\nu_i}+g_L^{S_p^-\psi e_j^c}g_L^{S_p^+e_k\nu_i}\big]\,,\nonumber\\
&C[{\cal S}^{\nu e\,R}_{ijk}](\mu_0)\stackrel[]{!}{=}\tfrac{2}{M_W^2}g_L^{W\psi e_k}g_L^{We_j^c\nu_i}+\tfrac{1}{m_{S_p^0}^2}g_L^{S_p^0\psi \nu_i}g_R^{S_p^0e_j^ce_k}\,,\nonumber\\
&C[{\cal V}^{\nu e\,L}_{ijk}](\mu_0)\stackrel[]{!}{=}-\tfrac{1}{M_Z^2}g_L^{Z\psi \nu_i}g_L^{Ze_j^ce_k}-\tfrac{1}{M_W^2}g_L^{W\psi e_k}g_L^{We_j^c\nu_i}+\tfrac{1}{2m_{S_p^{\pm}}^2}g_R^{S_p^-\psi e_j^c}g_L^{S_p^+e_k\nu_i}\,,\nonumber\\
&C[{\cal V}^{\nu e\,R}_{ijk}](\mu_0)\stackrel[]{!}{=}-\tfrac{1}{M_Z^2}g_L^{Z\psi \nu_i}g_R^{Ze_j^ce_k}-\tfrac{1}{2m_{S_p^{\pm}}^2}g_R^{S_p^+\psi e_k}g_L^{S_p^-e_j^c\nu_i}\,,\\
&C[{\cal T}^{\nu e}_{ijk}](\mu_0)\stackrel[]{!}{=}\tfrac{1}{2m_{S_p^{\pm}}^2}\Big[g_L^{S_p^-\psi e_j^c}g_L^{S_p^+e_k\nu_i}-g_L^{S_p^+\psi e_k}g_L^{S_p^-e_j^c\nu_i}\Big]\,.\nonumber
\end{align}

The decay amplitudes $\psi\to\bar{\nu}_iq_j\bar{q}_k$ ($q=u,d$) and $\psi\to\bar{e}_iu_j\bar{d}_k$ allow the identification of the Wilson coefficients 
of the semi-leptonic operators:
\begin{align}\label{eq:matchselep}
&C[{\cal S}^{\nu q\,L}_{ijk}](\mu_0)\stackrel[]{!}{=}\tfrac{1}{m_{S_p^0}^2}g_L^{S_p^0\psi\nu_i}g_L^{S_p^0q_j^cq_k}-\tfrac{1}{2m^2_{\tilde{Q}_p}}
\Big[g_L^{\tilde{Q}_p^*\psi q_k}g_L^{\tilde{Q}_pq^c_j\nu_i}+g_L^{\tilde{Q}_p\psi q_j^c}g_L^{\tilde{Q}_p^*q_k\nu_i}\Big]\,,\nonumber\\
&C[{\cal S}^{\nu q\,R}_{ijk}](\mu_0)\stackrel[]{!}{=}\tfrac{1}{m_{S_p^0}^2}g_L^{S_p^0\psi\nu_i}g_R^{S_p^0q_j^cq_k}\,,\nonumber\\
&C[{\cal V}^{\nu q\,L}_{ijk}](\mu_0)\stackrel[]{!}{=}-\tfrac{1}{M_Z^2}g_L^{Z\psi \nu_i}g_L^{Zq_j^cq_k}+\tfrac{1}{2m^2_{\tilde{Q}_p}}g_R^{\tilde{Q}_p
\psi q^c_j}g_L^{\tilde{Q}_p^*q_k\nu_i}\,,\nonumber\\
&C[{\cal V}^{\nu q\,R}_{ijk}](\mu_0)\stackrel[]{!}{=}-\tfrac{1}{M_Z^2}g_L^{Z\psi \nu_i}g_R^{Zq_j^cq_k}-\tfrac{1}{2m^2_{\tilde{Q}_p}}g_R^{\tilde{Q}_p^*
\psi q_k}g_L^{\tilde{Q}_pq^c_j\nu_i}\,,\\
&C[{\cal T}^{\nu q}_{ijk}](\mu_0)\stackrel[]{!}{=}\tfrac{1}{2m^2_{\tilde{Q}_p}}\Big[g_L^{\tilde{Q}_p\psi q_j^c}g_L^{\tilde{Q}_p^*q_k\nu_i}-g_L^{\tilde{Q}
_p^*\psi q_k}g_L^{\tilde{Q}_pq^c_j\nu_i}\Big]\,,\nonumber\\
&C[{\cal S}^{eq\,LL}_{ijk}](\mu_0)\stackrel[]{!}{=}\tfrac{1}{m^2_{S_p^{\pm}}}g_L^{S_p^+\psi e_i}g_L^{S_p^-d_j^cu_k}-\tfrac{1}{2m^2_{\tilde{U}_p}}g_L^{\tilde{U}_p^*\psi u_k}g_L^{\tilde{U}_pd_j^ce_i}-\tfrac{1}{2m^2_{\tilde{D}_p}}g_L^{\tilde{D}_p\psi d_j^c}g_L^{\tilde{D}^*_pu_ke_i}\,,\nonumber\\
&C[{\cal S}^{eq\,LR}_{ijk}](\mu_0)\stackrel[]{!}{=}\tfrac{1}{m^2_{S_p^{\pm}}}g_L^{S_p^+\psi e_i}g_R^{S_p^-d_j^cu_k}\,,\nonumber\\
&C[{\cal V}^{eq\,LL}_{ijk}](\mu_0)\stackrel[]{!}{=}-\tfrac{1}{M_W^2}g_L^{W\psi e_i}g_L^{Wd_j^cu_k}+\tfrac{1}{2m^2_{\tilde{D}_p}}g_R^{\tilde{D}_p\psi d_j^c}
g_L^{\tilde{D}^*_pu_ke_i}\,,\nonumber\\
&C[{\cal V}^{eq\,LR}_{ijk}](\mu_0)\stackrel[]{!}{=}-\tfrac{1}{M_W^2}g_L^{W\psi e_i}g_R^{Wd_j^cu_k}-\tfrac{1}{2m^2_{\tilde{U}_p}}g_R^{\tilde{U}_p^*\psi u_k}
g_L^{\tilde{U}_pd_j^ce_i}\,,\nonumber\\
&C[{\cal T}^{eq\,L}_{ijk}](\mu_0)\stackrel[]{!}{=}-\tfrac{1}{2m^2_{\tilde{U}_p}}g_L^{\tilde{U}_p^*\psi u_k}g_L^{\tilde{U}_pd_j^ce_i}+\tfrac{1}{2m^2_{\tilde{D}_p}}
g_L^{\tilde{D}_p\psi d_j^c}g_L^{\tilde{D}^*_pu_ke_i}\,,\nonumber\\
&C[{\cal S}^{eq\,RL}_{ijk}](\mu_0)\stackrel[]{!}{=}\tfrac{1}{m^2_{S_p^{\pm}}}g_R^{S_p^+\psi e_i}g_L^{S_p^-d_j^cu_k}\,,\nonumber\\
&C[{\cal S}^{eq\,RR}_{ijk}](\mu_0)\stackrel[]{!}{=}\tfrac{1}{m^2_{S_p^{\pm}}}g_R^{S_p^+\psi e_i}g_R^{S_p^-d_j^cu_k}-\tfrac{1}{2m^2_{\tilde{U}
_p}}g_R^{\tilde{U}_p^*\psi u_k}g_R^{\tilde{U}_pd_j^ce_i}-\tfrac{1}{2m^2_{\tilde{D}_p}}g_R^{\tilde{D}_p\psi d_j^c}g_R^{\tilde{D}^*_pu_ke_i}\,,\nonumber\\
&C[{\cal V}^{eq\,RL}_{ijk}](\mu_0)\stackrel[]{!}{=}-\tfrac{1}{M_W^2}g_R^{W\psi e_i}g_L^{Wd_j^cu_k}-\tfrac{1}{2m^2_{\tilde{U}_p}}g_L^
{\tilde{U}_p^*\psi u_k}g_R^{\tilde{U}_pd_j^ce_i}\,,\nonumber\\
&C[{\cal V}^{eq\,RR}_{ijk}](\mu_0)\stackrel[]{!}{=}-\tfrac{1}{M_W^2}g_R^{W\psi e_i}g_R^{Wd_j^cu_k}+\tfrac{1}{2m^2_{\tilde{D}_p}}g_L
^{\tilde{D}_p\psi d_j^c}g_R^{\tilde{D}^*_pu_ke_i}\,,\nonumber\\
&C[{\cal T}^{eq\,R}_{ijk}](\mu_0)\stackrel[]{!}{=}\tfrac{1}{2m^2_{\tilde{D}_p}}g_R^{\tilde{D}_p\psi d_j^c}g_R^{\tilde{D}^*_pu_ke_i}-\tfrac{1}{2m^2_{\tilde{U}_p}}
g_R^{\tilde{U}_p^*\psi u_k}g_R^{\tilde{U}_pd_j^ce_i}\,.\nonumber
\end{align}

We already derived the Wilson coefficients of the hadronic operators in Ref.\,\cite{Chamoun:2020aft}. For completeness, with $j\leq k$ (by assumption):
\begin{align}\label{eq:matchhad}
&C[{\cal H}^{LL}_{ijk}](\mu_0)\stackrel[]{!}{=}\tfrac{1}{m_{\tilde{U}_p}^2}g_L^{\tilde{U}_p\psi u^c_i}g_L^{\tilde{U}_p^*d_j^cd_k^c}\,,& &C[{\cal H}^{RL}_{ijk}]
(\mu_0)\stackrel[]{!}{=}\tfrac{1}{m_{\tilde{U}_p}^2}g_R^{\tilde{U}_p\psi u^c_i}g_L^{\tilde{U}_p^*d_j^cd_k^c}\,,\nonumber\\
&C[\widetilde{\cal H}^{LL}_{ijk}](\mu_0)\stackrel[]{!}{=}\tfrac{1}{m_{\tilde{D}_p}^2}g_L^{\tilde{D}_p\psi d^c_j}g_L^{\tilde{D}_p^*u_i^cd_k^c}\,,& &C[\widetilde
{\cal H}^{RL}_{ijk}](\mu_0)\stackrel[]{!}{=}\tfrac{1}{m_{\tilde{D}_p}^2}g_R^{\tilde{D}_p\psi d^c_j}g_L^{\tilde{D}_p^*u_i^cd_k^c}\,,\\
&C[\widetilde{\cal H}^{LL}_{ikj}](\mu_0)\stackrel[]{!}{=}\tfrac{1}{m_{\tilde{D}_p}^2}g_L^{\tilde{D}_p\psi d^c_k}g_L^{\tilde{D}_p^*u_i^cd_j^c}\,,& &C[\widetilde
{\cal H}^{RL}_{ikj}](\mu_0)\stackrel[]{!}{=}\tfrac{1}{m_{\tilde{D}_p}^2}g_R^{\tilde{D}_p\psi d^c_k}g_L^{\tilde{D}_p^*u_i^cd_j^c}\,.\nonumber
\end{align}
As a consequence of the redundancy $\widetilde{\cal H}^{LL}_{ikj}=\widetilde{\cal H}^{LL}_{ijk}+{\cal H}^{LL}_{ijk}$ that we mentioned earlier, $C[\widetilde
{\cal H}^{LL}_{ikj}](\mu_0)$ could be distributed onto the other Wilson coefficients. In addition, $g_R^{\tilde{U}_p^*d_j^cd_k^c}\stackrel[]{!}{=}0\stackrel[]{!}
{=}g_R^{\tilde{D}_p^*u_i^cd_k^c}$ in the RpV MSSM implies that the remaining effective couplings vanish at tree-level order: $C[{\cal H}^{LR}_{ijk}](\mu_0)
=0=C[{\cal H}^{RR}_{ijk}](\mu_0)=C[\widetilde{\cal H}^{LR}_{ijk}](\mu_0)=C[\widetilde{\cal H}^{RR}_{ijk}](\mu_0)$.

The neutralino decay into antineutrino and photon, $\psi\to\gamma\bar{\nu}_i$, is a one-loop process in the RpV MSSM and it might be misleading to include 
it in this analysis while other channels are considered at tree-level. Nevertheless, it was calculated under various approximations in 
\textit{e.g.}~Refs.~\cite{Hall:1983id,Dawson:1985vr,Mukhopadhyaya:1999gy} and can also be viewed as a radiative neutralino decay \cite{Haber:1988px}. Let us 
parametrize the decay amplitude as follows:
\begin{equation}
{\cal A}^{\text{MSSM}}[\psi(p_{\psi})\to\gamma(p_{\gamma})\bar{\nu}_i(p_{\nu_i})]\equiv-\frac{e}{16\pi^2}(p_{\gamma})_{\mu}\varepsilon^*(p_{\gamma})_
{\nu}\big(\bar{v}_{\psi}(p_{\psi})\Sigma^{\mu\nu}P_Lv_{\nu_i}(p_{\nu_i})\big)C^{\text{MSSM}}[{\cal E}_i]\,.
\end{equation}
We compute $C^{\text{MSSM}}[{\cal E}_i]$, \textit{cf.} Eq.~(\ref{eq:elmdipop}), in the Feynman gauge. At one-loop order, it involves vertex diagrams of 
$W$-bosons (or charged a Goldstone) and chargino/leptons ($\chi_j^{\pm}$), as well as loops of scalar fields ($S$) with an associated fermion ($f$). The 
$Z-\gamma$ mixing diagrams ensure UV-finiteness with the triangle diagrams. Combining all contributions,
\begin{align}\label{eq:CMSSMelm}
C^{\text{MSSM}}[{\cal E}_i]=&\frac{1}{M_W^2}\Big[m_{\psi}\big(g_L^{W^+\psi\chi_j^-}g_L^{W^-\chi_j^+\nu}-g_L^{W^-\psi\chi_j^+}g_L^{W^+\chi_j^-\nu}\big){\cal I}^V_L\big(\tfrac{m^2_{\psi}}{M_W^2},\tfrac{m^2_{\chi_j^{\pm}}}{M_W^2}\big)\nonumber\\
&\null\hspace{1.5cm}+m_{\chi_j}\big(g_R^{W^+\psi\chi_j^-}g_L^{W^-\chi_j^+\nu}-g_R^{W^-\psi\chi_j^+}g_L^{W^+\chi_j^-\nu}\big){\cal I}^V_R\big(\tfrac{m^2_{\psi}}{M_W^2},\tfrac{m^2_{\chi_j^{\pm}}}{M_W^2}\big)\Big]\\
&+\frac{Q_f}{m_S^2}\Big[m_f \big(g_L^{S^*\psi f}g_L^{Sf^c\nu}-g_L^{S\psi f^c}g_L^{S^*f\nu}\big){\cal I}^S_L\big(\tfrac{m^2_{\psi}}{m_S^2},\tfrac{m^2_{f}}{m_S^2}\big)\nonumber\\
&\hspace{3.6cm}+m_{\psi}\big(g_R^{S^*\psi f}g_L^{Sf^c\nu}-g_R^{S\psi f^c}g_L^{S^*f\nu}\big){\cal I}^S_R\big(\tfrac{m^2_{\psi}}{m_S^2},\tfrac{m^2_{f}}{m_S^2}\big)\Big]\,.\nonumber
\end{align}
The implicit sum over $\chi_j^{\pm}$ runs over the five corresponding mass eigenstates. For the scalar diagrams, contained in the second set of brackets,
the implicit sum runs over all $(S,f,Q_f)$'s in the following list: $\big\{(S_m^-,\chi_j^-,-1), (\tilde{U}_m,u,\tfrac{2}{3}), (\tilde{D}_m,d,-\tfrac{1}{3})\big\}$, including color 
factors for the quark/squark contributions. The loop integrals ${\cal I}^{V,S}_{L,R}$ are provided in Appendix~\ref{ap:loopfunc}.

In the EFT, the amplitude corresponding to the $\psi\to\gamma\bar{\nu}_i$ transition receives contributions from ${\cal E}_i$ at tree-level, but also (at 
competitive order) from one-loop diagrams involving the operators ${\cal T}^{\nu f\,(L)}_{ijj}$, $f=e,q$, $j=1,2,3$. In particular, it is necessary to decouple 
step by step the heavy fermions ($t,b$) that are integrated out at low-energy but not at the scale $\mu_0$. The amplitude reads:
\begin{multline}\label{eq:nngcorr}
{\cal A}^{\text{EFT}}[\psi(p_{\psi})\to\gamma(p_{\gamma})\bar{\nu}_i(p_{\nu_i})]=-\frac{e}{16\pi^2}(p_{\gamma})_{\mu}\varepsilon^*(p_{\gamma})_{\nu}\big(
\bar{v}_{\psi}(p_{\psi})\Sigma^{\mu\nu}P_Lv_{\nu_i}(p_{\nu_i})\big)\\
\times\Bigg\{C[{\cal E}_i]+\delta C[{\cal E}_i]+\sum_{f_j}2 Q_{f_j} N_c^{f_j} m_{f_j} C[{\cal T}^{\nu f\,(L)}_{ijj}]\bigg(\frac{1}{2-\tfrac{D}{2}}-\gamma_E+\ln
\frac{4\pi\mu_0^2}{m_{f_j}^2}\bigg)\Bigg\}(\mu_0)\,,
\end{multline}
where $D$ represents the dimension used for the regularization of the loop integrals, $\gamma_E$ is the Euler-Mascheroni constant, $\mu_0$ is the 
renormalization scale (as already implied), $Q_{f_j}$ is the electric charge of the fermion species $f_j$, $N_c^{f_j}=1$ or $3$ for color singlets and triplets, 
respectively. $\delta C[{\cal E}_i]$ denotes the counterterm associated with ${\cal E}_i$, which is needed for the renormalization of the amplitude. Therefore, 
the matching condition connecting $C[{\cal E}_i](\mu_0)$ and $C^{\text{MSSM}}[{\cal E}_i]$ is not completely trivial, as the ${\cal T}^{\nu f\,(L)}_{ijj}$ continue 
to contribute below $\mu_0$: the renormalization of the EFT will clarify the meaning of this decomposition for the bookkeeping of logarithmic corrections.

Under the assumption of subdominant mixing and negligible Yukawa couplings, it is possible to cast the contributions to the Wilson coefficients in a much 
simpler form, where trilinear RpV couplings dominate the phenomenology associated with the light bino-like neutralino: see Appendix~\ref{ap:simpWilsonC}. 
Comparison with Refs.\,\cite{deVries:2015mfw,Dey:2020juy} for the contributions to semi-leptonic operators revealed probable issues in these references with 
effects of tensor type.

\subsection{Renormalization group evolution under QCD}
While the short-distance effects that generate the contributions to the operators mediating the neutralino decay are characterized by the high-energy scale 
$\mu_0$, the physical transition itself (the neutralino decay) takes place at a low energy determined by $m_{\psi}$. As perturbative QCD fails at low-energy 
scales, we will employ an infrared scale $\mu\gsim m_c$. Given the large hierarchy between $\mu_0$ and $\mu$, large QCD logarithms of UV type are 
expected to develop and it thus appears meaningful to resum them through the renormalization group evolution within the EFT. To this end, 
we renormalize all the parameters of the EFT in the $\overline{\text{MS}}$ scheme (following the traditional procedure, described in 
\textit{e.g.}~Ref.~\cite{Chalons:2013mya} in the context of $B$ decays), thus deriving the Callan-Symanzik equations associated with the Wilson coefficients at 
leading logarithmic order in QCD.

For commodity, we introduce the strong coupling constant $\alpha_s\equiv\tfrac{g_s^2}{4\pi}$, the coefficient of the quadratic Casimir operator $C_2(3)\equiv
\tfrac{4}{3}$, the leading contribution to the QCD beta function $\beta_0[n_f]\equiv\tfrac{11N_c-2n_f}{3}$, with $N_c\equiv3$ the number of colors, and $n_f$ 
the number of active quark flavors. We also employ the following notation for the ratios of the $\alpha_s$'s at various scales: $\eta_{\mu}\equiv\tfrac{\alpha_s
(\mu)}{\alpha_s(m_b)}$, $\eta_{b}\equiv\tfrac{\alpha_s(m_b)}{\alpha_s(m_t)}$, and $\eta_{t}\equiv\tfrac{\alpha_s(m_t)}{\alpha_s(\mu_0)}$. The quark masses 
$m_{t,b}$ appearing in these equations correspond to decoupling thresholds and should be understood as the $\overline{\text{MS}}$ masses $m_{t,b}(m_{t,b})$.
The RGE's associated with the hadronic operators were previously derived in Ref.~\cite{Abbott:1980zj} and lead to the following evolution (for $m_c\lsim\mu
<m_b$):
\begin{equation}
C[{\cal H}^{JK}_{ijk}](\mu)=\eta_{\mu}^{2/\beta_0[4]}\,\eta_{b}^{2/\beta_0[5]}\,\eta_{t}^{2/\beta_0[6]}\,C[{\cal H}^{JK}_{ijk}](\mu_0)\,,
\end{equation}
and similarly for the operators of $\widetilde{\cal H}$-type. To appreciate the order of magnitude of this correction, the choice ($\mu\stackrel[]{!}{=}2$\,GeV,\ 
$\mu_0\stackrel[]{!}{=}1$\,TeV) returns an enhancement factor of about $1.4$ at low energy.

The semi-leptonic operators ($\ell=\nu,e$) receive the following scaling with energy:
\begin{align}\label{eq:RGEselep}
&C[{\cal S}^{\ell q\,JK}_{ijk}](\mu)=\eta_{\mu}^{3C_2(3)/\beta_0[4]}\,\eta_{b}^{3C_2(3)/\beta_0[5]}\,\eta_{t}^{3C_2(3)/\beta_0[6]}\,C[{\cal S}^{\ell q\,JK}_{ijk}](\mu_0)\,,\nonumber\\[1.5mm]
&C[{\cal V}^{\ell q\,JK}_{ijk}](\mu)=C[{\cal V}^{\ell q\,JK}_{ijk}](\mu_0)\,,\\[1.5mm]
&C[{\cal T}^{\ell q\,J}_{ijk}](\mu)=\eta_{\mu}^{-C_2(3)/\beta_0[4]}\,\eta_{b}^{-C_2(3)/\beta_0[5]}\,\eta_{t}^{-C_2(3)/\beta_0[6]}\,C[{\cal T}^{\ell q\,J}_{ijk}](\mu_0)\,.\nonumber
\end{align}
We here recover the usual scaling of scalar semi-leptonic operators, similar to a quark mass, while the vector operators have a vanishing anomalous dimension. 
However, we remind the reader that these simple scaling conditions narrowly depend on the choice of operator basis. For instance, replacing ${\cal S}^{\ell q\,J}
_{ijk}$ and ${\cal T}^{\ell q\,J}_{ijk}$ by $\hat{\cal S}^{\ell q\,JK}_{ijk}\equiv{\cal S}^{\ell q\,JK}_{ijk}$ and $\widetilde{\cal S}^{\ell q\,J}_{ijk}\equiv(\bar{\Psi}P_JQ_k)
(\bar{Q}_jP_JE_i)=-\tfrac{1}{2}{\cal T}^{\ell q\,J}_{ijk}-\tfrac{1}{2}{\cal S}^{\ell q\,JJ}_{ijk}$ would lead to a hybrid running for $C[\hat{\cal S}^{\ell q\,JJ}_{ijk}]$ (even 
though $\hat{\cal S}^{\ell q\,JJ}_{ijk}={\cal S}^{\ell q\,JJ}_{ijk}$). The choice ($\mu\stackrel[]{!}{=}2$\,GeV, $\mu_0\stackrel[]{!}{=}1$\,TeV) implies an enhancement 
factor of about $1.9$ in the case of scalar coefficients, while tensor coefficients are suppressed by about $0.8$, in this example.

The leptonic operators do not receive QCD corrections at leading order. In the case of the electric dipole operator, the corrections from the ${\cal T}^{\nu f}_{ijj}$ operators --- see 
Eq.~(\ref{eq:nngcorr}) --- require the renormalization of ${\cal E}_i$ and imply running conditions for the Wilson coefficient:
\begin{align}
&C[{\cal E}_i](\mu)=C[{\cal E}_i](\mu_0)-\sum_{e_j}2Q_{e_j}m_{e_j}C[{\cal T}^{\nu e}_{ijj}](\mu_0)\,\ln\frac{\text{Max}[\mu^2,m_{e_j}^2]}{\mu_0^2}\nonumber\\
&\null\hspace{0.8cm}-\sum_{q_j}\frac{8\pi N_c Q_{q_j}}{\beta_0[6]-2C_2[3]}\Big[\eta_t^{2C_2(3)/\beta_0[6]-1}-1\Big]\Big(\tfrac{m_{q_j}}{\alpha_s}C[{\cal T}^{\nu q\,L}_{ijj}]\Big)(\mu_0)\\
&\null\hspace{0.8cm}-\sum_{q_j\neq t}\frac{8\pi N_c Q_{q_j}}{\beta_0[5]-2C_2[3]}\eta_t^{2C_2(3)/\beta_0[6]-1}\Big[\eta_b^{2C_2(3)/\beta_0[5]-1}-1\Big]\Big(\tfrac{m_{q_j}}{\alpha_s}C[{\cal T}^{\nu q\,L}_{ijj}]\Big)(\mu_0)\nonumber\\
&\null\hspace{0.8cm}-\sum_{q_j\neq t,b}\frac{8\pi N_c Q_{q_j}}{\beta_0[4]-2C_2[3]}\eta_t^{2C_2(3)/\beta_0[6]-1}\eta_b^{2C_2(3)/\beta_0[5]-1}\Big[\eta_{\mu}^{2C_2(3)/\beta_0[4]-1}-1\Big]\Big(\tfrac{m_{q_j}}{\alpha_s}C[{\cal T}^{\nu q\,L}_{ijj}]\Big)(\mu_0)\,.\nonumber
\end{align}
The significance of these corrections is made more explicit after expressing $C[{\cal E}_i](\mu_0)$ in terms of the corresponding coefficient in the RpV MSSM --- see Eq.~(\ref{eq:nngcorr}):
\begin{align}
&\hspace{0.3cm}C[{\cal E}_i](\mu)=C^{\text{MSSM}}[{\cal E}_i]-\sum_{e_j}2Q_{e_j}m_{e_j}C[{\cal T}^{\nu e}_{ijj}](\mu_0)\,\ln\frac{\text{Max}[\mu^2,m_{e_j}^2]}{m_{e_j}^2}\nonumber\\
&\null\hspace{0.1cm}-\sum_{q_j}8\pi N_c Q_{q_j}\bigg[\tfrac{\eta_t^{2C_2(3)/\beta_0[6]-1}-1}{\beta_0[6]-2C_2[3]}+\tfrac{\alpha_s(\mu_0)}{4\pi}\ln\tfrac{\mu_0^2}{m_t^2}\bigg]\Big(\tfrac{m_{q_j}}{\alpha_s}C[{\cal T}^{\nu q\,L}_{ijj}]\Big)(\mu_0)\\
&\null\hspace{0.1cm}-\sum_{q_j\neq t}8\pi N_c Q_{q_j}\Big[\eta_t^{2C_2(3)/\beta_0[6]-1}\tfrac{\eta_b^{2C_2(3)/\beta_0[5]-1}-1}{\beta_0[5]-2C_2[3]}+\tfrac{\alpha_s(\mu_0)}{4\pi}\ln\tfrac{m_t^2}{m_b^2}\Big]\Big(\tfrac{m_{q_j}}{\alpha_s}C[{\cal T}^{\nu q\,L}_{ijj}]\Big)(\mu_0)\nonumber\\
&\null\hspace{0.05cm}-\!\!\sum_{q_j\neq t,b}8\pi N_c Q_{q_j}\Big[\eta_t^{2C_2(3)/\beta_0[6]-1}\eta_b^{2C_2(3)/\beta_0[5]-1}\tfrac{\eta_{\mu}^{2C_2(3)/\beta_0[4]-1}-1}{\beta_0[4]-2C_2[3]}+\tfrac{\alpha_s(\mu_0)}{4\pi}\ln\tfrac{m_b^2}{m_{q_j}^2}\Big]\Big(\tfrac{m_{q_j}}{\alpha_s}C[{\cal T}^{\nu q\,L}_{ijj}]\Big)(\mu_0).\nonumber
\end{align}
It is thus evident that the scale evolution re-inserts in the Wilson coefficient those pieces that were subtracted at the level of the matching for being captured by 
the semi-leptonic operators. However, quark loops are now included in a form resumming leading QCD logarithms between $\mu_0$ and $\mu$. Finally, we 
stress that the decay amplitude at the scale $\mu$ adds a further factor $2N_c^{f_j}Q_{f_j}\big(m_{f_j}C[{\cal T}^{\nu f}_{ijj}]\big)(\mu)\,\ln\frac{\mu^2}{m_{f_j}^2}$, 
for all active fermion fields $f_j$ at the scale $\mu$, to the decay amplitude at this scale.

This step completes the determination of the Wilson coefficients at low energy, hence the definition of the EFT for phenomenological use.

\section{Decay Widths of the Neutralino}

After determining the Wilson coefficients at high-energy through matching, and running them down to the infrared scale, we have obtained a working partonic 
theory describing the low-energy phenomenology of the light neutralino. In order to make meaningful predictions of the decay widths, it is then necessary to 
translate the partonic operators into interactions involving the actual physical degrees of freedom at these scales, \textit{i.e.}~hadrons instead of quarks.

\subsection{Radiative and leptonic decays}
The decay widths in the radiative and leptonic channels can be derived in a straightforward fashion. In this paper, we systematically neglect interaction effects 
between final states, which, strictly speaking, is only valid in situations with enough kinetic energy. 

The two-body radiative decay is kinematically open at neutralino masses above the neutrino masses, \textit{i.e.}~essentially in all the considered range for 
$m_{\psi}$. However, the choice $\mu\stackrel[]{!}{=}m_{\psi}$ for the renormalization scale would be problematic for $m_{\psi}\lsim m_c$, as the perturbative 
description of the strong coupling loses significance. Maintaining $\mu\gsim m_c$ means that the resummation of QCD logarithms is not comprehensive for 
light quarks. Yet, given that the contributions of light fermions to the decay amplitude is suppressed in proportion to their mass, light quark contributions are 
non-essential. The decay width then reads (for $\mu\gsim m_c$, $i=1,2,3$):
\begin{equation}
\Gamma[\psi\to\gamma\bar{\nu}_i]=\frac{\alpha m_{\psi}^3}{4096\pi^4}\Big|C[{\cal E}_i](\mu)+\sum_{f_j\neq t,b,\ldots}2Q_{f_j}N_c^{f_j}m_{f_j}\ln\tfrac{\mu^2}{m^2_{f_j}}\,C[{\cal T}^{\nu\,f}_{ijj}](\mu)\Big|^2=\Gamma[\psi\to\gamma\nu_i]\,.
\end{equation}

The decays into three neutrinos are also accessible down to very low $m_{\psi}$. As we restricted ourselves to resumming QCD logarithms, the Wilson coefficients of leptonic operators do not run, and the choice of $\mu$ does not matter here. For the seven independent three-neutrino channels $\nu_i\nu_j\nu_k$ ($i\neq j=k$ and $(i,j,k)=(1,2,3)$):
\begin{align}\label{eq:3nudec}
&\Gamma[\psi\to\bar{\nu}_i\bar{\nu}_j\bar{\nu}_j]=\frac{m_{\psi}^5}{3072\pi^3}\big|C[\widetilde{\cal N}_{ijj}](\mu)\big|^2=\Gamma[\psi\to\nu_i\nu_j\nu_k]\,,\\[1.5mm]
&\Gamma[\psi\to\bar{\nu}_1\bar{\nu}_2\bar{\nu}_3]=\frac{m_{\psi}^5}{6144\pi^3}\Big\{\big|C[\widetilde{\cal N}_{123}](\mu)\big|^2+\big|C[\widetilde{\cal N}_{213}](\mu)\big|^2-\text{Re}\big[C[\widetilde{\cal N}_{123}](\mu)^*C[\widetilde{\cal N}_{213}](\mu)\big]\Big\}\nonumber\\[1.5mm]
&\null\hspace{2.45cm}=\Gamma[\psi\to\nu_1\nu_2\nu_3]\,.\nonumber
\end{align}
Similarly, the neutralino decays into two antineutrinos ($i\leq k$) and one neutrino ($j$) read:
\begin{equation}\label{eq:3nnbdec}
\Gamma[\psi\to\bar{\nu}_i\nu_j\bar{\nu}_k]=\frac{1+\delta_{ik}}{1536\pi^3}m_{\psi}^5\big|C[{\cal N}_{ijk}](\mu)\big|^2=\Gamma[\psi\to\nu_i\bar{\nu}_j\nu_k]\,,
\end{equation}
corresponding to $18$ channels (+ CP-conjugates).

The decays into an antineutrino and a pair of charged leptons is computed to be:
\begin{equation}\label{eq:lepkinint}
\Gamma[\psi\to\bar{\nu}_ie_j\bar{e}_k]=\frac{m_{\psi}^5}{256\pi^3}\sum_{\Omega,\Omega'}C[\Omega]C[\Omega']^*I_{\Omega,\Omega'}\big[\tfrac{m_{e_j}}{m_{\psi}},\tfrac{m_{e_k}}{m_{\psi}}\big]=\Gamma[\psi\to\nu_i\bar{e}_je_k]\,,
\end{equation}
where the sums over $\Omega$ and $\Omega'$ run over the list of leptonic $\nu e$ operators (see Eq.~(\ref{eq:leptop})) and the kinematic integrals $I_{\Omega,\Omega'}$ can be explicitly calculated: see Appendix~\ref{ap:lepkinintt}.

\subsection{Hadronic decays}
In the presence of baryon-number violation, the hadronic operators of Eq.~(\ref{eq:hadop}) generate interactions between $\psi$ and the 
baryons that are difficult to characterize on a quantitative basis. We will restrict ourselves to the chiral picture to shed some light at the 
qualitative level, considering only the baryon and meson octets of lowest energy. We refer the reader to Refs.\cite{Claudson:1981gh,JLQCD:1999dld} 
for a more detailed derivation of the formalism.

We introduce the light meson and baryon octets of the chiral approach in $3\times\bar{3}$ space of the vectorial $(u,d,s)^T$ $SU(3)$ symmetry:
\begin{align}\label{eq:chiralfields}
&M\equiv\begin{bmatrix}
\tfrac{\pi^0}{\sqrt{2}}+\tfrac{\eta^0_8}{\sqrt{6}}& \pi^+& K^+\\
\pi^-& -\tfrac{\pi^0}{\sqrt{2}}+\tfrac{\eta^0_8}{\sqrt{6}}& K^0\\
K^-& \bar{K}^0& -\sqrt{\tfrac{2}{3}}\eta^0_8
\end{bmatrix}\sim\begin{bmatrix}
u^{\alpha}\bar{u}^{\alpha}& u^{\alpha}\bar{d}^{\alpha}& u^{\alpha}\bar{s}^{\alpha}\\
d^{\alpha}\bar{u}^{\alpha}& d^{\alpha}\bar{d}^{\alpha}& d^{\alpha}\bar{s}^{\alpha}\\
s^{\alpha}\bar{u}^{\alpha}& s^{\alpha}\bar{d}^{\alpha}& s^{\alpha}\bar{s}^{\alpha}
\end{bmatrix}\,,\\[2mm]
&B\equiv\begin{bmatrix}
\tfrac{\Sigma^0}{\sqrt{2}}+\tfrac{\Lambda^0}{\sqrt{6}}& \Sigma^+& p^+\\
\Sigma^-& -\tfrac{\Sigma^0}{\sqrt{2}}+\tfrac{\Lambda^0}{\sqrt{6}}& n^0\\
\Xi^-& \Xi^0& -\sqrt{\tfrac{2}{3}}\Lambda^0
\end{bmatrix}\sim\varepsilon_{\alpha\beta\gamma}\begin{bmatrix}
u^{\alpha}(d^{\beta}s^{\gamma})& u^{\alpha}(s^{\beta}u^{\gamma})& u^{\alpha}(u^{\beta}d^{\gamma})\\
d^{\alpha}(d^{\beta}s^{\gamma})& d^{\alpha}(s^{\beta}u^{\gamma})& d^{\alpha}(u^{\beta}d^{\gamma})\\
s^{\alpha}(d^{\beta}s^{\gamma})& s^{\alpha}(s^{\beta}u^{\gamma})& s^{\alpha}(u^{\beta}d^{\gamma})
\end{bmatrix}\,.\nonumber
\end{align}
The properties of the hadronic operators ${\cal H}$ and $\widetilde{\cal H}$ under the $SU(3)_L\times SU(3)_R$ symmetry (as well as parity) may 
then be exploited to convert them into effective terms of the chiral Lagrangian depending on only two low-energy constants $\tilde{\alpha}$ [for 
$(3_L,\bar{3}_R$) or ($\bar{3}_L,3_R)$ terms] and $\tilde{\beta}$ [for $(8_L,1_R)$ or $(1_L,8_R)$ terms]:
{\small\begin{align}\label{eq:psiChPT}
{\cal L}_{\psi}^{\chi\text{PT}}\ni&\!-\!\tilde{\alpha}\,\bar{\Psi}P_R\text{Tr}\bigg\{\Big(C[{\cal H}^{LR}_{112}]^*E_{11}\!-C[\widetilde{\cal H}^{LR}_{111}]^*E_{32}\!
+C[\widetilde{\cal H}^{LR}_{112}]^*E_{22}\!-C[\widetilde{\cal H}^{LR}_{121}]^*E_{33}\!+C[\widetilde{\cal H}^{LR}_{122}]^*E_{23}\Big)\xi^{\dagger}B\xi^{\dagger}
\bigg\}\nonumber\\
&\!-\!\tilde{\alpha}\,\bar{\Psi}P_L\text{Tr}\bigg\{\Big(C[{\cal H}^{RL}_{112}]^*E_{11}\!-C[\widetilde{\cal H}^{RL}_{111}]^*E_{32}\!+C[\widetilde{\cal H}^{RL}_{112}]
^*E_{22}\!-C[\widetilde{\cal H}^{RL}_{121}]^*E_{33}\!+C[\widetilde{\cal H}^{RL}_{122}]^*E_{23}\Big)\xi B\xi\bigg\}\nonumber\\
&\!-\!\tilde{\beta}\,\bar{\Psi}P_R\text{Tr}\bigg\{\Big(C[{\cal H}^{LL}_{112}]^*E_{11}\!-C[\widetilde{\cal H}^{LL}_{111}]^*E_{32}\!+C[\widetilde{\cal H}^{LL}_{112}]^*
E_{22}\!-C[\widetilde{\cal H}^{LL}_{121}]^*E_{33}\!+C[\widetilde{\cal H}^{LL}_{122}]^*E_{23}\Big)\xi^{\dagger}B\xi\bigg\}\nonumber\\
&\!-\!\tilde{\beta}\,\bar{\Psi}P_L\text{Tr}\bigg\{\Big(C[{\cal H}^{RR}_{112}]^*E_{11}\!-C[\widetilde{\cal H}^{RR}_{111}]^*E_{32}\!+C[\widetilde{\cal H}^{RR}_{112}
]^*E_{22}\!-C[\widetilde{\cal H}^{RR}_{121}]^*E_{33}\!+C[\widetilde{\cal H}^{RR}_{122}]^*E_{23}\Big)\xi B\xi^{\dagger}\bigg\}\nonumber\\
&+\text{h.c.}.
\end{align}}\noindent
Here $\xi\equiv\exp\big[\tfrac{i}{f_{\chi}}M\big]$, and $f_{\chi}\simeq131$\,MeV is the chiral-symmetry breaking scale (pion decay constant).
The $E_{ij}$'s represent the elements of the canonical basis of $3\times3$ real matrices (\textit{i.e.}~the entries of $E_{ij}$ are $0$ except 
for the one in the $i^{\text{th}}$ line and $j^{\text{th}}$ column, which is $1$). The low-energy constants $\tilde{\alpha}\approx-0.0144$\,GeV$^3$ 
and $\tilde{\beta}\approx0.0144$\,GeV$^3$ have been determined on the lattice \cite{Aoki:2017puj}, but could also be identified from individual 
hadronic matrix elements of proton decays into mesons, leading to a wider range of variation.

Eq.~(\ref{eq:psiChPT}) highlights two physical phenomena. First, it generates mass-mixing terms between the light neutralino and the neutral 
baryons $n^0$, $\Sigma^0$, $\Lambda^0$ and $\Xi^0$. This effect is analogous to that appearing in the case of a light pseudoscalar mixing 
with the mesons --- see \textit{e.g.}~Refs.\cite{Domingo:2016unq,Domingo:2016yih}. A sterile $\psi$ close in mass to a neutral baryon could 
then decay into lighter baryons via its subleading baryonic component, leading to a spectrum of resonances when varying $m_{\psi}$. However, 
the hadronic decays of hyperons ($\Sigma^0$, $\Lambda^0$ and $\Xi^0$; see \textit{e.g.}~Ref.\,\cite{Borasoy:1998ku}) are highly suppressed, 
because they are mediated by the weak interaction and typically involve an $s\to d$ transition. As a result, the impact of the baryon-neutralino 
mixing on the neutralino decays remains extremely narrow, as long as we focus on the lightest baryonic octet. That we do not attempt to describe 
the interactions with heavier and broader baryons, such as the nucleon states at $\sim1.35$\,GeV, also restricts the validity of our analysis of 
baryon-number violating decays to $m_{\psi}\lsim1.3$\,GeV. As a complementary feature, the mixing between neutralino and hadrons implies 
baryon-antibaryon oscillations for the baryon-dominated states, which, in view of the experimental limits on $n^0-\bar{n}^0$ oscillations, 
constrains the corresponding mixing to comparatively weak values. We discuss this feature in Appendix~\ref{ap:nnbarosc}. Furthermore, if 
lepton-number is violated, the subleading $\psi$ component of the baryon-dominated state could mediate baryon- (and lepton)-number violating 
decays, \textit{e.g.}~of the neutron.

The second effect  in Eq.~(\ref{eq:psiChPT}) is the explicit emergence of $\psi$-baryon-meson couplings, which allow for neutralino 
decays into lighter baryon+meson pairs, even though the mixing effect remains negligible. The tree level calculation is presented in 
Appendix~\ref{ap:haddecwi}. Expressing the effective couplings between the neutralino $\psi$, the mesons $M_i\in\{\pi^0,\pi^+,\pi^-,K^0,\bar{K}
^0,K^+,K^-,\eta^0_8\}$ and the baryons $B_j\in\{\Sigma^0,\Sigma^+,\Sigma^-,n^0,\Xi^0,p^+,\Xi^-,\Lambda^0\}$ as $i g_{L,R}^{\psi M_i B_j}M_i
\bar{\Psi}P_{L,R}B_j$, we can derive the decay width for the process $\psi\to\bar{M}_i\bar{B}_j$, if kinematically open:
\begin{multline}\label{eq:haddecwi}
\Gamma[\psi\to\bar{M}_i\bar{B}_j]=\frac{m_{\psi}}{32\pi}\left\{\Big(1+\tfrac{m_{B_j}^2-m^2_{M_i}}{m_{\psi}^2}\Big)\Big[\big|g_{L}^{\psi M_i B_j}
\big|^2+\big|g_{R}^{\psi M_i B_j}\big|^2\Big]+4\tfrac{m_{B_j}}{m_{\psi}}\text{Re}\Big[\big(g_{L}^{\psi M_i B_j}\big)^*g_{R}^{\psi M_i B_j}\Big]\right\}\\
\cdot\Big(1-2\tfrac{m_{B_j}^2+m^2_{M_i}}{m_{\psi}^2}+\tfrac{(m_{B_j}^2-m^2_{M_i})^2}{m_{\psi}^4}\Big)^{1/2}=\Gamma[\psi\to M_iB_j]\,.
\end{multline}
We stress that this description would need to be extended and include the impact of additional (and broad) baryonic states, such as the heavy 
nucleons at $1.35$\,GeV or even the $\Delta$'s at $\sim1.2$~GeV, as soon as these become kinematically accessible.

Further decays, such as the radiative $\psi\to\gamma B_j$ (for $B_j\in\{\Sigma^0,n^0,\Xi^0,\Lambda^0\}$) could also be assessed in the chiral 
description. Finally, for $m_{\psi}<m_p$, hadronic decays of the neutralino are kinematically forbidden, as no hadronic fermion can be reached 
below the proton mass: it is then the baryons which may decay into $\psi$ and a meson \cite{Chamoun:2020aft}.

\subsection{Semi-leptonic decays}
The semi-leptonic operators produce a collection of neutralino decay channels into mesons and a lepton. The simplest final states, already 
considered in \textit{e.g.}~Ref.\cite{deVries:2015mfw}, consist of two bodies, one lepton ($\ell_i$) and one meson $M$. The decay amplitude 
may be schematically written as:
\begin{align}
{\cal A}[\psi\to\bar{\ell}_iM]=&\ i\, C[{\cal S}_{ijk}^{\ell q\, JK}]\,\left<M(p_M)\right|\bar{Q}_jP_KQ_k\left|0\right>\,\bar{v}_{\psi}(p_{\psi})P_Jv_{\ell_i}(p_{\ell_i})\\
&+i\, C[{\cal V}_{ijk}^{\ell q\, JK}]\,\left<M(p_M)\right|\bar{Q}_j\gamma^{\mu}P_KQ_k\left|0\right>\,\bar{v}_{\psi}(p_{\psi})\gamma_{\mu}P_Jv_{\ell_i}(p_{\ell_i})\nonumber\\
&+i\, C[{\cal T}_{ijk}^{\ell q\, J}]\,\left<M(p_M)\right|\bar{Q}_j\tfrac{\Sigma^{\mu\nu}}{2}P_JQ_k\left|0\right>\,\bar{v}_{\psi}(p_{\psi})\tfrac{\Sigma_{\mu\nu}}{2}P_Jv_{\ell_i}(p_{\ell_i})\,,\nonumber
\end{align}
where the scalar, vector and tensor form factors of the meson $M$ depend on its nature. At this level, the interactions of the final-state lepton
with the meson have been neglected.

Focusing first on the pseudoscalar mesons of Eq.~(\ref{eq:chiralfields}), to which we may add the $\eta^0_1\sim\tfrac{1}{\sqrt{3}}(u^{\alpha}\bar{u}
^{\alpha}+d^{\alpha}\bar{d}^{\alpha}+s^{\alpha}\bar{s}^{\alpha})$ and account for $\eta-\eta'$ mixing, the pseudoscalar form factors in the chiral 
limit are all connected to the parameter $B_0$ which relates meson masses to quark mass parameters, \textit{e.g.}~$B_0\approx\tfrac{f_{\chi}}
{\sqrt{2}m_s}(m_{K^{\pm}}^2+m_{K^0}^2-m_
{\pi^0}^2)\sim0.5$\,GeV$^2$:
\begin{equation}
\left<M(p_M)\right|\bar{Q}_jP_RQ_k\left|0\right>=-\frac{i B_0}{2\sqrt{2}}{\cal R}_{M}[q_j\bar{q}_k]=-\left<M(p_M)\right|\bar{Q}_jP_LQ_k\left|0\right>\,.
\end{equation}
Here ${\cal R}_{M}$ represents the overlap of the $q_j\bar{q}_k$ pair with the valence quarks of $M(p_M)$; for instance ${\cal R}_{\pi^0}[u\bar{u}]=\tfrac{1}{\sqrt{2}}=-{\cal R}_{\pi^0}[d\bar{d}]$ and all other ${\cal R}_{\pi^0}[q_j\bar{q}_k]$ vanish. The axial form factors then relate to the above through the Dirac equation:
\begin{equation}
\left<M(p_M)\right|\bar{Q}_j\gamma^{\mu}P_RQ_k\left|0\right>=-\frac{i\,B_0}{2\sqrt{2}}\frac{m_{q_j}+m_{q_k}}{m_{M}^2}\,{\cal R}_{M}[q_j\bar{q}_k]\,p_{M}^{\mu}=-\left<M(p_M)\right|\bar{Q}_j\gamma^{\mu}P_LQ_k\left|0\right>\,.
\end{equation}
In turn, $(m_{q_j}+m_{q_k})B_0$ can be written in terms of $f_{\chi}$ and meson masses. There is no antisymmetric tensor associated with 
scalars so that $\left<M(p_M)\right|\bar{Q}_j\tfrac{\Sigma^{\mu\nu}}{2}P_{L,R}Q_k\left|0\right>=0$. In order to account for effects beyond the 
chiral approximation, it is possible to re-parametrize the pseudoscalar and axial form-factors in terms of the decay constants of the meson 
$f_{M}^{q_j\bar{q}_k}$, as already suggested in Ref.\cite{deVries:2015mfw}: $\left<M(p_M)\right|\bar{Q}_j\gamma^{\mu}\gamma^5Q_k\left|
0\right>=-i\,f_{M}^{q_j\bar{q}_k}\,p_{M}^{\mu}$. We finally arrive at the decay widths (with implicit sum over the quark indices):
\begin{align}\label{eq:SL2bdecwi}
& \Gamma[\psi\to \bar{\ell}_iM]=\frac{m_{\psi}}{32\pi}\Big\{\big(1+\tfrac{m_{\ell_i}^2-m_{M}^2}{m_{\psi}^2}\big)\big(|g_L^{\psi\bar{M}\ell_i}|^2+|g_R^{\psi\bar{M}\ell_i}|^2\big)+4\tfrac{m_{\ell_i}}{m_{\psi}}\text{Re}\big[(g_L^{\psi\bar{M}\ell_i})^*g_R^{\psi\bar{M}\ell_i}\big]\Big\}\nonumber\\
&\null\hspace{4.5cm}\times\Big[1-2\tfrac{m_{M}^2+m_{\ell_i}^2}{m_{\psi}^2}+\tfrac{(m_{M}^2-m_{\ell_i}^2)^2}{m_{\psi}^4}\Big]^{1/2}\,,\\
&g_L^{\psi\bar{M}\ell_i}\equiv
\tfrac{f_{M}^{q_j\bar{q}_k}}{2}\Big\{\tfrac{B_0}{\sqrt{2}f_{\chi}}\Big(\tfrac{\sqrt{2}f_{\chi}m^2_{M}}{B_0(m_{q_j}+m_{q_k})}\Big)\big(C[{\cal S}_{ijk}^{\ell q\, LR}]-C[{\cal S}_{ijk}^{\ell q\, LL}]\big)\nonumber\\
&\null\hspace{4.5cm}-
\Big[m_{\psi}\big(C[{\cal V}_{ijk}^{\ell q\, LR}]-C[{\cal V}_{ijk}^{\ell q\, LL}]\big)-m_{\ell_i}\big(C[{\cal V}_{ijk}^{\ell q\, RR}]-C[{\cal V}_{ijk}^{\ell q\, LR}]\big)\Big]\Big\}\,,\nonumber\\
&g_R^{\psi\bar{M}\ell_i}\equiv
\tfrac{f_{M}^{q_j\bar{q}_k}}{2}\Big\{\tfrac{B_0}{\sqrt{2}f_{\chi}}\Big(\tfrac{\sqrt{2}f_{\chi}m^2_{M}}{B_0(m_{q_j}+m_{q_k})}\Big)\big(C[{\cal S}_{ijk}^{\ell q\, RR}]-C[{\cal S}_{ijk}^{\ell q\, RL}]\big)\nonumber\\
&\null\hspace{4.5cm}-
\Big[m_{\psi}\big(C[{\cal V}_{ijk}^{\ell q\, RR}]-C[{\cal V}_{ijk}^{\ell q\, RL}]\big)-m_{\ell_i}\big(C[{\cal V}_{ijk}^{\ell q\, LR}]-C[{\cal V}_{ijk}^{\ell q\, LL}]\big)\Big]\Big\}\,.\nonumber
\end{align}
We consider the $19$ such channels involving light leptons (plus their $19$ CP-conjugate).

Turning to the light vector mesons $M^*_{\mu}\sim(\bar{q}_j\gamma_{\mu}q_k)$,
\begin{equation}
M^*_{\mu}\equiv\begin{bmatrix}
\tfrac{\rho^0_{\mu}}{\sqrt{2}}+\tfrac{\phi^0_{8\,\mu}}{\sqrt{6}}+\tfrac{\phi^0_{1\,\mu}}{\sqrt{3}}& \rho_{\mu}^+& K^{*\,+}_{\mu}\\
\rho_{\mu}^-& -\tfrac{\rho^0_{\mu}}{\sqrt{2}}+\tfrac{\phi^0_{8\,\mu}}{\sqrt{6}}+\tfrac{\phi^0_{1\,\mu}}{\sqrt{3}}& K^{*\,0}_{\mu}\\
K^{*\,-}_{\mu}& \bar{K}^{*\,0}_{\mu}& -\sqrt{\tfrac{2}{3}}\phi^0_{8\,\mu}+\tfrac{\phi^0_{1\,\mu}}{\sqrt{3}}
\end{bmatrix}\,,
\end{equation}
and after mixing $\phi^0_8$ and $\phi^0_1$ to account for the $\phi$ and $\omega$, one can define the vector and tensor form-factors:
\begin{align}
&\left<M^*(p_M)\right|\bar{Q}_j\gamma^{\mu}Q_k\left|0\right>=f_{M^*}^V\,m_{M^*}\,{\cal R}_{M^*}[q_j\bar{q}_k]\,\epsilon^{*\,\mu}_{M^*}(p_M)\,,\\[1.5mm]
&\left<M^*(p_M)\right|\bar{Q}_j\Sigma^{\mu\nu}Q_k\left|0\right>=-i f_{M^*}^T\,{\cal R}_{M^*}[q_j\bar{q}_k]\,\big[p_M^{\mu}\epsilon^{*\,\nu}_{M^*}(p_M)-
p_M^{\nu}\epsilon^{*\,\mu}_{M^*}(p_M)\big]\,.\nonumber
\end{align}
Then, we may parametrize the decay amplitude as:
\begin{align}
&\null\hspace{1cm}{\cal A}[\psi\to M^*\bar{\ell}_i]=i\bar{v}_{\psi}(p_{\psi})\Big\{g_{V\,L,R}^{M^*\psi\ell_i}\slashed{\varepsilon}^*_{M^*}(p_{M^*})P_{L,R}+2\,p_{\psi}\cdot\varepsilon^*_{M^*}(p_{M^*})\,g_{S\,L,R}^{M^*\psi\ell_i}P_{L,R}\Big\}v_{\ell_i}(p_{\ell_i})\,,\\
&g_{V\,J}^{M^*\psi\ell_i}\equiv \tfrac{1}{2}{\cal R}_{M^*}[q_j\bar{q}_k]\Big\{m_{M^*}f_{M^*}^V\big(C[{\cal V}_{ijk}^{\ell q\, JL}]+\!C[{\cal V}_{ijk}^{\ell q\, JR}]\big)-\!f_{M^*}^T\big(m_{\psi}C[{\cal T}_{ijk}^{\ell q\, J}]+m_{\ell_i}C[{\cal T}_{ijk}^{\ell q\, J+1}]\big)\Big\}\,,\nonumber\\
&g_{S\,J}^{M^*\psi\ell_i}\equiv -\tfrac{f_{M^*}^T}{2}{\cal R}_{M^*}[q_j\bar{q}_k]C[{\cal T}_{ijk}^{\ell q\, J}]\,,\nonumber
\end{align}
leading to the decay width:
\begin{align}\label{eq:SL2bVdecwi}
&\Gamma[\psi\to M^*\bar{\ell}_i]=\frac{m_{\psi}^3}{32\pi m^2_{M^*}}\Big[1-2\tfrac{m_{M^*}^2+m_{\ell_i}^2}{m_{\psi}^2}+\tfrac{(m_{M^*}^2-m_{\ell_i}^2)^2}{m_{\psi}^4}\Big]^{1/2}\times\\
&\null\hspace{0.3cm}\Bigg\{\Big(|g_{V\,L}^{M^*\psi\ell_i}|^2+|g_{V\,R}^{M^*\psi\ell_i}|^2\Big)\Big[1+\tfrac{m^2_{M^*}-2m^2_{\ell_i}}{m^2_{\psi}}-\tfrac{2m^4_{M^*}-m^2_{\ell_i}(m^2_{M^*}+m^2_{\ell_i})}{m_{\psi}^4}\Big]\nonumber \\
&-12\frac{m^2_{M^*}m_{\ell_i}}{m^3_{\psi}}\text{Re}\big[g_{V\,L}^{M^*\psi\ell_i\,*}g_{V\,R}^{M^*\psi\ell_i}\big]
+m^2_{\psi}\Big[1-2\tfrac{m_{M^*}^2+m_{\ell_i}^2}{m_{\psi}^2}+\tfrac{(m_{M^*}^2-m_{\ell_i}^2)^2}{m_{\psi}^4}\Big]\nonumber\\
&\Big[\big(|g_{S\,L}^{M^*\psi\ell_i}|^2+|g_{S\,R}^{M^*\psi\ell_i}|^2\big)\big(1+\tfrac{m^2_{\ell_i}-m^2_{M^*}}{m^2_{\psi}}\big)+4\frac{m_{\ell_i}}{m_{\psi}}\text{Re}\big[g_{S\,L}^{M^*\psi\ell_i\,*}g_{S\,R}^{M^*\psi\ell_i}\big]\Big]\nonumber\\
&\null\hspace{1cm}-2m_{\psi}\Big[1-2\tfrac{m_{M^*}^2+m_{\ell_i}^2}{m_{\psi}^2}+\tfrac{(m_{M^*}^2-m_{\ell_i}^2)^2}{m_{\psi}^4}\Big]\text{Re}\big[g_{V\,L}^{M^*\psi\ell_i\,*}g_{S\,L}^{M^*\psi\ell_i}+g_{V\,R}^{M^*\psi\ell_i\,*}g_{S\,R}^{M^*\psi\ell_i}\big]\nonumber\\
&\null\hspace{1.cm}-2m_{\ell_i}\Big[1-2\tfrac{m_{\ell_i}^2}{m_{\psi}^2}+\tfrac{(m_{M^*}^2-m_{\ell_i}^2)^2}{m_{\psi}^4}\Big]\text{Re}\big[g_{V\,L}^{M^*\psi\ell_i\,*}g_{S\,R}^{M^*\psi\ell_i}+g_{V\,R}^{M^*\psi\ell_i\,*}g_{S\,L}^{M^*\psi\ell_i}\big]\Bigg\}\,.\nonumber
\end{align}
However, the production of vector mesons can be seen as a partial contribution to the more general final states with two (or three) pseudoscalar 
mesons, see for example Ref.~\cite{Daub:2012mu} in the R-parity violating context. For this reason, it is preferable to dismiss the two-body decay 
formalism for vector mesons and directly exploit the more complete form-factors for meson pairs, \textit{i.e.}~three-body semi-leptonic decay 
channels, as we perform below. Yet, in the case of the $\omega$, mainly decaying into three pseudoscalar mesons, the two-body formalism remains 
sufficient at low energy.

The properties of scalar and tensor mesons being ill-known (and broad), an application of the two-body formalism with such particles in the final 
state would provide limited insight. Instead, it is again more meaningful to regard these states as mediators in the three-body decay widths.

Decays into one lepton and two pseudoscalar mesons are relevant as soon as enough phase-space is available. In view of the CP-conservation 
observed by the strong interaction, we need not consider pseudoscalar and axial-vector mediations and the decay amplitude schematically reads 
(still neglecting the interactions between leptons and mesons in the final state):
\begin{align}\label{eq:semilepamp}
{\cal A}[\psi\to\bar{\ell}_iM_1M_2]=&\ \,\ \ \tfrac{i}{2}\, C[{\cal S}_{ijk}^{\ell q\, JK}]\,\left<M_1(p_1)M_2(p_2)\right|\bar{Q}_jQ_k\left|0\right>\,\bar{v}_
{\psi}(p_{\psi})P_Jv_{\ell_i}(p_{\ell_i})\\
&+\tfrac{i}{2}\, C[{\cal V}_{ijk}^{\ell q\, JK}]\,\left<M_1(p_1)M_2(p_2)\right|\bar{Q}_j\gamma^{\mu}Q_k\left|0\right>\,\bar{v}_{\psi}(p_{\psi})\gamma_
{\mu}P_Jv_{\ell_i}(p_{\ell_i})\nonumber\\
&+\tfrac{i}{2}\, C[{\cal T}_{ijk}^{\ell q\, J}]\,\left<M_1(p_1)M_2(p_2)\right|\bar{Q}_j\Sigma^{\mu\nu}Q_k\left|0\right>\,\bar{v}_{\psi}(p_{\psi})
\tfrac{\Sigma_{\mu\nu}}{2}P_Jv_{\ell_i}(p_{\ell_i})\,.\nonumber
\end{align}
We then introduce the scalar, vector and tensor form-factors (with $s\equiv(p_1+p_2)^2$):
\begin{align}\label{eq:semilepformfac}
&\left<M_1(p_1)M_2(p_2)\right|\bar{Q}_jQ_k\left|0\right>\equiv F^{M_1M_2}_{S\,jk}(s)\,,\\
&\left<M_1(p_1)M_2(p_2)\right|\bar{Q}_j\gamma^{\mu}Q_k\left|0\right>\equiv (p_1+p_2)^{\mu}\tfrac{m_j-m_k}{s}F^{M_1M_2}_{S\,jk}(s)\nonumber \\
&\hspace{5.1cm}+\big[p_1^{\mu}-p_2^{\mu}-\tfrac{m_{M_1}^2-m_{M_2}^2}{s}(p_1+p_2)^{\mu}\big]F^{M_1M_2}_{V\,jk}(s)\,,\nonumber\\
&\left<M_1(p_1)M_2(p_2)\right|\bar{Q}_j\Sigma^{\mu\nu}Q_k\left|0\right>\equiv i(p_1^{\mu}p_2^{\nu}-p_2^{\mu}p_1^{\nu})F^{M_1M_2}_{T\,jk}(s)\,.\nonumber
\end{align}
Here $F^{M_2M_1}_{S\,jk}(s)=F^{M_1M_2}_{S\,jk}(s)$, while $F^{M_2M_1}_{V\,jk}(s)=-F^{M_1M_2}_{V\,jk}(s)$ and  $F^{M_2M_1}_{T\,jk}(s)
=-F^{M_1M_2}_{T\,jk}(s)$.
The two-meson form factors automatically include contributions from scalar, vector or tensor resonances. They can be computed 
from chiral perturbation theory in the low-energy range: we refer the reader to Ref.\cite{Shi:2020rkz} for a recent and comparatively 
exhaustive derivation. We note in passing that radiative corrections (including resonance effects) significantly modify the leading 
order contribution in chiral perturbation theory, so that no simple approximation is viable in this regime. 

However, theoretical approaches to the two-meson form factors fail for energies somewhat above $1$\,GeV because they rely on the 
two-body approximation for intermediate states in the scattering, this condition not being reliable any longer at higher energies. 
Phenomenological parametrizations exploiting experimental data from tau decays or $e^+e^-$ scattering can be employed essentially 
for vector form factors --- see \textit{e.g.} Ref.~\cite{Pich:2013lsa}. The case of scalar form-factors, determinant for the RpV 
MSSM, is far more difficult to systematically address without trusting a specific model \cite{Ropertz:2018stk,VonDetten:2021rax}. Here, 
we employ the form factors provided by Ref.\cite{Shi:2020rkz}, which can be used up to an invariant mass of $1.2$\,GeV for the di-meson 
system. Beyond this energy, we simply extrapolate the form factors with a rational function $A/(s-M^2)$, mimicking the mediation by the 
principal resonance: this is a coarse approximation for this higher-energy regime. For the vector and tensor currents, it amounts to 
neglecting the impact of secondary resonances, which we believe is a reasonable approximation. The situation for scalar currents is more 
complicated as the various form factors at $1.2$\,GeV have not yet reached a `tail' behavior. As the scalar operators contain the main 
contributions in the RpV MSSM, this means that the description of semi-leptonic decays at masses $m_{\psi}\gsim1.2$\,GeV remains purely 
qualitative. In addition, it is likely that four-body decay channels become relevant already at energies below the tau mass, so that a partonic 
description of the inclusive hadronic width at $m_{\psi}\approx m_{\tau}$ may become preferable to a channel-by-channel approach.

Once the form-factors for the two-meson channels are numerically defined, we can derive the corresponding $48$ (+ $48$ CP-conjugate) 
decay widths for the neutralino:
\begin{equation}\label{eq:SL3bdecwi}
\Gamma[\psi\to\bar{\ell}_iM_1M_2]=\frac{m_{\psi}^3B_0^2}{512(1+\delta_{M_1M_2})\pi^3f_{\chi}^2}\sum_{\Omega,\Omega'}C[\Omega]
C[\Omega']^*\,G^{\bar{\ell}_iM_1M_2}[\Omega,\Omega']\,,
\end{equation}
where the sum extends over all relevant pairs of semi-leptonic operators $(\Omega,\Omega')$. The kinematic integrals $G^{\bar{\ell}_iM_1
M_2}[\Omega,\Omega']$ are defined in Appendix~\ref{ap:semilepdec} and \textit{a priori} depend on all the masses of the process.

Given the considered range of neutralino mass, we will assume that the available semi-leptonic final states are dominated by the one- and 
two-meson channels, without need of addressing the three-body channels (beyond the resonant contribution resulting from single meson 
production). At higher masses $\sim m_{\tau}$, the partonic description --- see Appendix~\ref{ap:partdecwi} --- should become applicable.

\section{About the Production of the Light Neutralino}
While we focus on the decays of the light exotic fermion, which can be conveniently classified through the listing of the low energy 
operators, the phenomenology of such a particle also crucially depends on its production modes. The latter can be extremely diverse and 
model-dependent since they may occur at any scale, depending on the beam energy of the collider. If the light bino-like particle is 
sufficiently long-lived, it may even possess thermal relics, allowing for tests in Dark-Matter detection experiments. Here, we restrict 
ourselves to a short discussion of typical production channels at colliders.

\subsection{R-parity conserving production modes}
Contrary to its decays, the dominant production channels of the light exotic fermion may not necessarily violate the $\mathbb{Z}_2$ 
symmetry distinguishing it from the SM matter (R-parity in the MSSM). Thus, $\psi$ may appear in cascade decays of heavy resonances 
produced in high-energy collisions. It then becomes necessary to assess the abundance of these $\mathbb{Z}_2$-conserving cascades 
with respect to the competition of a direct $\mathbb{Z}_2$-violating decay of the heavy resonances into SM matter.

If we turn to the direct production of the light fermion from SM sources (\textit{i.e.}~not involving further new physics states), then 
$\mathbb{Z}_2$ conservation implies a production in pairs. The most obvious `portals' with SM matter then involve the $Z$- or Higgs 
bosons via renormalizable operators of the form $Z_{\mu}\,\bar{\psi}\bar{\sigma}^{\mu}\psi$ and $h^0\,\psi\psi$. Limits on invisible (or exotic) 
$Z$ and Higgs decays place strict limits on such production mechanisms. In the case of the RpV MSSM, the couplings of $\psi$ to the 
$Z$-boson or the SM-like Higgs would involve the subleading higgsino components of this state (while the dominant bino component remains 
inert), hence providing a straightforward suppression mechanism. However, precision experiments, \textit{e.g.}~at a linear collider or a 
$Z$-factory, might allow to detect small exotic decay widths of the neutral SM bosons. This possibility was examined in 
\textit{e.g.}~Refs.\,\cite{Dercks:2018wum,Wang:2019orr}.

At low energy, pair production would proceed through non-renormalizable operators of the form (restricting ourselves to dimension $6$ and 
omitting the hermitian conjugates; $f=e,u,d$; $F=(f,\bar{f^c})$; $j,k=1,2,3$; $J=L,R$):
\begin{equation}\label{eq:chipairop}
\dot{\cal S}_{jk}^{f\,J}\equiv(\bar{\Psi}P_L\Psi)(\bar{F}_jP_JF_k)\,,\hspace{1.5cm}\dot{\cal V}_{jk}^{f\,J}\equiv(\bar{\Psi}\gamma^{\mu}P_L
\Psi)(\bar{F}_j\gamma_{\mu}P_JF_k)\,,
\end{equation}
satisfying the $U(1)_{\text{em}}\times SU(3)_c$ gauge symmetry (and possibly additional discrete symmetries depending on the model) --- 
the tensor operator $\dot{\cal T}_{jk}^{f}\equiv\frac{1}{4}(\bar{\Psi}\Sigma^{\mu\nu}P_L\Psi)(\bar{F}_j\Sigma_{\mu\nu}P_LF_k)$ identically 
vanishes. In particular, such operators involving quarks open up the possibility of $\psi$ production in the decays (or collisions) of hadrons 
(of various spin). Let us illustrate this feature with an example: we consider the low-energy operators $(\bar{\Psi}\,P_{L} \Psi)(\bar{B}\,P_
{L,R}S)$. Then, $\psi$ pairs can be produced in a direct $B_s^0$ decay (calculable via the decay constant $\left<0\right|\bar{B}\gamma_5
S\left|B_s^0\right>$), or in semi-leptonic $B$ decays, such as $B\to K \psi\psi$ (calculable with the $B\to K$ form factor) and more generally 
$B\to X_s \psi\psi$ (calculable using the heavy quark expansion). These individual decay modes might be testable at a $B$ factory.\footnote{See also
Ref.~\cite{Dreiner:2009er} for the related decay $K\to\pi\psi\psi$.} Similarly, 
the $(\bar{\Psi}\,\gamma^{\mu}P_{L} \Psi)(\bar{C}\,\gamma^{\mu}P_{L,R}C)$ or $(\bar{\Psi}\,\gamma^{\mu}P_{L} \Psi)(\bar{B}\,\gamma^{\mu}
P_{L,R}B)$ operators would open up the possibility of production from vector quarkonia, such as the $J/\psi$ or the $\Upsilon$. For meson decays to two light neutralinos see 
Ref.~\cite{Dreiner:2009er}. The situation 
is more complicated if the hadronic initial state produced in the experiment is not clearly identified (because many hadrons can potentially 
decay via the low-energy operator that couples them to the $\psi$ pair), so that $\psi$ production in a $b$-jet, for instance, would have to be 
modeled together with the hadronization string for a quantitative estimate. Leptonic four-fermion operators $(\dot{\cal S},\dot{\cal V},\dot{\cal 
T})^{e\,(J)}_{jk}$ may also be exploited to search for $\psi$'s in lepton-antilepton decays or, in the presence of lepton-flavor violation, in leptonic 
tau (or muon) decays.

Contributions to the operators of Eq.~(\ref{eq:chipairop}) depend on the UV-completion. In the RpV MSSM at tree level, typical `s-channel' 
mediators would be the $Z$- and Higgs bosons. In view of the limits on invisible / exotic decays of the $Z$ or the SM-like Higgs, these two 
mediators are probably not competitive. Heavy Higgs doublets would be legitimate mediators (at least in couplings to third-generation fermions) 
provided their coupling to the bino-dominated state is not exceedingly suppressed. Sfermion exchange provides another type of `t-channel' 
contribution. In fact, the matching conditions for the $(\dot{\cal S},\dot{\cal V},\dot{\cal T})^{f\,(J)}_{jk}$ operators in the RpV MSSM at tree-level 
are virtually identical to those of the $({\cal S,V,T})^{\nu f J(K)}_{ijk}$ operators ($f=e,u,d$) provided in Eqs.(\ref{eq:matchnulep},\ref{eq:matchselep}), 
up to the substitution of $\nu_i$ by $\psi$ and the generalization of the indices $j$ and $k$, so as to include $\tau$, $c$ or $b$. At the loop 
level, additional contributions from \textit{e.g.}~SUSY box diagrams are also expected. The evolution under the renormalization group is also 
unchanged, see Eq.~(\ref{eq:RGEselep}), although the infrared scale for production should be adjusted to the corresponding center-of-mass energy.

\subsection{R-parity violating production modes}
Violation of the $\mathbb{Z}_2$ symmetry allows for single production of particles in the `exotic' sector. Again, $\psi$ might then occur as the 
end product of the $\mathbb{Z}_2$-conserving cascade decay of a singly-produced heavy resonance, provided the competition of 
$\mathbb{Z}_2$-violating decays does not overshadow such modes. In the RpV MSSM, a heavy Higgs state might also exploit its subleading 
slepton component to decay into $\psi$ and a lepton.

Direct production of $\psi$ from SM particles can proceed through low-energy operators of the type displayed in 
Eqs.(\ref{eq:elmdipop}-\ref{eq:hadop}), but extended to heavier fermions. The matching conditions at tree-level in the RpV MSSM are again a 
simple generalization of Eqs.(\ref{eq:matchnulep}-\ref{eq:matchhad}). Lepton decays would then appear as a first potential lepton-number 
violating source of the light exotic fermion. For instance, the decay widths associated with $\tau\to\psi\nu_i\bar{\nu_j}$, $\tau\to\psi\ell_i\bar
{\ell_j}$, $\tau\to\psi M_i$, or $\tau\to\psi M_iM_j$ can be computed similarly to the leptonic and semi-leptonic decays of the light neutralino [see 
Eqs.(\ref{eq:3nudec},\ref{eq:3nnbdec},\ref{eq:lepkinint},\ref{eq:SL2bdecwi},\ref{eq:SL2bVdecwi},\ref{eq:SL3bdecwi})]. For a very light fermion, 
a partonic approach may efficiently describe the inclusive $\tau$ decays into $\psi$ and light hadrons, since the large phase space suggests a 
possible saturation of the hadronic width. Individual channels have been considered in Ref.~\cite{Dey:2020juy}. The limited phase space in 
muon or electron decays restricts the possibility of $\psi$ production to leptonic channels.

Operators of semi-leptonic and hadronic types also suggest single production of the exotic fermion in hadron decays. Exchanging the fermions 
in the initial and final states in Eq.~(\ref{eq:haddecwi}) opens the path to a lepton-number conserving disintegration of baryons and, in particular, 
of nucleons \cite{Chamoun:2020aft}: this is a very constrained production channel, but constantly investigated by experiments. Now turning to 
operators including a bottom (generated in the RpV MSSM via \textit{e.g.}~$\lambda''_{113}$) or charm quark (via \textit{e.g.}~$\lambda''_{212}$), 
one can expect single production from beauty- or charm-flavored hadrons. The weight of a few inclusive channels, such as $B,D\to X\psi$, may be 
estimated from the partonic picture. On the other hand, exclusive channels, such as $B^+,D^+\to p^+ \psi$ or $\Lambda_{b,c}^0\to\pi^0\psi$, 
are much more difficult to assess since they would require the knowledge of baryon-number and flavor violating form factors ($B^+\to p^+$ or 
$\Lambda^0_b\to\pi^0$ in our example). As in the case of pair production, the $\psi$ content of a $b$- (or $c$-) jet due to the presence of 
baryon-number violating operators would probably require detailed modeling. The situation is comparable when considering semi-leptonic 
operators, even though it should be possible to theoretically describe a few individual exclusive channels testable at $B$-factories: the 
corresponding form-factors (\textit{e.g.}~$B\to K$) are no longer baryon-number violating, hence more easily accessible. A few such processes 
have been exploited in \textit{e.g.}~Refs.\cite{deVries:2015mfw,Helo:2018qej,Dercks:2018wum}.

To summarize this short discussion, the potential production channels of a light neutralino-like particle are even more diverse than its decay 
modes. Many transitions may be difficult to describe accurately with the usual techniques, as they would call upon unknown form-factors or 
require a dedicated hadronization model. The relative magnitude of pair and single production channels should also be systematically compared, 
as these sources are largely independent from one another. Given that our focus in this paper are the decays of the neutralino-like particle, we 
will leave this discussion of the production modes at this qualitative level. A few formulae for the production of a light neutralino in the decays of 
SM particles are provided for reference in Appendix~\ref{ap:prod}.

\section{Numerical Applications}

The formalism described in the previous sections is collected within a Mathematica package, which allows for the computation of all the considered 
decay widths of a light neutralino for the most general MSSM input. Numerical comparison of a few individual semi-leptonic widths with a private 
code, derived from Ref.\,\cite{deVries:2015mfw} and subsequently employed in simulations of far-detector sensitivities, returned discrepancies 
by a factor $10$ (underestimating the pseudoscalar meson channel) to $100$ (overestimating the vector meson channel). These differences 
were traced back to the impact of QCD running --- see Eq.~(\ref{eq:RGEselep}) ---, the issues in matching raised at the very end of 
Section~\ref{sec:matching} and numerical bugs. We here demonstrate a few possible applications of these calculations at the numerical level.

\subsection{Baryon-number violation}

\begin{figure}[p]
\begin{center}
	\includegraphics[width=0.9\linewidth]{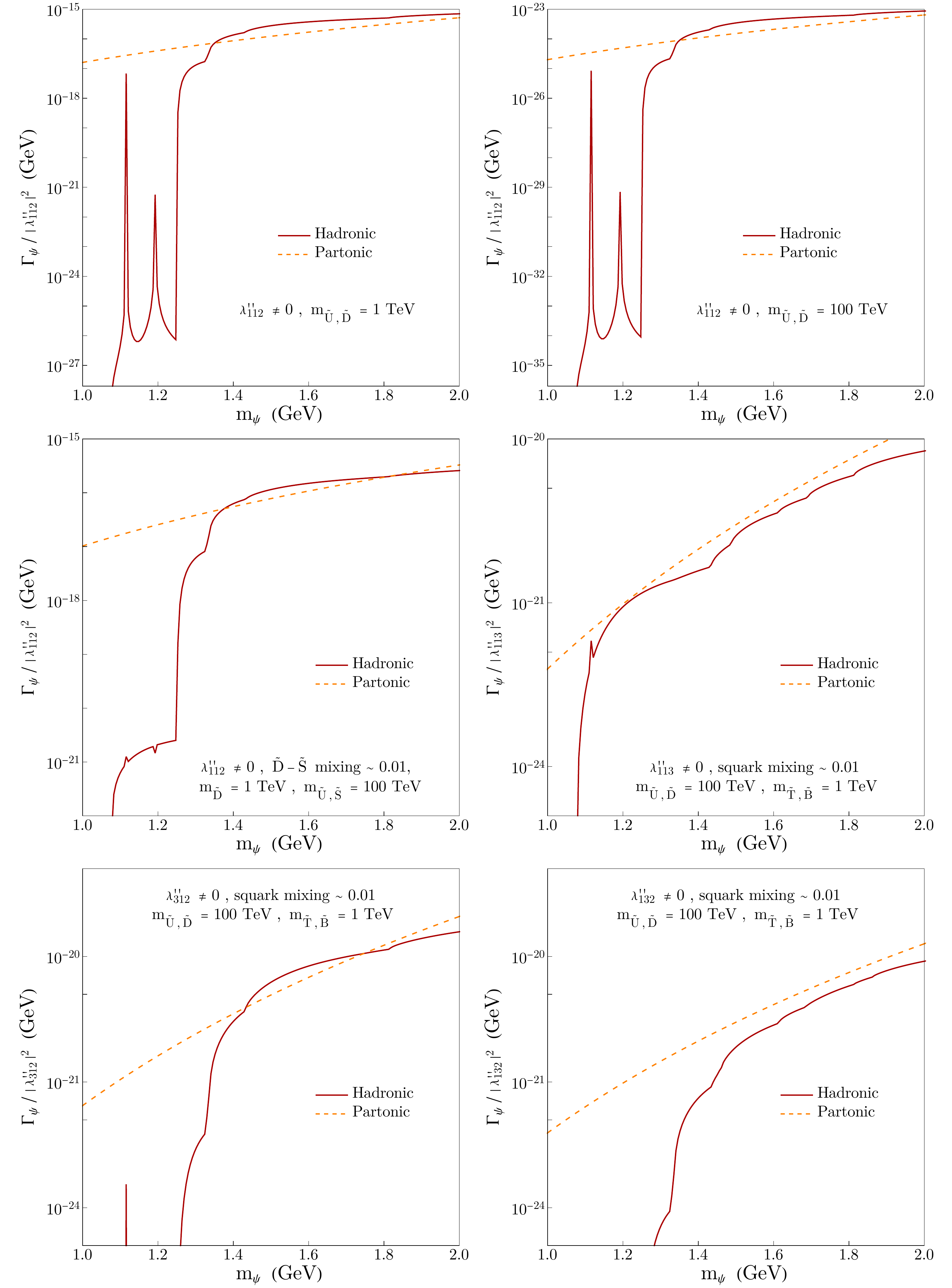}
	\caption{\small  Hadronic decay width of the neutralino in various scenarios.
		\textit{First row}: No squark mixing; the neutralino decays are mediated by the (right-handed) squarks of first and second generation 
		with mass at $1$\,TeV (Left) and $100$\,TeV (Right).
		\textit{Second row, left}: Mixing of magnitude $10^{-2}$ between right-handed light down-type ($1$\,TeV) and heavy strange-type 
		squarks ($100$\,TeV).
		\textit{Second (right) and third row}: Mixing of $10^{-2}$ between the heavy squarks of first generations ($100$\,TeV) and the light 
		squarks of third generation ($1$\,TeV); the dominant coupling of {\sc udd}-type is chosen differently in the three cases: $\lambda''_{113}$, 
		$\lambda''_{312}$ and $\lambda''_{132}$.
		The inclusive hadronic width, normalized to the dominant coupling of {\sc udd}-type squared is displayed in red. The partonic 
		approximation is shown in dashed orange.
		\label{fig:Hadplot}}
\end{center}
\end{figure}
We first assume that lepton number is conserved. Neutralino decays are then driven by baryon-number violating couplings. We stress again 
that hadronic decays are kinematically forbidden for a fermion with mass below the proton mass, and in fact, as long as we neglect the radiative 
hadronic decay $\psi\to n^0\gamma$, below the threshold for nucleon+pion production, \textit{i.e.}~$\sim1.075$\,GeV. Then, our description for 
such channels is at best of qualitative value, due first to the limited validity of the chiral description, especially in the higher mass range, and 
also to the restriction to the fundamental baryon octet, beyond which further hadronic states become available almost from the opening 
of the kinematic window for hadronic decays.

If flavor mixing is negligible in the sfermion sector, only one RpV parameter is liable to mediate $\psi$ decays at tree-level: $\lambda''_{112}=-
\lambda''_{121}$. Decay widths will then scale like $|\lambda''_{112}|^2$. Yet, such a coupling cannot lead to decays into nucleons and pions (the 
kinematically available channels at low mass) without a source of $s\to d$ transition, and the kinematic threshold is pushed further up to the 
opening of $\Lambda^0+\pi^0$ production (at $\sim1.25$\,GeV). Nevertheless, the weak interaction is always available as a source of flavor 
violation and can mediate neutralino decays into nucleons and pions at low mass via the mixings of $\psi$ with the hyperons: the recourse to the 
weak Lagrangian considerably reduces the widths, however.\footnote{From a high-energy model building point of view one might expect the
coupling $\lambda''_{112}$ not to appear alone, but those couplings related by CKM mixing to also be present in the theory, correspondingly 
suppressed. See for example \cite{Dreiner:1991pe,Agashe:1995qm}.} The total neutralino width in such a setup is presented in the first row of 
Fig.\,\ref{fig:Hadplot} for squark masses of $1$\,TeV (left) and $100$\,TeV (right). The solid red line represents the actual hadronic width, while the 
dashed orange curve corresponds to the partonic three-body approximation (hence ignores hadronic thresholds). The scaling of the Wilson 
coefficients as the inverse of the mass-squared of the squark mediator amounts to a $10^{-8}$ suppression at the level of the decay width in the 
scenario with the heavier sfermions --- QCD logarithms also produce a prefactor of ${\cal O}(1)$. In the mass range accessible only to the 
nucleon+pion final states, the decay width is significantly suppressed (as compared to the partonic width) and clearly dominated by the resonant 
behavior caused by the $\psi$ mixing with $\Lambda^0$ and $\Sigma^0$. Above $\sim1.25$\,GeV, the hadronic neutralino decays can proceed 
without recourse to the weak interaction and promptly reach a magnitude comparable with the partonic width.\footnote{For a related comparison of
partonic decay widths to those incorporating hadronic thresholds in the case of sterile neutrinos see Ref.~\cite{Bondarenko:2018ptm}.} The latter 
may actually offer a quantitatively more reliable description of the (un-charmed) inclusive hadronic decay width for $m_{\psi}\to2$\,GeV.

The presence of squark mixing may substantially alter the picture presented above, in particular with respect to the relative importance of 
nucleon+pion channels as opposed to the strange-violating ones. In the second row of Fig.\,\ref{fig:Hadplot}, on the left, we allow for a mixing of 
magnitude $10^{-2}$ between a light right-handed sdown at $1$\,TeV and the heavy sstrange at $100$\,TeV. Then the nucleon+pion decays only 
receive a mixing squared suppression ($10^{-4}$) with respect to the partonic width, instead of the $\sim10^{-9}$ suppression observed in the 
scenarios without squark mixing of the first row. Squark mixing also allows for other couplings of {\sc udd}-type, beyond $\lambda''_{112}$, to 
contribute to the neutralino decays: such effects will always be suppressed in proportion to the needed mixing to couple squark mediators to the 
light quarks. The last three plots of Fig.\,\ref{fig:Hadplot} illustrate this setup with various choices of dominant {\sc udd}-coupling ($\lambda''_{113}$, 
$\lambda''_{312}$ and $\lambda''_{132}$) and a mixing of magnitude $10^{-2}$ between heavy squarks of first and second generations ($100$\,TeV) 
and light squarks of third generation ($1$\,TeV). Again, the relative importance of the decays into nucleons and pions depends on the possibility 
to generate a contribution to the $\widetilde{\cal H}_{111}^{JK}$ operators via squark mixing. In this respect, the scenario with dominant $\lambda
''_{113}$ results in no suppression of the strange-number conserving channels with respect to the strange-violating ones, since all require the same 
degree of mixing of the squarks of third generation with those of first and second generations (this mixing was made symmetrical for both 
generations in our example). On the other hand, strange-number conserving decays in the scenarios with dominant $\lambda''_{312}$ or $\lambda''
_{132}$ require one additional degree of flavor mixing as compared to the strange-violating ones, explaining the suppression of the nucleon+lepton 
channels in the corresponding plots.

We note the comparative agreement of the hadronic and partonic widths in the higher mass range (once enough decay channels are kinematically 
accessible) for all scenarios of Fig.\,\ref{fig:Hadplot}. This situation may be largely coincidental since the chiral description is not expected to be 
predictive (or even exhaustive) in this regime, while the partonic description is insensitive to hadronic thresholds, hence only valid at sufficiently high 
mass. However, this agreement allows for a comparatively smooth transition between the two models at $m_{\psi}\approx1.5$\,GeV. Finally, we 
stress that the hadronic description of the baryon-number violating scenario should also be put into perspective with its consequences for the 
phenomenology of baryons. Correspondingly, constraints on nucleon decays \cite{Chamoun:2020aft} apply for $m_{\psi}\lsim m_{p}$ while limits on 
neutron-antineutron oscillations --- see Appendix~\ref{ap:nnbarosc} --- restrict the acceptable magnitude of contributions to the $\widetilde{\cal H}_
{111}^{JK}$ operators (those relevant for strange-number conserving neutralino decays).

\subsection{LQ$\bar{\mathrm{D}}$ couplings and semi-leptonic decays}

\begin{figure}[th!]
\begin{center}
	\includegraphics[width=0.98\linewidth]{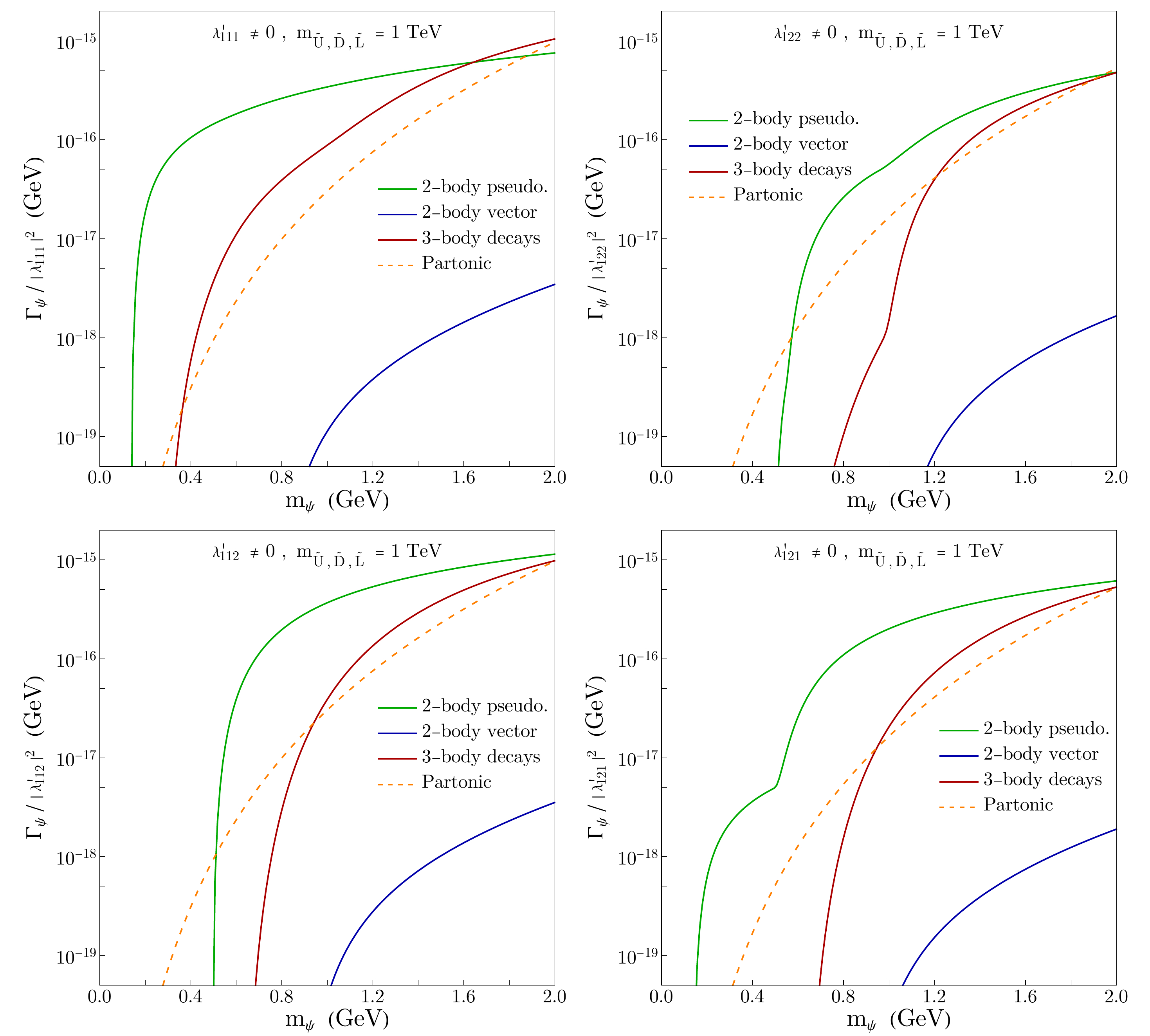}
	\caption{\small Semi-Leptonic decay width of the neutralino in various scenarios without sfermion mixing. Both slepton and squark masses 
	are set to $1$\,TeV.
	\textit{First row}: quark-flavor conserving scenarios; lepton-number violation is mediated by the $\lambda'_{111}$ coupling (left) or the 
	$\lambda'_{122}$ coupling (right).
	\textit{Second row}: the lepton-number violating coupling also violates the strange-number; it is chosen as $\lambda'_{112}$ in the left plot 
	and as $\lambda'_{121}$ in the right one.
	The green, blue and red curves respectively represent the two-body decay width into one lepton and a pseudoscalar meson, the two-body 
	decay width into one lepton and a vector meson and the three-body decay width into one lepton and two pseudoscalar mesons.
	\label{fig:SeLepplot1}}
\end{center}
\end{figure}

We now turn to the lepton-number violating case and consider the decay channels driven by the couplings of {\sc lqd}-type. In the absence of RpV 
bilinear terms, the semi-leptonic decay amplitudes are all expected to scale linearly with a $\lambda'$ coupling and receive a quadratic suppression 
with the mass of the mediating sfermions.

In Fig.\,\ref{fig:SeLepplot1}, we investigate the neutralino decays in the mass-range $[0,2]$\,GeV in several scenarios without sfermion mixing. 
Correspondingly, the squark (and slepton) mediators always belong to the first or second generation. In the upper left-hand quadrant, we consider 
the decays proceeding through $\lambda'_{111}\neq0$. Wilson coefficients of scalar-type dominate the considered regime, resulting in suppressed 
vector-meson production (mostly driven by tensor operators here; shown in blue) as compared to the channels with pseudoscalar mesons. 
Unsurprisingly, the two-body decays with one pseudoscalar meson in the final state (shown in green) primarily involve (neutral or charged) pions, 
and secondarily, $\eta$ ($\eta'$): these are the channels coupling to the $\bar{U}D$ and $\bar{D}D$ pseudoscalar sources opened by $\lambda'_
{111}$. In addition, despite the kinematic suppression, three-body decays (shown in red) are quite important and eventually dominate the 
semi-leptonic width at high mass. At $m_{\psi}\approx1.2$\,GeV (the upper limit for fully reliable form factors), they already amount to about $50\%$ 
of the magnitude of the two-body widths: the radiation of an extra pion indeed costs little phase space at such energies and the relevant channels 
thus involve two pions or pion+$\eta$ in their final state, which also couple to the $\bar{U}D$ and $\bar{D}D$ scalar sources. 

In the upper right-hand 
quadrant of Fig.\,\ref{fig:SeLepplot1}, we turn to the case of a dominant $\lambda'_{122}$, hence switching on the $\bar{S}S$ scalar and pseudoscalar 
sources (charm production is kinematically forbidden). As a result, the two-body decays of the neutralino are dominated by $\eta$ or $\eta'$ production. 
The two-pion channels dominate the three-body decay widths: while $K-\bar{K}$ is allowed, these channels only open in the higher range of mass. In 
addition, the Cabibbo Kobayashi Maskawa (CKM) mixing between first and second generation ($V^{\text{CKM}}_{us}\approx0.225$) opens a hardly 
suppressed $\bar{U}S$ source which controls the channels with charged leptons in the final state and leads to substantial kaon (and $K+\pi$) 
production. In any case, this scenario is characterized by a higher threshold for neutralino decays, since favored final states are all more massive than 
in the scenario with dominant $\lambda'_{111}$.

In the second row of Fig.\,\ref{fig:SeLepplot1}, we consider {\sc lqd} couplings that violate the strange number: $\lambda'_{112}$ (left) and $\lambda'_{121}$ (right). The resulting $\bar{U}S$, $\bar{D}S$ or $\bar{S}D$ (pseudo)scalar sources trigger kaon production in semi-leptonic neutralino decays. Three-body channels are again competitive in the higher mass window, involving the radiation of one extra pion. A noteworthy difference between the two scenarios emerges from CKM mixing, as the $V^{\text{CKM}}_{us}$ allows for pion+electron final states, already accessible in the lower mass range, in the case where the second-generation index is carried by the $SU(2)_L$ doublet, but not if it is carried by the $SU(2)_L$ singlet.

\begin{figure}[tbh!]
	\begin{center}
		\includegraphics[width=0.98\linewidth]{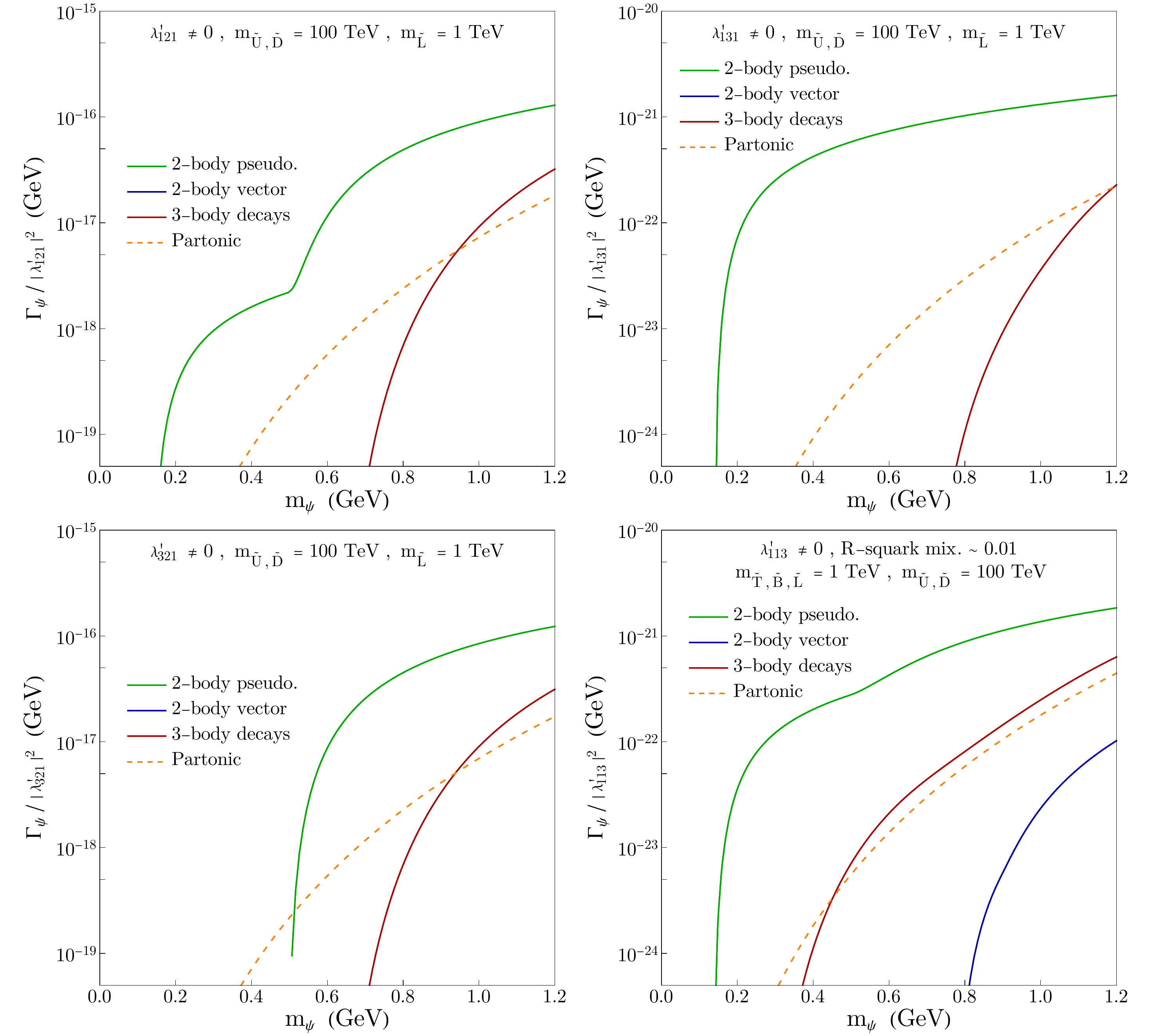}
		\caption{\small Sfermion effects in semi-Leptonic decay width of the neutralino in scenarios. The squark masses of the first two generations are set to $100$\,TeV, while the third generation squarks and the sleptons remain at about $1$\,TeV. 
			\textit{Left}: {\sc lqd} couplings violating the strange-number are considered, $\lambda'_{121}$ (up) and $\lambda'_{321}$ (down) in the presence of heavy squarks;
			\textit{Right}: {\sc lqd} couplings involving the third generation squarks, $\lambda'_{131}$ or $\lambda'_{113}$, are considered without (up) or in the presence of flavor-violating squark mixing (down).
			The color conventions are similar to those of Fig.\,\ref{fig:SeLepplot1}.
			\label{fig:SeLepplot2}}
	\end{center}
\end{figure}

In all the scenarios of Fig.\,\ref{fig:SeLepplot1}, we have displayed the inclusive partonic width as a dashed orange curve. Such a description is obviously unreliable in the lower mass-regime, as it ignores thresholds and tends to underestimate the actual semi-leptonic decay width by orders of magnitude (due to the three-body phase-space suppression). Yet, as the semi-leptonic width becomes increasingly controlled by the production of meson pairs for $m_{\psi}\gsim1.5$\,GeV, the partonic approach also starts providing an inclusive width of comparable magnitude. In fact, the partonic description of uncharmed semi-leptonic decays may be quantitatively more meaningful at $m_{\psi}\approx2$\,GeV, where the mesonic form factors are largely speculative.

In Fig.\,\ref{fig:SeLepplot2}, we study the impact of the sfermion sector on the neutralino decays. We restrict ourselves to masses $m_{\psi}\leq1.2$, where the two-meson form factors are fully reliable. In the plots on the left-hand side, we linger on scenarios with strange-number violation and no squark mixing, but consider very massive squarks of first and second generations at $100$\,TeV, instead of $1$\,TeV in the previous set of plots --- sleptons continue to be comparatively light, at $1$\,TeV. The mediation of squarks in low-energy processes is thus inhibited. However, in the upper plot, we observe that the suppression of squark mediation hardly affects the magnitude of neutralino decay channels driven by (pseudo)scalar $\bar{Q}Q$ currents --- a factor $\sim0.5$ as compared to the corresponding scenario in Fig.\,\ref{fig:SeLepplot1}. Indeed, these operators also receive contributions from slepton mediators, which remain effective independently from the squark sector. On the other hand, the tensor (and vector) currents receive only contributions from squark mediators, so that the widths involving vector mesons in the final state are even more suppressed than in the previous scenarios. In the lower plot, we vary the lepton index of the {\sc lqd} coupling, setting it to $\lambda'_{321}$: as tau production is kinematically inaccessible in the decays of a light neutralino, only decay channels involving a neutrino in the final state are allowed, which contrasts with the scenario of the upper plot.

On the right-hand side of Fig.\,\ref{fig:SeLepplot2}, we consider mixing effects in the quark or squark sectors. We already stressed, when discussing Fig.\,\ref{fig:SeLepplot1}, that $\lambda'_{121}$ opened electron+pion decays through the CKM mixing, but not $\lambda'_{112}$. The same applies to $\lambda'_{131}$, considered in the upper plot, with an additional suppression of $|V^{\text{CKM}}_{ub}/V^{\text{CKM}}_{us}|^2\approx3\cdot10^{-4}$. On the other hand, while strange-number violating decays are allowed with $\lambda'_{121}$ (see plot in the upper left-hand corner), this is no longer possible with third generation quarks, so that the only allowed decay channels in the scenario with dominant $\lambda'_{131}$ are indeed these CKM-rotated ones involving electrons and pions. In the lower plot, we consider a dominant $\lambda'_{113}$: this coupling cannot funnel the decays of a light neutralino except in the presence of squark mixing involving the right-handed sbottom. Therefore, we introduce a mixing of $\mathcal{O}(10^{-2})$ between down-type squarks of the two first generations and of the third one, resulting in neutralino decays into leptons and pions, $\eta$ or kaons. As the squark mediation is crucial in the neutralino widths employing squark mixing, we observe that the decays into vector mesons recover their relative, though subleading, importance with respect to the channels involving a single pseudoscalar meson. Finally, we stress that similar effects are triggered by slepton mixing and that, in the scenario with dominant $\lambda'_{321}$ (in the lower left-end corner of Fig.\,\ref{fig:SeLepplot2}), decays involving electrons or muons in the final state could be restored in this fashion.

To summarize this discussion, we have observed that the semi-leptonic decays of the neutralino driven by {\sc lqd}-couplings are controlled by scalar and pseudoscalar contributions. In particular, the production of a single meson dominates the low-mass window, with meson pair production becoming competitive as soon as pion radiation costs little relative energy, at $m_{\psi}\approx1$\,GeV. A partonic description may become reliable around the tau mass. Furthermore, it would be relevant to place flavor-violating effects in neutralino decays into the perspective of known constraints on flavor transitions.

\subsection{Lepton-number violating bilinear couplings}
The interplay of trilinear couplings of {\sc lle}-type with leptonic neutralino decays offers limited novelty. Instead, we here consider the decays triggered by RpV bilinear terms. We observe that such terms are generically present in the Lagrangian and generated by the renormalization group as soon as lepton number is violated. There exist two categories of bilinear couplings in the RpV MSSM:
\begin{enumerate}
\item the mass terms $\mu_i$, $i=1,2,3$, introduced in the superpotential in particular produce lepton-electrowikino mixing, opening the possibility of breaking lepton number directly at the level of the external legs of the operators mediating neutralino decays;
\item the mass-squared terms $B_i$, $i=1,2,3$, introduced in the soft SUSY-breaking Lagrangian result in admixtures between the heavy Higgs doublets and the slepton sector: lepton number may then be violated in neutralino decays by the double nature of the scalar mediators.
\end{enumerate}

\begin{figure}[b!]
\begin{center}
\includegraphics[width=0.98\linewidth]{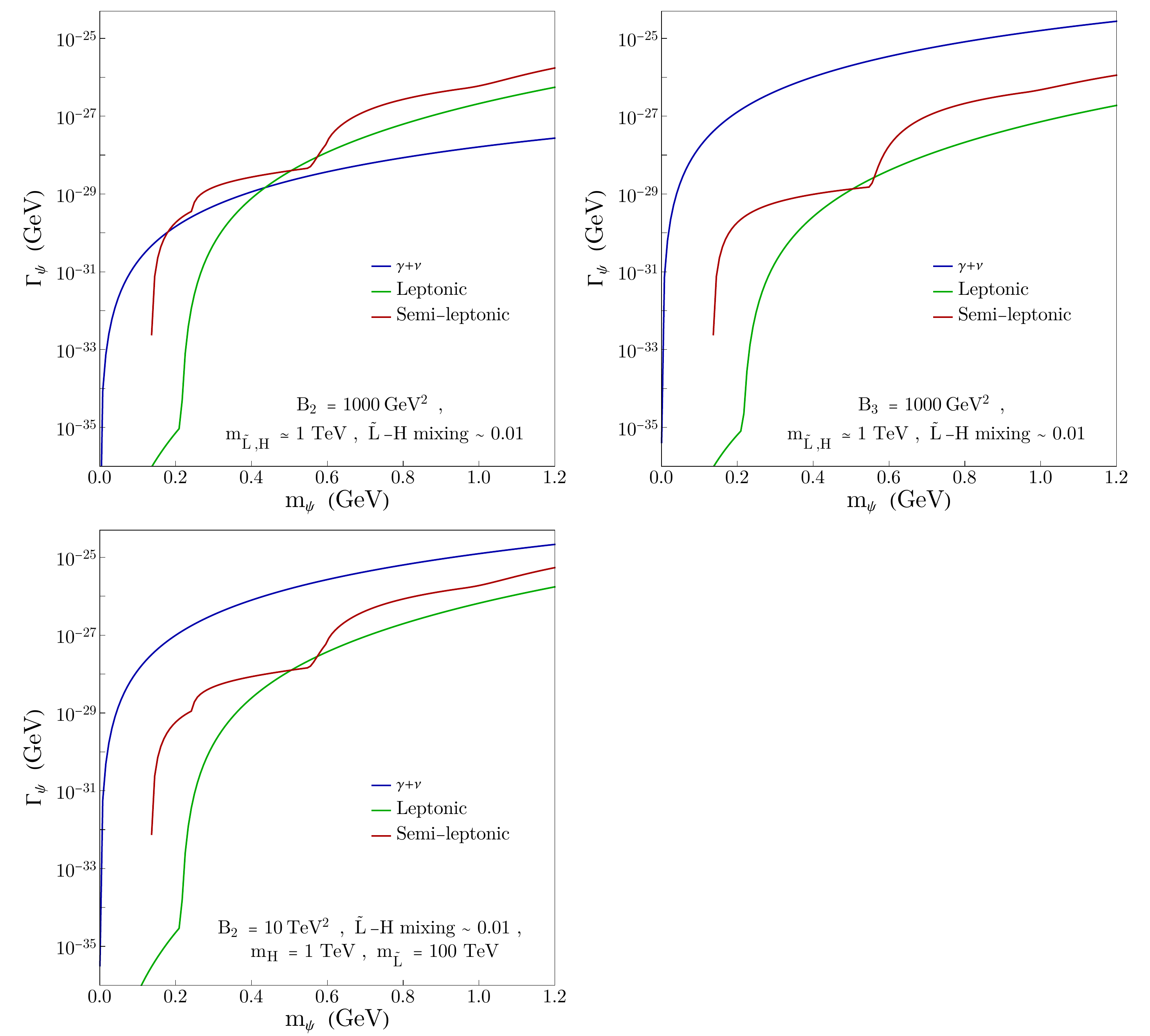}
\caption{\small Neutralino decays mediated by bilinear scalar terms. Higgs-slepton mixing is maintained at the level of $\sim10^{-2}$.
	\textit{Left}: neutralino decays employ the non-vanishing $B_2$, resulting in Higgs-smuon mixing; all these scalars take mass in the TeV range in the upper plot while sleptons are made very heavy ($100$\,TeV) in the lower plot.
	\textit{Right}: $B_3$, generating Higgs-stau mixing is chosen to dominate the decays.
	The blue, green and red curves respectively represent the inclusive photon+neutrino, leptonic and semi-leptonic widths.
	\label{fig:ScBilplot}}
\end{center}
\end{figure}

We first focus on scenarios with non-vanishing scalar bilinears $B_i$. Considering that it is necessary to break lepton number in diagrams mediating 
neutralino decays, hence play on the double nature of the slepton-Higgs admixtures, contributions to four-fermion operators then necessarily exploit Higgs 
couplings to SM fermions: this implies a Yukawa suppression for all the decay amplitudes involving light fermions, adding to the scale and mixing 
suppression factors. Consequently, the neutralino partial widths conveyed by the $B_i$ terms are typically very narrow. On the other hand, contrarily to the 
trilinear terms, the individual bilinear couplings do not specify the nature of the ensuing (lepton-number violating) decays, so that semi-leptonic channels 
dominated by the strange Yukawa coexist with leptonic channels triggered by the muon Yukawa and electromagnetic decays conveyed by chargino / 
slepton-Higgs loops. The latter escape the Yukawa suppression while the loop-factor suppression might be alleviated by the logarithmic dependence in case 
of hierarchical spectra.

We illustrate this discussion about neutralino decays mediated by the bilinear scalar terms with Fig.\,\ref{fig:ScBilplot}. In the upper left-hand quadrant, through $B_2\neq0$, we generate a mixing of magnitude $\sim10^{-2}$ between heavy Higgs and smuons taking mass in the TeV range. The higher mass-window is dominated by semi-leptonic neutralino decays conveyed by the strange Yukawa coupling, in particular the $\eta^{(\prime)}\nu_{\mu}$ and, to a lesser extent (after $V^{\text{CKM}}_{us}$ suppression), the $K^+\mu^-$ channels. Below threshold for these, the $\pi^0\nu_{\mu}$ and $\pi^+\mu^-$ decays can still contribute, though at a reduced rate. Similarly to the previous subsection, channels involving meson pairs in the final state contribute at a sizable level in the upper mass range: the $\bar{S}S$ operator mediates $\pi\pi$ as well as $KK$ decays, while $\bar{u}s$ conveys charged $K\pi$ decays. We also observe that the specific pattern imprinted by the $B_i$ couplings favors semi-leptonic scalar contributions over vector and tensor, so that decays into vector mesons are virtually absent. Leptonic decay channels are of comparable magnitude, which can be related to the comparative size of the muon and strange Yukawa couplings, and are mostly represented by the $\nu_{\mu}\mu^+\mu^-$ channel: it is remarkable that this specific decay is not straightforwardly generated with the antisymmetric {\sc lle} couplings, unless slepton mixing is present. $\nu_ee^+\mu^-$ and $\nu_{\mu}e^+e^-$ contribute to the neutralino decays at a far lesser extent. Finally, the radiative decays $\psi\to\gamma\nu_{\mu},\gamma\bar{\nu}_{\mu}$, generated by chargino and Higgs-smuon loops, control the low mass-range.

In the upper right-hand corner of Fig.\,\ref{fig:ScBilplot}, we turn to the case of a dominant $B_3$, resulting in Higgs-stau mixing. The size of leptonic and semi-leptonic decays is not fundamentally modified by this replacement, with all leading channels now involving $\nu_{\tau}$ instead of $\nu_{\mu}$ (while the charged channel are kinematically forbidden by the replacement $\mu\to\tau$). On the other hand, the electromagnetic dipole contributions are greatly enhanced by the additional mediation of $\tau$ + Higgs-stau loops, involving a sizable Yukawa coupling ($\tan\beta$ is set to the typical value of $10$) as well as a large $\ln\tfrac{M_{\text{SUSY}}}{m_{\tau}}$. As a consequence, the $\psi\to\gamma\nu_{\tau},\gamma\bar{\nu}_{\tau}$ widths dominate the neutralino decays in all the considered mass-range.

In the lower plot of Fig.\,\ref{fig:ScBilplot}, we return to the case of a dominant $B_2$, but now with a decoupling SUSY spectrum, albeit the Higgs-slepton mixing is kept at the level of $10^{-2}$. Leptonic and semi-leptonic decay channels receive contributions of comparable magnitude to their counterparts of the upper plot, since the Higgs-dominated mediator in the TeV range has about unchanged properties. On the other hand, the importance of the radiative channels $\psi\to\gamma\nu_{\mu},\gamma\bar{\nu}_{\mu}$ is increased by the mass-splitting between scalar states, as cancellations between the smuon and Higgs loops no longer occur.

\begin{figure}[tbh!]
\begin{center}
	\includegraphics[width=0.98\linewidth]{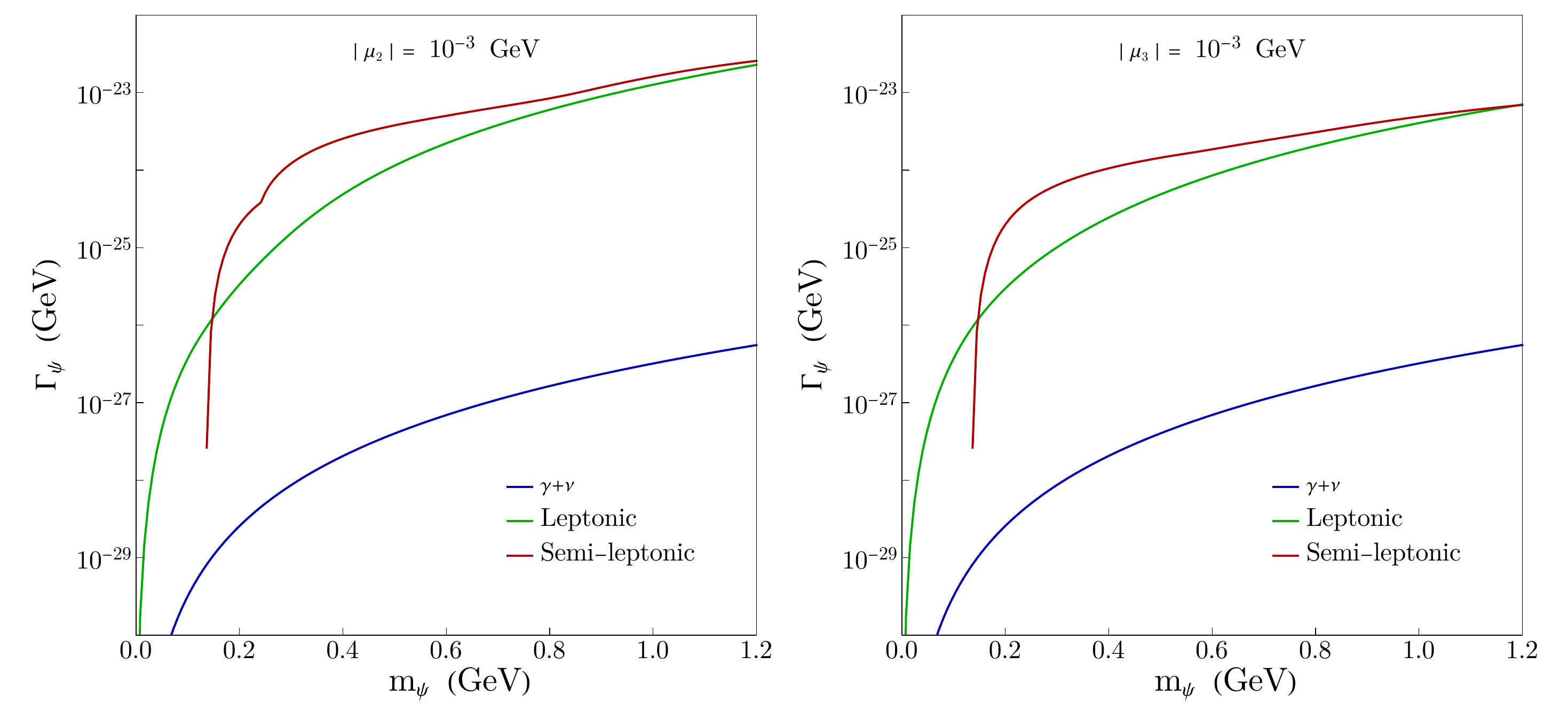}
	\caption{Neutralino decays mediated by bilinear `fermionic' terms..
		\textit{Left}: neutralino decays employ the non-vanishing $\mu_2=10^{-3}$, resulting in chargino-muon and neutralino-$\nu_{\mu}$ mixing.
		\textit{Right}: a non-vanishing $\mu_3$ replaces $\mu_2$.
		The blue, green and red curves respectively represent the inclusive photon+neutrino, leptonic and semi-leptonic widths.
		\label{fig:FeBilplot}}
\end{center}
\end{figure}

In Fig.\,\ref{fig:FeBilplot}, we consider the impact of the supersymmetric bilinear mixing parameters $\mu_i$ on the decays of a light neutralino. 
The main effect rests with the appearance of chargino-lepton and neutralino-neutrino admixtures and the RpV / lepton-number breaking does 
not involve the mediation of heavy fields. Indeed, electroweak gauge bosons then become the main contributors to the dimension $6$ operators 
of Eqs.(\ref{eq:elmdipop}-\ref{eq:seleptop}) and the corresponding neutralino decays hardly depend on the mass of the sfermion fields. 
Nevertheless, large values of the $\mu_i$ parameters would generate unacceptable neutrino masses \cite{Allanach:2003eb}. For small $\mu_i$, 
the latter depend on secondary mixing of the neutrinos with higgsinos and gauginos, so that small mass values are easily explained. In practice, 
we will consider $|\mu_i|=10^{-3}$\,GeV, which typically produces neutrino masses under $10^{-10}$\,GeV in our scenarios. We note that the 
bilinear `fermionic' parameters also generate mixing of the heavy Higgs doublets with the right-handed sleptons (similarly to $B_i$ with the 
left-handed sleptons). Yet, as the effect is proportional to the lepton masses (and the small $\mu_i$), it hardly has a chance to reach 
phenomenological relevance for scalar fields in the TeV range or beyond.

In the plot on the left-hand side of Fig.\,\ref{fig:FeBilplot}, we set $\mu_2$ to $10^{-3}$\,GeV, thus generating mixings of the chargino states with the 
muon and of the neutralino states with $\nu_{\mu}$. The mediation of the neutralino decays by electroweak gauge bosons means that, contrarily to 
the previously considered scenarios, the vector operators dominate in the current setup. Consequently, the production of vector mesons (mainly 
$\rho$'s) controls the semi-leptonic decays in the higher mass window, generating large decays into pion pairs + muons or neutrinos. Decay 
channels involving single pions are still significant and control the lower mass range, where kinematically allowed. Leptonic decays are also sizable, 
which can be traced back to the gauge nature of the mediation. Channels violating muon number, such as $\nu_{\mu}\ell^+\ell^-$ or $\nu_{\mu}\bar
{\nu}_{\ell}\nu_{\ell}$, are determinant  in our example, with decays involving a neutrino+antineutrino triplet controlling the low mass-range. 
Electromagnetic decays are subleading in this scenario and proceed essentially through a muon / $W$ loop. The plot on the right-hand side, 
corresponding to the decays caused by $\mu_3=10^{-3}$\,GeV, offers limited novelty, except for the disappearance of the channels with a charged 
lepton, since $\tau$ production is kinematically forbidden. The mediation through gauge fields ensures that the decay widths remain of comparable 
magnitude with respect to the previous case.

We thus observe that the choice of lepton-number violating bilinear parameter controlling the decays of the light bino-like state leaves a 
distinguishable imprint on the leading channels. This variety simply reflects the diversity of the accessible low-energy operators on which the lepton- 
and baryon-number violating parameters of the UV theory (in our case the RpV MSSM) can project.

\section{Discussion and Conclusions}
In this paper, we have systematically studied the baryon- or lepton-number violating decays of a light exotic fermion from the perspective of an EFT 
and connected this analysis to the specific case of a SUSY RpV UV completion. Here the light fermion is a light bino $\psi$, a well motivated scenario, 
with mass $m_\psi\leq m_\tau$. The interactions of the exotic particle with hadronic matter represent the main difficulty in controlling its phenomenology, 
but are of paramount importance in order to provide testable predictions for searches at collider and derive reliable limits.

In this respect, our study, though marking a step forward, still leaves ample room for improvement. In particular, baryon-number violating decays of the 
light fermion are at best qualitatively described by our chiral approach restricting to the fundamental baryon octet: going beyond this approximation 
certainly represents a challenge for hadronic physics, however. Turning to the semi-leptonic channels, our analysis greatly benefits from the inclusion 
of the two-meson + lepton final states through the form-factors of Ref.\,\cite{Shi:2020rkz}: these decay modes indeed tend to overtake the channels 
with a single meson  at energies of about $1$\,GeV. Corresponding effects can be approximated by the production of vector resonances only in 
scenarios where operators of vector type dominate, which, in the RpV context, only happens when the phenomenology is determined by the bilinear 
parameters $\mu_i$. Extending our predictions for masses above $1.2$\,GeV however rests on uncertain theoretical ground, especially in the case of 
the determinant scalar operators. For all channels, we have neglected (non-hadronic) interactions between final state particles, which is 
certainly an illegitimate approximation in the immediate vicinity of decay thresholds.

Further improvements could also be considered from the perspective of the low-energy EFT itself, such as the inclusion of next-to-leading QCD 
logarithms in the group evolution. Higher-orders in the matching to the RpV MSSM would likewise help ascertain how much of the associated 
phenomenology can be absorbed within the definition of the tree-level couplings. The definition of multiple thresholds may also be justified in scenarios 
with hierarchical exotic sectors.

Another desirable feature would rest with a systematic assessment of the low-energy phenomenological constraints applying to the tested scenario. In 
the case of baryon-number violation, we have already studied the bounds resulting from nucleon decays in Ref.\,\cite{Chamoun:2020aft} and sketched 
how to apply limits from neutron-antineutron oscillations in the presence of a light exotic fermion in Appendix~\ref{ap:nnbarosc}. Further limits from 
di-nucleon decays \cite{Calibbi:2016ukt} are also applicable. Nevertheless, associated constraints strongly depend on the modelization of the 
interactions of the new physics field with hadronic matter and require a good control over flavor-violating effects in corresponding observables, both in 
and beyond the SM. Concerning the lepton-number violating case, the various scenarios need to be put into the perspective of \textit{e.g.}~neutrino 
mass predictions or lepton-flavor transitions, often requiring the inclusion of loop effects.

Extension of our analysis for masses above the $\tau$ mass \textit{a priori} appears as a difficult undertaking, since the interactions with charmed 
hadrons then need to be reliably described. A few individual channels, such as semi-leptonic decays involving a single $D^{(*)}$ meson, can be 
predicted (and have been considered in \textit{e.g.}~Ref.\,\cite{deVries:2015mfw}), but would generally prove insufficient as soon as the phase space 
allows for further meson radiation.

Finally, quantitatively testing the existence of a light exotic field requires to control the production modes of this particle in experiment, which may be 
difficult when this production employs hadronic channels. A few reference formulae for calculable processes can be found in the appendix. 

\section*{Acknowledgements}
We are grateful to B.~Kubis for helpful discussions, as well as to Y.~Shi for providing us with the two-meson form-factors of Ref.\,\cite{Shi:2020rkz}. 
We thank J.~G\"unther for numerical comparisons. We acknowledge partial financial support by the Deutsche Forschungsgemeinschaft (DFG, German
Research Foundation)  through the funds provided to the Sino-German Collaborative Research Center TRR110 ``Symmetries and the Emergence of 
Structure in QCD" (DFG Project ID 196253076 - TRR 110.

\appendix

\section{Dimension \boldmath $6$ operators for neutralino decay in two-component spinor notation\label{ap:opin2comp}}
For clarity, we express all the operators listed in Eqs.~(\ref{eq:elmdipop}-\ref{eq:hadop}), which potentially mediate a neutralino decay in two-component spinor notation.
\begin{itemize}
	\item electromagnetic dipoles: 
	\begin{equation}\label{eq:elmdipopbis}
		{\cal E}_i\equiv\frac{e}{16\pi^2}(\psi\sigma^{\mu\nu}\nu_i)F_{\mu\nu}\,,\ (i=1,2,3)\,;
	\end{equation}
	\item leptonic operators:
	\begin{align}\label{eq:leptopbis}
		&\widetilde{\cal N}_{ijk}\equiv(\psi\nu_{i})(\nu_j\nu_k)\,,\ \nonumber \\ 
		&(i,j,k)\in\{(1,2,2),(1,3,3),(1,2,3),(2,1,1),(2,3,3),(2,1,3),(3,1,1),(3,2,2)\}\,;\nonumber\\[2mm]
		&\null\hspace{2cm}{\cal N}_{ijk}\equiv(\bar{\psi}\bar{\sigma}^{\mu}\nu_i)(\bar{\nu}_j\bar{\sigma}_{\mu}\nu_k)\,,\hspace{2.7cm}(1\leq i\leq k\leq3)\,;\\
		&\null\hspace{2cm}{\cal S}_{ijk}^{\nu e\,L}\equiv(\psi\nu_i)(e^c_je_k)\,,\hspace{3.3cm}{\cal S}_{ijk}^{\nu e\,R}\equiv(\psi\nu_i)(\bar{e}_j\bar{e}^c_k)\,,\nonumber\\
		&\null\hspace{2cm}{\cal V}_{ijk}^{\nu e\,L}\equiv(\bar{\psi}\bar{\sigma}^{\mu}\nu_i)(\bar{e}_j\bar{\sigma}_{\mu}e_k)\,,\hspace{2.5cm}{\cal V}_{ijk}^{\nu e\,R}\equiv(\bar{\psi}\bar{\sigma}^{\mu}\nu_i)(e^c_j\sigma_{\mu}\bar{e}^c_k)\,,\nonumber\\
		&\null\hspace{2cm}{\cal T}_{ijk}^{\nu e}\equiv(\psi\sigma^{\mu\nu}\nu_i)(e^c_j\sigma_{\mu\nu}e_k)\,,\hspace{2.5cm}(i=1,2,3\,;\ j,k=1,2)\,;\nonumber
	\end{align}
	\item semi-leptonic operators:
	\begin{align}\label{eq:seleptopbis}
		&{\cal S}_{ijk}^{eq\,LL}\equiv(\psi e_i)(d^c_ju_k)\,,& &{\cal S}_{ijk}^{eq\,LR}\equiv(\psi e_i)(\bar{d}_j\bar{u}^c_k)\,,\nonumber\\
		&{\cal S}_{ijk}^{eq\,RL}\equiv(\bar{\psi}\bar{e}^c_i)(d^c_ju_k)\,,& &{\cal S}_{ijk}^{eq\,RR}\equiv(\bar{\psi}\bar{e}^c_i)(\bar{d}_j\bar{u}^c_k)\,,\\
		&{\cal V}_{ijk}^{eq\,LL}\equiv(\bar{\psi}\bar{\sigma}^{\mu}e_i)(\bar{d}_j\bar{\sigma}_{\mu}u_k)\,,& &{\cal V}_{ijk}^{eq\,LR}\equiv(\bar{\psi}\bar{\sigma}^{\mu}e_i)(d^c_j\sigma_{\mu}\bar{u}^c_k)\,,\nonumber\\
		&{\cal V}_{ijk}^{eq\,RL}\equiv(\psi\sigma^{\mu}\bar{e}^c_i)(\bar{d}_j\bar{\sigma}_{\mu}u_k)\,,& &{\cal V}_{ijk}^{eq\,RR}\equiv(\psi\sigma^{\mu}\bar{e}^c_i)(d^c_j\sigma_{\mu}\bar{u}^c_k)\,,\nonumber\\
		&{\cal T}_{ijk}^{eq\,L}\equiv(\psi\sigma^{\mu\nu}e_i)(d^c_j\sigma_{\mu\nu}u_k)\,,& &{\cal T}_{ijk}^{eq\,R}\equiv(\bar{\psi}\bar{\sigma}^{\mu\nu}\bar{e}^c_i)(\bar{d}_j\bar{\sigma}_{\mu\nu}\bar{u}^c_k)\,,\nonumber\\
		& & & (i,j=1,2\,,\ k=1)\,;\nonumber\\
		& {\cal S}_{ijk}^{\nu u\,L}\equiv(\psi\nu_i)(u^c_ju_k)\,,& & {\cal S}_{ijk}^{\nu u\,R}\equiv(\psi\nu_i)(\bar{u}_j\bar{u}^c_k)\,,\nonumber\\
		& {\cal V}_{ijk}^{\nu u\,L}\equiv(\bar{\psi}\bar{\sigma}^{\mu}\nu_i)(\bar{u}_j\bar{\sigma}_{\mu}u_k)\,,& & {\cal V}_{ijk}^{\nu u\,R}\equiv(\bar{\psi}\bar{\sigma}^{\mu}\nu_i)(u^c_j\sigma_{\mu}\bar{u}^k_k)\,,\nonumber\\
		&{\cal T}^{\nu u}_{ijk}\equiv(\psi\sigma^{\mu\nu}\nu_i)(u^c_j\sigma_{\mu\nu}u_k)\,,& & (i=1,2,3\,,\ j,k=1)\,;\nonumber\\
		& {\cal S}_{ijk}^{\nu d\,L}\equiv(\psi\nu_i)(d^c_jd_k)\,,& & {\cal S}_{ijk}^{\nu d\,R}\equiv(\psi\nu_i)(\bar{d}_j\bar{d}^c_k)\,,\nonumber\\
		& {\cal V}_{ijk}^{\nu d\,L}\equiv(\bar{\psi}\bar{\sigma}^{\mu}\nu_i)(\bar{d}_j\bar{\sigma}_{\mu}d_k)\,,& & {\cal V}_{ijk}^{\nu d\,R}\equiv(\bar{\psi}\bar{\sigma}^{\mu}\nu_i)(d^c_j\sigma_{\mu}\bar{d}^k_k)\,,\nonumber\\
		&{\cal T}^{\nu d}_{ijk}\equiv(\psi\sigma^{\mu\nu}\nu_i)(d^c_j\sigma_{\mu\nu}d_k)\,,& & (i=1,2,3\,,\ j,k=1,2)\,;\nonumber
	\end{align}
	\item hadronic operators:
	\begin{align}\label{eq:hadopbis}
		&{\cal H}_{ijk}^{LL}\equiv\varepsilon_{\alpha\beta\gamma}(\psi u^{c\,\alpha}_i)(d^{c\,\beta}_jd^{c\,\gamma}_k)\,, & &{\cal H}_{ijk}^{LR}\equiv\varepsilon_{\alpha\beta\gamma}(\psi u^{c\,\alpha}_i)(\bar{d}^{\beta}_j\bar{d}^{\gamma}_k)\,,\nonumber\\
		&{\cal H}_{ijk}^{RL}\equiv\varepsilon_{\alpha\beta\gamma}(\bar{\psi} \bar{u}^{\alpha}_i)(d^{c\,\beta}_jd^{c\,\gamma}_k)\,, & &{\cal H}_{ijk}^{RR}\equiv\varepsilon_{\alpha\beta\gamma}(\bar{\psi} \bar{u}^{\alpha}_i)(\bar{d}^{\beta}_j\bar{d}^{\gamma}_k)\,,\nonumber\\
		& & & (i=1\,,\ 1\leq j<k\leq2)\,;\nonumber\\ 
		&\widetilde{\cal H}_{ijk}^{LL}\equiv\varepsilon_{\alpha\beta\gamma}(\psi d^{c\,\beta}_j)(u^{c\,\alpha}_id^{c\,\gamma}_k)\,,& & \widetilde{\cal H}_{ijk}^{LR}\equiv\varepsilon_{\alpha\beta\gamma}(\psi d^{c\,\beta}_j)(\bar{u}^{\alpha}_i\bar{d}^{\gamma}_k)\,,\nonumber\\
		&\widetilde{\cal H}_{ijk}^{RL}\equiv\varepsilon_{\alpha\beta\gamma}(\bar{\psi} \bar{d}^{\beta}_j)(u^{c\,\alpha}_id^{c\,\gamma}_k)\,,& & \widetilde{\cal H}_{ijk}^{RR}\equiv\varepsilon_{\alpha\beta\gamma}(\bar{\psi} \bar{d}^{\beta}_j)(\bar{u}^{\alpha}_i\bar{d}^{\gamma}_k)\,,\nonumber\\
		& & & (i=1\,,\ j,k=1,2)\,.\nonumber
	\end{align}
\end{itemize}

\section{Feynman Rules}\label{ap:couplingdef}
Below, we list the relevant MSSM RpV couplings of Eq.~(\ref{eq:couplingdef}). Mixing matrices are defined in appendix~\ref{ap:mixing}.
\subsection{Electroweak gauge couplings}
\begin{subequations}
\begin{align}
&g_L^{Z\chi_i^0\chi_j^0}=-\frac{\sqrt{g_1^2+g_2^2}}{2}\,\big[N_{id}N_{jd}^*-N_{iu}N_{ju}^*+N_{i\nu_l}N_{j\nu_l}^*\big]=-\Big(g_R^{Z\chi_j^0\chi_i^0}\Big)^*,\\
&g_L^{Z\chi_i^+\chi_j^-}=\sqrt{g_1^2+g_2^2}\,\big[c_W^2\,U_{iw}U_{jw}^*+\tfrac{c_W^2-s_W^2}{2}(U_{id}U_{jd}^*+U_{ie_l}U_{je_l}^*)\big]=-g_R^{Z\chi_j^-\chi_i^+},\\
&g_R^{Z\chi_i^+\chi_j^-}=\sqrt{g_1^2+g_2^2}\,\big[c_W^2\,V_{iw}^*V_{jw}+\tfrac{c_W^2-s_W^2}{2}V_{iu}^*V_{ju}-s_W^2\,V_{ie_l}^*V_{je_l}\big]=-g_L^{Z\chi_j^-\chi_i^+},\\
&g_L^{Zu_i^cu_j}=-\frac{\sqrt{g_1^2+g_2^2}}{2}\,\big(1-\tfrac{4}{3}s_W^2\big)\delta_{ij}=-g_R^{Zu_ju^c_i},\\
&g_R^{Zu_i^cu_j}=\frac{2}{3}\sqrt{g_1^2+g_2^2}\,s_W^2\,\delta_{ij}=-g_L^{Zu_ju^c_i},\\
&g_L^{Zd_i^cd_j}=\frac{\sqrt{g_1^2+g_2^2}}{2}\,\big(1-\tfrac{2}{3}s_W^2\big)\delta_{ij}=-g_R^{Zd_jd^c_i},\\
&g_R^{Zd_i^cd_j}=-\frac{\sqrt{g_1^2+g_2^2}}{3}\,s_W^2\,\delta_{ij}=-g_L^{Zd_jd^c_i},\\
&g_L^{W^+\chi_i^0\chi_j^-}=-g_2\big[N_{iw}U_{jw}^*+\tfrac{1}{\sqrt{2}}N_{id}U_{jd}^*+\tfrac{1}{\sqrt{2}}N_{i\nu_l}U_{je_l}^*\big]\\
&\null\hspace{1.6cm}=\Big(g_L^{W^-\chi_j^+\chi_i^0}\Big)^*=-g_R^{W^+\chi_j^-\chi_i^0}=-\Big(g_R^{W^-\chi_i^0\chi_j^+}\Big)^*,\nonumber\\
&g_R^{W^+\chi_i^0\chi_j^-}=-g_2\big[N_{iw}^*V_{jw}-\tfrac{1}{\sqrt{2}}N_{iu}^*V_{ju}\big]=\Big(g_R^{W^-\chi_j^+\chi_i^0}\Big)^*=-g_L^{W^+\chi_j^-\chi_i^0}
=-\Big(g_L^{W^-\chi_i^0\chi_j^+}\Big)^*,\\
&g_L^{W^-d_i^cu_j}=-\frac{g_2}{\sqrt{2}}\,V^{\text{CKM}\,*}_{ji}=-g_R^{W^-u_jd_i^c}=\Big(g_L^{W^+u_j^cd_i}\Big)^*=-\Big(g_R^{W^+d_iu_j^c}\Big)^*,\\
&g_R^{W^-d_i^cu_j}=0=-g_L^{W^-u_jd_i^c}=\Big(g_R^{W^+u_j^cd_i}\Big)^*=-\Big(g_L^{W^+d_iu_j^c}\Big)^*.
\end{align}
\end{subequations}

\subsection{Scalar couplings}
\begin{subequations}
\begin{align}
&g_L^{S_m^0\chi_i^0\chi_j^0}=\frac{1}{2}(g_1N_{ib}^*-g_2N_{iw}^*)\big[X_{md}^*N_{jd}^*-X_{mu}^*N_{ju}^*+X_{m\tilde{N}_l}^*N_{j\nu_l}^*\big]+(i\leftrightarrow j)=\Big(g_R^{S_m^0\chi_j^0\chi_i^0}\Big)^*,\\
&g_L^{S_m^0\chi_i^+\chi_j^-}=\frac{Y_e^l}{\sqrt{2}}V_{ie_l}^*\big[X_{m\tilde{N}_l}U_{jd}^*-X_{md}U_{je_l}^*\big]-\frac{g_2}{\sqrt{2}}\big[X_{md}^*V_{iw}^*U_{jd}^*+X_{mu}^*V_{iu}^*U_{jw}^*+X_{m\tilde{N}_l}^*V_{iw}^*U_{je_l}^*\big]\nonumber\\
&\null\hspace{1.6cm}-\frac{\lambda_{lnp}}{\sqrt{2}}X_{m\tilde{N}_l}V_{ie_p}^*U_{je_n}^*=\Big(g_R^{S_m^0\chi_j^+\chi_i^-}\Big)^*=g_L^{S_m^0\chi_j^-\chi_i^+}=\Big(g_R^{S_m^0\chi_i^-\chi_j^+}\Big)^*,\\
&g_L^{S_m^0u_i^cu_j}=-\frac{Y_u^i}{\sqrt{2}}X_{mu}\delta_{ij}=\Big(g_R^{S_m^0u_j^cu_i}\Big)^*=g_L^{S_m^0u_ju_i^c}=\Big(g_R^{S_m^0u_iu_j^c}\Big)^*,\\
&g_L^{S_m^0d_i^cd_j}=-\frac{Y_d^i}{\sqrt{2}}X_{md}\delta_{ij}-\frac{\lambda'_{lji}}{\sqrt{2}}X_{m\tilde{N}_l}=\Big(g_R^{S_m^0d_j^cd_i}\Big)^*=g_L^{S_m^0d_jd_i^c}=\Big(g_R^{S_m^0d_id_j^c}\Big)^*,\\
&g_L^{S_m^-\chi_i^+\chi^0_j}=-\frac{g_1}{\sqrt{2}}(X^{C\,*}_{mu}V^*_{iu}+2X^{C\,*}_{mE_R^l}V^*_{ie_l})N_{jb}^*-\frac{g_2}{\sqrt{2}}X^{C\,*}_{mu}(V^*_{iu}N_{jw}^*+\sqrt{2}V^*_{iw}N_{ju}^*)\\
&\null\hspace{1.9cm}-Y_e^lV_{ie_l}^*(X^{C\,*}_{mE_L^l}N^*_{jd}-X^{C\,*}_{md}N^*_{j\nu_l})-\lambda_{lnp}X^{C\,*}_{mE_L^n}V_{ie_p}^*N^*_{j\nu_l}\nonumber\\
&\null\hspace{1.4cm}=\Big(g_R^{S_m^+\chi_i^-\chi^0_j}\Big)^*=g_L^{S_m^-\chi^0_j\chi_i^+}=\Big(g_R^{S_m^+\chi^0_j\chi_i^-}\Big)^*,\nonumber\\
&g_R^{S_m^-\chi_i^+\chi^0_j}=\frac{1}{\sqrt{2}}(X^{C\,*}_{md}U_{id}+X^{C\,*}_{mE_L^l}U_{ie_l})(g_1N_{jb}+g_2N_{jw})-g_2U_{iw}(X^{C\,*}_{md}N_{jd}+X^{C\,*}_{mE_L^l}N_{j\nu_l})\\
&\null\hspace{1.83cm}-Y_e^lX^{C\,*}_{mE_R^l}(U_{ie_l}N_{jd}-U_{id}N_{j\nu_l})-\lambda_{lnp}^*X^{C\,*}_{mE_R^p}U_{ie_n}N_{j\nu_l}\nonumber\\
&\null\hspace{1.4cm}=\Big(g_L^{S_m^+\chi_i^-\chi^0_j}\Big)^*=g_R^{S_m^-\chi^0_j\chi_i^+}=\Big(g_L^{S_m^+\chi^0_j\chi_i^-}\Big)^*,\nonumber\\
&g_L^{S_m^-d_i^cu_j}=Y_d^iV^{\text{CKM}\,*}_{ji}X^{C\,*}_{md}+\lambda'_{lni}V^{\text{CKM}*}_{jn}X^{C\,*}_{mE_L^l}=\Big(g_R^{S_m^+u_j^cd_i}\Big)^*=g_L^{S_m^-u_jd_i^c}=\Big(g_R^{S_m^+d_iu_j^c}\Big)^*,\\
&g_R^{S_m^-d_i^cu_j}=Y_u^jV^{\text{CKM}\,*}_{ji}X^{C\,*}_{mu}=\Big(g_L^{S_m^+u_j^cd_i}\Big)^*=g_R^{S_m^-u_jd_i^c}=\Big(g_L^{S_m^+d_iu_j^c}\Big)^*,\\
&g_L^{U_mu_i^c\chi_j^0}=-Y_u^iX_{mU_L^i}^*N_{ju}^*+\frac{2\sqrt{2}}{3}g_1X_{mU_R^i}^*N_{jb}^*=\Big(g_R^{U_m^*\chi_j^0u_i}\Big)^*=g_L^{U_m\chi_j^0u_i^c}=\Big(g_R^{U_m^*u_i\chi_j^0}\Big)^*,\hspace{-0.2cm}\\
&g_R^{U_mu_i^c\chi_j^0}=-Y_u^iX_{mU_R^i}^*N_{ju}-\frac{1}{\sqrt{2}}X_{mU_L^i}^*(\tfrac{g_1}{3}N_{jb}+g_2N_{jw})\nonumber \\
&\null\hspace{1.4cm}  =\Big(g_L^{U_m^*\chi_j^0u_i}\Big)^*=g_R^{U_m\chi_j^0u_i^c}=\Big(g_L^{U_m^*u_i\chi_j^0}\Big)^*,\\
&g_L^{D_md_i^c\chi_j^0}=-Y_d^iX_{mD_L^i}^*N_{jd}^*-\frac{\sqrt{2}}{3}g_1X_{mD_R^i}^*N_{jb}^*-\lambda'_{lni}X_{mD_L^n}^*N_{j\nu_l}^*\\
&\null\hspace{1.4cm}=\Big(g_R^{D_m^*\chi_j^0d_i}\Big)^*=g_L^{D_m\chi_j^0d_i^c}=\Big(g_R^{D_m^*d_i\chi_j^0}\Big)^*,\nonumber\\
&g_R^{D_md_i^c\chi_j^0}=-Y_d^iX_{mD_R^i}^*N_{jd}-\frac{1}{\sqrt{2}}X_{mD_L^i}^*(\tfrac{g_1}{3}N_{jb}-g_2N_{jw})-\lambda'^*_{lin}X_{mD_R^n}^*N_{j\nu_l}\\
&\null\hspace{1.4cm}=\Big(g_L^{D_m^*\chi_j^0d_i}\Big)^*=g_R^{D_m\chi_j^0d_i^c}=\Big(g_L^{D_m^*d_i\chi_j^0}\Big)^*\nonumber\\
&g_L^{U_md_i^c\chi_j^-}=Y_d^iV^{\text{CKM}\,*}_{ni}X_{mU_L^n}^*U_{jd}^*+\lambda'_{lpi}V^{\text{CKM}\,*}_{np}X_{mU_L^n}^*U_{je_l}^*\nonumber \\
&\null\hspace{1.4cm} =\Big(g_R^{U_m^*\chi_j^+d_i}\Big)^*=g_L^{U_m\chi_j^-d_i^c}=\Big(g_R^{U_m^*d_i\chi_j^+}\Big)^*,\hspace{-0.3cm}\\
&g_R^{U_md_i^c\chi_j^-}=Y_u^nV^{\text{CKM}\,*}_{ni}X_{mU_R^n}^*V_{ju}-g_2V^{\text{CKM}\,*}_{ni}X_{mU_L^n}^*V_{jw}\nonumber \\
&\null\hspace{1.4cm}  =\Big(g_L^{U_m^*\chi_j^+d_i}\Big)^*=g_R^{U_m\chi_j^-d_i^c}=\Big(g_L^{U_m^*d_i\chi_j^+}\Big)^*,\\
&g_L^{D^*_mu_i\chi_j^-}=Y_d^nV^{\text{CKM}\,*}_{in}X_{mD_R^n}U_{jd}^*-g_2V^{\text{CKM}\,*}_{in}X_{mD_L^n}U_{jw}^*+\lambda'_{lnp}V^{\text{CKM}\,*}_{in}X_{mD_R^p}U_{je_l}^*\\
&\null\hspace{1.4cm}=\Big(g_R^{D_m\chi_j^+u^c_i}\Big)^*=g_L^{D^*_m\chi_j^-u_i}=\Big(g_R^{D_mu^c_i\chi_j^+}\Big)^*,\nonumber\\
&g_R^{D^*_mu_i\chi_j^-}=Y_u^iV^{\text{CKM}\,*}_{in}X_{mD_L^n}V_{ju}=\Big(g_L^{D_m\chi_j^+u^c_i}\Big)^*=g_R^{D^*_m\chi_j^-u_i}=\Big(g_L^{D_mu^c_i\chi_j^+}\Big)^*,\\
&g_L^{U^*_md^c_id_j^c}=-\lambda''_{nij}X_{mU_R^n}=\Big(g_R^{U_md_id_j}\Big)^*=-g_L^{U^*_md^c_jd_i^c}=-\Big(g_R^{U_md_jd_i}\Big)^*\ \ \ ; \ \ \ g_R^{U^*_md^c_id_j^c}=0,\\
&g_L^{D^*_mu^c_id_j^c}=-\lambda''_{inj}X_{mD_R^n}=\Big(g_R^{D_md_ju_i}\Big)^*=g_L^{D^*_md^c_ju_i^c}=\Big(g_R^{D_mu_id_j}\Big)^*\ \ \ ; \ \ \ g_R^{D^*_mu^c_id_j^c}=0.
\end{align}
\end{subequations}

\section{Mixing in the RpV MSSM}\label{ap:mixing}
\subsection{Mixing in the chargino / lepton sector} 
We work in the field basis where the sneutrino fields do not take a vacuum expectation value (vev) \cite{Grossman:1998py}. Then the 
Dirac mass-terms in the chargino/lepton sector read 
\begin{equation}
-{\cal L}\ni
(\tilde{w}^+,\tilde{h}_u^+,e_R^{c\,f}){\cal M}_{\tilde{C}}(\tilde{w}^-,\tilde{h}_d^-,e_L^g)^T+h.c.,
\end{equation}
with the $5\times5$ matrix:
\begin{equation}\label{eqn:Cmat}
{\cal M}_{\tilde{C}}=\begin{bmatrix}
M_2 & g_2 v_d & 0\\ g_2 v_u & \mu & \mu_j\\ 0 & 0 & Y_e^i\delta_{ij}v_d
\end{bmatrix}\,.
\end{equation}
Here $i,j$ correspond to the flavor indices. $\mu_j$ is the RpV bilinear coupling. For the mass matrix of Eq.~(\ref{eqn:Cmat}) we perform
a singular value decomposition \cite{Dreiner:2008tw} with a pair of unitary matrices such that: ${\cal M}_{\tilde{C}}=V^T\text{diag}(m_
{\chi^{\pm}_i})U$, from which we deduce the mass eigenstates: $\chi_i^+=V^*_{iw}\tilde{w}^++V^*_{ih}\tilde{h}_u^++V^*_{ie_j}e_R^{c\,j}$, 
$\chi_i^-=U^*_{iw}\tilde{w}^-+U^*_{ih}\tilde{h}_d^-+U^*_{ie_j}e_L^{j}$.

In the hierarchical context $|M_2|,|\mu|,\big||M_2|-|\mu|\big|\gg gv,|\mu_j|,Y_e^gv_d$, we can expand the mixing matrices according to 
$W=\mathbf{I}+W^{(1)}+\ldots$, for $W=U,V$. The non-trivial elements of $W^{(1)}$ read:
\begin{align}
& U_{1d}^{(1)}=\frac{g_2(v_dM_2^*+v_u\mu)}{|M_2|^2-|\mu|^2}\,, & & U_{2w}^{(1)}=-\frac{g_2(v_dM_2+v_u\mu^*)}{|M_2|^2-|\mu|^2}\,, & & U_{2e_l}^{(1)}=\frac{\mu_l}{\mu}\,,\\
& V_{1u}^{(1)}=\frac{g_2(v_d\mu+v_uM_2^*)}{|M_2|^2-|\mu|^2}\,, & &  V_{2w}^{(1)}=-\frac{g_2(v_d\mu^*+v_uM_2)}{|M_2|^2-|\mu|^2}\,, & & U_{3_ld}^{(1)}=-\frac{\mu^*_l}{\mu^*}\,.\nonumber
\end{align}
where the mass-indices $1$, $2$, $3_l$ respectively refer to mostly wino, higgsino and lepton states. Here $e_l+e\,\mu,\,\tau$ for $l=1,2,3$.

\subsection{Mixing in the neutralino / neutrino sector} 
Similarly to the chargino/lepton sector, electroweak-symmetry breaking and RpV mixing produce the $7\times7$ Majorana mass-matrix between neutral gauginos, higgsinos and neutrinos, in the basis $(\tilde{b}^0,\tilde{w}^0,\tilde{h}_d^0,\tilde{h}_u^0,\nu_L^f)$:
\begin{equation}\label{eqn:Nmat}
{\cal M}_{\tilde{N}}=\begin{bmatrix}
M_1 & 0 & -\tfrac{g_1}{\sqrt{2}} v_d & \tfrac{g_1}{\sqrt{2}} v_u & 0\\
0 & M_2 & \tfrac{g_2}{\sqrt{2}} v_d & -\tfrac{g_2}{\sqrt{2}} v_u & 0\\ 
-\tfrac{g_1}{\sqrt{2}} v_d &\tfrac{g_2}{\sqrt{2}} v_d  & 0 & -\mu & 0\\ 
\tfrac{g_1}{\sqrt{2}} v_u & -\tfrac{g_2}{\sqrt{2}} v_u & -\mu & 0 & -\mu_g\\ 
0 & 0 & 0 & -\mu_g & 0
\end{bmatrix}.
\end{equation}
This matrix is Takagi diagonalized \cite{Dreiner:2008tw} via a $7\times7$ unitary matrix $N$ according to ${\cal M}_{\tilde{N}}=N^T\text{diag}(m_{\chi^0})N$, 
from which we deduce the mass eigenstates $\chi_i^0=N^*_{ib}\tilde{b}+N^*_{iw}\tilde{w}^0+N^*_{ih_d}\tilde{h}_d^0+N^*
_{ih_u}\tilde{h}_u^0+N^*_{i\nu_f}\nu_L^f$, $i=1,\ldots,7$. Again, in a hierarchical context, the mixing matrix can be expanded as $N=N^{(0)}+N^{(1)}+N^{(2)}+\ldots$, with:
\begin{equation}
N^{(0)}\!=\!\begin{bmatrix}
1 & 0 & 0 & 0 & 0\\
0 & 1 & 0 & 0 & 0\\
0 & 0 & \tfrac{1}{\sqrt{2}} & \tfrac{1}{\sqrt{2}} & 0\\
0 & 0 & -\tfrac{1}{\sqrt{2}} & \tfrac{1}{\sqrt{2}} & 0\\
0 & 0 & 0 & 0 & 1
\end{bmatrix}
\ \ ,\ \ \begin{array}{ll}
N_{1h_d}^{(1)}=-\tfrac{g_1}{\sqrt{2}}\frac{M_1^*v_d+\mu v_u}{|M_1|^2-|\mu|^2} & N_{1h_u}^{(1)}=\tfrac{g_1}{\sqrt{2}}\frac{M_1^*v_u+\mu v_d}{|M_1|^2-|\mu|^2},\\[1.7mm]
N_{2h_d}^{(1)}=\tfrac{g_2}{\sqrt{2}}\frac{M_2^*v_d+\mu v_u}{|M_2|^2-|\mu|^2}, &
N_{2h_u}^{(1)}=-\tfrac{g_2}{\sqrt{2}}\frac{M_2^*v_u+\mu v_d}{|M_2|^2-|\mu|^2},\\[1.7mm]
N_{3b}^{(1)}=\tfrac{g_1}{2}\tfrac{(v_d-v_u)(M_1-\mu^*)}{|M_1|^2-|\mu|^2}, &
N_{3w}^{(1)}=\tfrac{g_2}{2}\tfrac{(v_u-v_d)(M_2-\mu^*)}{|M_2|^2-|\mu|^2},\\[1.7mm]
N_{4b}^{(1)}=-\tfrac{g_1}{2}\tfrac{(v_d+v_u)(\mu^*+M_1)}{|M_1|^2-|\mu|^2}, &
N_{4w}^{(1)}=\tfrac{g_2}{2}\tfrac{(v_d+v_u)(\mu^*+M_2)}{|M_2|^2-|\mu|^2},\\[1.7mm]
N_{3\nu_l}^{(1)}=\tfrac{\mu_l}{\sqrt{2}\mu}=-N_{4\nu_l}^{(1)}, & N_{5_lh_d}^{(1)}=-\tfrac{\mu_l^*}{\mu^*}\,,
\end{array}
\end{equation}
and all other elements of $N^{(1)}$ are trivial. As we only consider bino- or neutrino-dominated fields in the external legs, we only provide the bino-neutrino mixing of second order:
\begin{equation}
N^{(2)}_{1\nu_l}=-\frac{g_1}{\sqrt{2}}\frac{\mu_l}{M_1}\frac{M_1v_u+\mu^*v_d}{|M_1|^2-|\mu|^2}\ \ ;\ \ N^{(2)}_{5_lb}=-\frac{g_1}{\sqrt{2}}\frac{\mu_l^*v_d}{M_1^*\mu^*}\,.
\end{equation}
In this description, the indices $1$, $2$, $3$, $4$ and $5_l$ refer to mostly bino, wino, a pair of higgsino and
neutrino-like mass eigenstates.  

\subsection{Mixing in the Higgs / slepton sector}
The neutral Higgs and sneutrino states mix at tree-level through the soft SUSY-breaking RpV parameters $B_i\equiv B_i^R+i\,B_i^I$ (after 
simplifying other mixing parameters via the minimization equations of the scalar potential). The $5\times5$ mass-blocks ${\cal M}^2_S$ 
and ${\cal M}_P^2$ for the CP-even and CP-odd states in the gauge eigenbases $(h_d^0,h_u^0,\tfrac{1}{\sqrt{2}}\text{Re}[\tilde{N}_i])$ 
and $(a_d^0,a_u^0,\tfrac{1}{\sqrt{2}}\text{Im}[\tilde{N}_i])$ (respectively) read:
\begin{eqnarray}
{\cal M}^2_S&=&\begin{bmatrix}
M_A^2s^2_{\beta}+M_Z^2c^2_{\beta} & -(M_A^2+M_Z^2)s_{\beta}c_{\beta} & (B_i^R)^Tt_{\beta}\\[1mm]
-(M_A^2+M_Z^2)s_{\beta}c_{\beta} & M_A^2c^2_{\beta}+M_Z^2s^2_{\beta} & -(B_i^R)^T\\[1mm]
(B_i^R)t_{\beta} & -(B_i^R) & \text{Re}[M^2_{\tilde{N}}]
\end{bmatrix} \,, \\ [1.7mm]
{\cal M}^2_P&=&\begin{bmatrix}
M_A^2s^2_{\beta} & M_A^2s_{\beta}c_{\beta} & (B_i^R)^Tt_{\beta}\\[1mm]
M_A^2s_{\beta}c_{\beta} & M_A^2c^2_{\beta} & (B_i^R)^T\\[1mm]
(B_i^R)t_{\beta} & (B_i^R) & \text{Re}[M^2_{\tilde{N}}]
\end{bmatrix}\,,
\end{eqnarray}
where $(B_i)$ should be understood as a column $3$-vector and $M^2_{\tilde{N}}=\big(m^2_{L\,mn}+\mu^*_m\mu_n+M_Z^2\tfrac{c_{2\beta}}{2}\delta_{mn}\big)$ is the $3\times3$ mass subblock corresponding to the sneutrinos. $\text{Im}[M^2_{\tilde{L}}]$ and $B_i^I$ mix CP-even and CP-odd degrees of freedom and the mass eigenstates can be decomposed as $S_p^0=X^R_{pd}\,h_d^0+X^R_{pu}\,h_u^0+\tfrac{X^R_{p\tilde{N}_i}}{\sqrt{2}}\text{Re}[\tilde{N}_i]+X^I_{pd}\,a_d^0+X^I_{pu}\,a_u^0+\tfrac{X^I_{p\tilde{N}_i}}{\sqrt{2}}\text{Im}[\tilde{N}_i]$, where $(X^R_{p\cdot},X^I_{p\cdot})$ forms an orthonormal eigenvector of the mass-matrix. We also define the notation $X\equiv X^R+i\,X^I$.

Assuming $M_A^2,M^2_{\tilde{N}\,ii}\gg M_Z^2,|B_i|,|M^2_{\tilde{N}\,ij}|$ ($j\neq i$), the mass eigenstates can be approximated by:
\begin{align}\label{eq:HSNstates}
&h^0\equiv c_{\beta}h_d^0+s_{\beta}h_u^0\,,\ \ m_{h^0}^2\approx M_Z^2c_{2\beta}^2\,,& & G^0\equiv -c_{\beta}a_d^0+s_{\beta}a_u^0\,,\ \ m_{G^0}^2=0\,, \\
&H^0\equiv s_{\beta}h_d^0-c_{\beta}h_u^0\,,\ \ m_{H^0}^2\approx M_A^2+M_Z^2s_{2\beta}^2\,,& & A^0\equiv s_{\beta}a_d^0+c_{\beta}a_u^0\,,\ \ m_{A^0}^2=M_A^2\,, \nonumber\\
&\tilde{N}_p^R\equiv\tfrac{1}{\sqrt{2}}\text{Re}[\tilde{N}_p]\,,\ \ \tilde{N}_p^I\equiv\tfrac{1}{\sqrt{2}}\text{Im}[\tilde{N}_p]\,,& &m_{\tilde{N}_p}^2\approx m^2_{L\,pp}+|\mu_p|^2+M_Z^2\tfrac{c_{2\beta}}{2}\,. \nonumber
\end{align}
First order corrections may be encoded as $S_p^0=(\delta_{pm}+X^{(1)}_{pm})(S_m^0)^{(0)}$ with $(S_m^0)^{(0)}$ corresponds to the states of Eq.~(\ref{eq:HSNstates}), where the non-trivial elements of $X^{(1)}$ are:
\begin{align}
&X^{(1)}_{h^0H^0}=-\frac{M_Z^2s_{4\beta}}{2M_A^2}=-X^{(1)}_{H^0h^0}\,, & & X^{(1)}_{\tilde{N}_p^R\tilde{N}_q^I}=-\frac{\text{Im}[M^2_{\tilde{N}}]_{pq}}{m_{\tilde{N}_p}^2-m_{\tilde{N}_q}^2}=-X^{(1)}_{\tilde{N}_q^I\tilde{N}_p^R}\,,\nonumber\\
&X^{(1)}_{\tilde{N}_p^R\tilde{N}_q^R}=\frac{\text{Re}[M^2_{\tilde{N}}]_{pq}}{m_{\tilde{N}_p}^2-m_{\tilde{N}_q}^2}=-X^{(1)}_{\tilde{N}_q^R\tilde{N}_p^R}\,, & &X^{(1)}_{\tilde{N}_p^I\tilde{N}_q^I}=\frac{\text{Re}[M^2_{\tilde{N}}]_{pq}}{m_{\tilde{N}_p}^2-m_{\tilde{N}_q}^2}=-X^{(1)}_{\tilde{N}_q^I\tilde{N}_p^I}\,,\\
&X^{(1)}_{H^0\tilde{N}_p^R}=\frac{B_p^R/c_{\beta}}{M_A^2-m_{\tilde{N}_p}^2}=-X^{(1)}_{\tilde{N}_p^RH^0}\,, & & X^{(1)}_{H^0\tilde{N}_p^I}=\frac{-B_p^I/c_{\beta}}{M_A^2-m_{\tilde{N}_p}^2}=-X^{(1)}_{\tilde{N}_p^IH^0}\,,\nonumber\\ &X^{(1)}_{A^0\tilde{N}_p^I}=\frac{B_p^R/c_{\beta}}{M_A^2-m_{\tilde{N}_p}^2}=-X^{(1)}_{\tilde{N}_p^IA^0}\,, & & X^{(1)}_{A^0\tilde{N}_p^R}=\frac{B_p^I/c_{\beta}}{M_A^2-m_{\tilde{N}_p}^2}=-X^{(1)}_{\tilde{N}_p^RA^0}\nonumber\,.
\end{align}
As a result, only the heavy Higgs doublets $H^0$, $A^0$ undergo RpV mixing with the sneutrinos at leading order.

Similarly, the mass matrix in the charged sector mixes charged Higgs and sleptons. In the gauge basis $(H_d^-,H_u^-,\tilde{E}_{L}^i,(\tilde{E}_R^{c\,i})^*)$,
\begin{eqnarray}
{\cal M}^2_C&=&\begin{bmatrix}
(M_A^2+M_W^2)s^2_{\beta} & (M_A^2+M_W^2)s_{\beta}c_{\beta} & (B_i)^Tt_{\beta} & (m_{e_i}\mu_i)t_{\beta}\\
(M_A^2+M_W^2)s_{\beta}c_{\beta} & (M_A^2+M_W^2)c^2_{\beta} & (B_i)^T & (m_{e_i}\mu_i)\\
(B_i)^*t_{\beta} & (B_i)^* & M^2_{\tilde{E}_L} & M^2_{\tilde{E}_{LR}}\\
(m_{e_i}\mu_i^*)t_{\beta} & (m_{e_i}\mu_i^*) & M^2_{\tilde{E}_{RL}} & M^2_{\tilde{E}_R}
\end{bmatrix}\,,\\
M^2_{\tilde{E}_L}&=&\big(m^2_{L\,mn}+\mu^*_m\mu_n+(m_{e_m}^2+\tfrac{g_1^2-g_2^2}{4}v^2c_{2\beta})\delta_{mn}\big)\,,\nonumber\\
M^2_{\tilde{E}_R}&=&\big(m^2_{E\,mn}+\mu^*_m\mu_n+(m_{e_m}^2-\tfrac{g_1^2}{4}v^2c_{2\beta})\delta_{mn}\big)\,,\nonumber\\
M^2_{\tilde{E}_{LR}}&=&(A^{e\,*}_{mn}v_d-m_{e_m}\,t_{\beta}\,\mu\delta_{mn}-\mu_k\lambda^*_{kmn}v_u)\,,\ \ M^2_{\tilde{E}_{RL}}=\big(M^2_{\tilde{E}_{LR}}\big)^{\dagger}\,. \nonumber
\end{eqnarray}
The mass eigenstates may then be written as $S_p^{-}=X^C_{pd}\,H_d^-+X^C_{pu}\,H_u^-+X^C_{p\tilde{E}^i_L}\,\tilde{E}^i_L+X^C_{p\tilde{E}^i_R}(\tilde{E}^{c\,i}_R)^*$. In a hierarchical context, they may be approximated by:
\begin{align}
&G^-=-c_{\beta}H_d^-+s_{\beta}H_u^-\,,\ \ m_{G^{\pm}}^2=0\,; & & H^{-}=s_{\beta}H_d^-+c_{\beta}H_u^-\,,\ \ m_{H^{\pm}}^2=M_A^2+M_W^2\,;\nonumber\\
&\tilde{E}_L^i\,,\ \  m_{\tilde{E}_L^i}^2=(M^2_{\tilde{E}_L})_{ii}\,; & & (\tilde{E}^{c\,i}_R)^*\,,\ \  m_{\tilde{E}_R^i}^2=(M^2_{\tilde{E}_R})_{ii}\,.
\end{align}
Then, the non-trivial mixings at first order are given by:
\begin{align}
&X^{C(1)}_{H^-\tilde{E}_L^p}=\frac{B_p/c_{\beta}}{M_A^2-m_{\tilde{E}_L^p}^2}=-X^{C(1)}_{\tilde{E}_L^pH^-}\,, & & X^{C(1)}_{H^-\tilde{E}_R^p}=\frac{m_{e_p}\mu_p/c_{\beta}}{M_A^2-m_{\tilde{E}_R^p}^2}=-\big(X^{C(1)}_{\tilde{E}_R^pH^-}\big)^*\,,\nonumber \\
&X^{C(1)}_{\tilde{E}_L^p\tilde{E}_L^q}=\frac{(M^2_{\tilde{E}_L})^*_{pq}}{m_{\tilde{E}_L^p}^2-m_{\tilde{E}_L^q}^2}=-\big(X^{C(1)}_{\tilde{E}_L^q\tilde{E}_L^p}\big)^*\,, & & X^{C(1)}_{\tilde{E}_R^p\tilde{E}_R^q}=\frac{(M^2_{\tilde{E}_R})^*_{pq}}{m_{\tilde{E}_R^p}^2-m_{\tilde{E}_R^q}^2}=-\big(X^{C(1)}_{\tilde{E}_R^q\tilde{E}_R^p}\big)^*\,,\\ &X^{C(1)}_{\tilde{E}_L^p\tilde{E}_R^q}=\frac{(M^2_{\tilde{E}_{LR}})^*_{pq}}{m_{\tilde{E}_L^p}^2-m_{\tilde{E}_R^q}^2}=-\big(X^{C(1)}_{\tilde{E}_R^q\tilde{E}_L^p}\big)^*\,. & &\nonumber
\end{align}

\subsection{Mixing in the squark sector} 
The sfermion mixing matrices can be written in the $(\tilde{F}_L,\tilde{F}_R^{c\,*})$ basis ($\tilde{F}=\tilde{U},\tilde{D}$) as:
\begin{equation}\label{eqn:Fmat}
{\cal M}^2_{\tilde{F}}=
\begin{bmatrix}
M^2_{\tilde{F}_L} & M^2_{\tilde{F}_{LR}}\\
M^2_{\tilde{F}_{RL}} & M^2_{\tilde{F}_R}
\end{bmatrix}\, ,\ \ \ \begin{cases}
M^2_{\tilde{F}_L}\equiv m_{F_L}^2+m_f^2+\frac{1}{2}\left(\frac{{\cal Y}_L^f}{2}g_1^2-I_3^fg_2^2\right)(v_u^2-v_d^2)\,,\\
M^2_{\tilde{F}_R}\equiv m_{F_R}^2+m_f^2+\frac{{\cal Y}_R^f}{4}g_1^2(v_u^2-v_d^2)\,,\\
M^2_{\tilde{F}_{LR}}\equiv A_f^*v_f-m_f\mu\frac{v_{f'}}{v_f}-\mu_i\lambda'^*_{ijk}v_u\delta_{\tilde{F}\tilde{D}}\ \ ;\ \ M^2_{\tilde{F}_{RL}}=M^{2\,\dagger}_{\tilde{F}_{LR}}\,,
\end{cases}
\end{equation}
where $f$ is the fermion corresponding to the sfermion $\tilde{F}$, while $f'$ is its $SU(2)_L$ partner. Then, ${\cal Y}_{L,R}^f$ are the associated hypercharges, $I_3^f$, the isospin. Finally, $v_f$ denotes the vev.\ of the Higgs doublet to which the fermion $f$ couples at tree-level.
Each entry should again be understood as a $3\times3$ block. We write the mass eigenstates as $\tilde{F}_p=X_{p\tilde{F}_L^i}\,\tilde{F}_L^i+X_{p\tilde{F}_R^i}\,\tilde{F}_R^i$. It is also possible to work under the simplifying assumption that inter-generation and left-right mixings are suppressed compared to the diagonal hierarchy.

\section{Loop Integrals for the \boldmath $\psi\to\bar{\nu}_i\gamma$ Transition}\label{ap:loopfunc}
The loop integrals ${\cal I}^{V,S}_{L,R}$ of Eq.~(\ref{eq:CMSSMelm}) can be written in terms of the standard Passarino-Veltman functions
\cite{tHooft:1978jhc,Passarino:1978jh}:
\begin{align}
&{\cal I}^{V}_{L}(\tfrac{m_{\psi}^2}{M_W^2},\tfrac{m_{\chi_j}^2}{M_W^2})=2\,M_W^2\Big([C_{21}-C_{23}-C_{12}](p_{\psi},-p_{\gamma},m^2_{\chi_j},M_W^2,M_W^2)\nonumber\\
&\null\hspace{4cm}+[C_{21}-C_{23}+2C_{11}-C_{12}+C_0](-p_{\psi},p_{\gamma},M_W^2,m^2_{\chi_j},m^2_{\chi_j})\Big)\,,\\
&{\cal I}^{V}_{R}(\tfrac{m_{\psi}^2}{M_W^2},\tfrac{m_{\chi_j}^2}{M_W^2})=4\,M_W^2\Big(C_{11}(p_{\psi},-p_{\gamma},m^2_{\chi_j},M_W^2,M_W^2)-[C_{11}+C_0](-p_{\psi},p_{\gamma},M_W^2,m^2_{\chi_j},m^2_{\chi_j})\Big)\,,\nonumber\\
&{\cal I}^{S}_{L}(\tfrac{m_{\psi}^2}{m_S^2},\tfrac{m_{f}^2}{m_S^2})=m_S^2\Big([C_{11}+C_0](p_{\psi},-p_{\gamma},m^2_{f},m_S^2,m_S^2)-C_{11}
(-p_{\psi},p_{\gamma},m_S^2,m^2_{f},m^2_{f})\Big)\,,\nonumber\\
&{\cal I}^{S}_{R}(\tfrac{m_{\psi}^2}{m_S^2},\tfrac{m_{f}^2}{m_S^2})=-m_S^2\Big([C_{21}-C_{23}+C_{11}-C_{12}](p_{\psi},-p_{\gamma},m^2_{f},m_S^2,m_S^2)\,,\nonumber\\
&\null\hspace{4.2cm}+[C_{21}-C_{23}+C_{11}-C_{12}](-p_{\psi},p_{\gamma},m_S^2,m^2_{f},m^2_{f})\Big)\,.\nonumber
\end{align}
In the regime $m_{\psi}^2\ll M_W^2,m_S^2$ that we consider here, it is useful (in terms of numerical stability) to replace these functions by their limit with vanishing first argument:
\begin{align}
	&{\cal I}^{V}_{L}(0,z)=-\tfrac{1}{1-z}\big[\tfrac{3}{2}-2z+(2z-1)f_{\gamma}(z)\big]\,,& &{\cal I}^{V}_{R}(0,z)=4\big(1-f_{\gamma}(z)\big)\,, & & \\
	&{\cal I}^{S}_{L}(0,z)=-f_{\gamma}(z)/z\,,& &{\cal I}^{S}_{R}(0,z)=\tfrac{1}{2(1-z)}\big(f_{\gamma}(z)-\tfrac{1}{2}\big)\,, & & \\
	&f_{\gamma}(z)\equiv-\tfrac{z}{1-z}\big(1+\tfrac{\ln z}{1-z}\big)\,.\nonumber
\end{align}

\section{Wilson Coefficients at Leading order in RpV / Mixing}\label{ap:simpWilsonC}
In this section, we express the Wilson coefficients at leading order in terms of the RpV parameters, in an expansion with respect to the mixing 
between gauge / flavor eigenstates and neglecting the Yukawa couplings. This operation provides a more intuitive view of the contributions of 
the RpV MSSM to the Wilson coefficients. We naturally assume that the light neutralino is bino-like. Nevertheless, the corresponding results do 
not accurately describe all arbitrary parameter choices of the model. In particular, bilinear RpV contributions (mixing leptons / sleptons and Higgs 
/ Higgsinos, hence involving higher orders in mixing or Yukawa couplings) are then systematically discarded. In regimes where $|\lambda^{(\prime),
(\prime\prime)}_{ijk}|\ll\big|\tfrac{\mu_{i'}}{\mu}\big|$ or if final states with charged leptons or hadrons are kinematically inaccessible, these 
bilinear parameters would however determine the decays of the light neutralino. For instance, decays into neutrinos are controlled by the following 
Wilson coefficients, involving two degrees of mixing:
\begin{equation}
C[{\mathcal N}_{ijk}]=-\frac{g_1 v_d(\mu_i\delta_{jk}+\mu_k\delta_{ij})}{4\sqrt{2}(1+\delta_{ik})\,\mu }\Big[\frac{1}{m^2_{\tilde{N}_j}}\Big(\frac{g_1^2}{M_1}+\frac{g_2^2}{M_2}\Big)+\frac{g_1^2+g_2^2}{M_1M_Z^2}\Big]\,.
\end{equation}
In addition, Yukawa contributions for the muon or the strange quarks may be competitive with mixing terms of higher order.

\subsection{Leptonic operators}
\begin{align}
&C[\widetilde{\cal N}_{ijk}](\mu_0)\approx0\,,\hspace{2.cm}C[{\cal N}_{ijk}](\mu_0)\approx0\,,\hspace{2.cm}C[{\cal S}_{ijk}^{\nu e\,R}](\mu_0)\approx 0\,,\nonumber\\
&C[{\cal S}_{ijk}^{\nu e\,L}](\mu_0)\approx-\frac{g_1}{\sqrt{2}}\left[\lambda_{ikj}\Big(\tfrac{1}{m^2_{\tilde{N}_i}}-\tfrac{1}{2m^2_{\tilde{E}_L^k}}+\tfrac{1}
{m^2_{\tilde{E}_R^j}}\Big)-\sum_{l\neq i}\frac{\lambda_{lkj}M^2_{\tilde{N}\,il}}{m^2_{\tilde{N}_i}m^2_{\tilde{N}_l}}+\sum_{l\neq k}\frac{\lambda_{ilj}M^2_{\tilde{E}_L\,lk}}{2m^2_{\tilde{E}_L^k}m^2_{\tilde{E}_L^l}}\right.\nonumber\\
&\left.\null\hspace{2.5cm}-\sum_{l\neq j}\frac{\lambda_{ikl}M^2_{\tilde{E}_R\,jl}}{m^2_{\tilde{E}_R^j}m^2_{\tilde{E}_R^l}}\right]\,,\nonumber\\
&C[{\cal V}_{ijk}^{\nu e\,L}](\mu_0)\approx\frac{g_1\lambda_{ikl}}{2\sqrt{2}}\frac{M^2_{\tilde{E}_{LR}\,jl}}{m^2_{\tilde{E}_L^j}m^2_{\tilde{E}_R^l}}\,,\hspace{2.cm}C[{\cal V}_{ijk}^{\nu e\,R}](\mu_0)\approx\frac{g_1\lambda_{ilj}}{\sqrt{2}}\frac{M^2_{\tilde{E}_{LR}\,lk}}{m^2_{\tilde{E}_L^l}m^2_{\tilde{E}_R^k}}\,,\\
&C[{\cal T}_{ijk}^{\nu e}](\mu_0)\approx\frac{g_1\lambda_{ikl}}{\sqrt{2}m^2_{\tilde{E}_R^j}}\Big[\delta_{jl}-(1-\delta_{jl})\frac{M^2_{\tilde{E}_R\,jl}}{m^2_{\tilde{E}_R^l}}\Big]+\frac{g_1\lambda_{ilj}}{2\sqrt{2}m^2_{\tilde{E}_L^k}}\Big[\delta_{kl}-(1-\delta_{kl})\frac{M^2_{\tilde{E}_L\,lk}}{m^2_{\tilde{E}_L^l}}\Big]\,.\nonumber
\end{align}
Below, we omit the $\displaystyle\sum_{l\neq\cdot}$ or $(1-\delta_{l\cdot})$ indicating sfermion mixing.

\subsection{Semi-leptonic operators}
\begin{align}
&C[{\cal S}_{ijk}^{\nu u\,L,R}](\mu_0)\approx0\approx C[{\cal V}_{ijk}^{\nu u\,L,R}](\mu_0)\approx C[{\cal T}_{ijk}^{\nu u}](\mu_0)\,,\hspace{2cm}C[{\cal S}_{ijk}^{\nu d\,R}](\mu_0)\approx0\,,\nonumber\\
&C[{\cal S}_{ijk}^{\nu d\,L}](\mu_0)\approx-\frac{g_1}{\sqrt{2}}\Big[\frac{\lambda'_{lkj}}{m^2_{\tilde{N}_i}}\big(\delta_{il}-\frac{M^2_{\tilde{N}\,il}}{m^2_{\tilde{N}_l}}\big)+\frac{\lambda'_{ilj}}{6m^2_{\tilde{D}_L^k}}\big(\delta_{kl}-\frac{M^2_{\tilde{D}_L\,lk}}{m^2_{\tilde{D}_L^l}}\big)+\frac{\lambda'_{ikl}}{3m^2_{\tilde{D}_R^j}}\big(\delta_{jl}-\frac{M^2_{\tilde{D}_R\,jl}}{m^2_{\tilde{D}_R^l}}\big)\Big]\,,\nonumber\\
&C[{\cal V}_{ijk}^{\nu d\,L}](\mu_0)\approx-\frac{g_1\lambda'_{ikl}M^2_{\tilde{D}_{LR}\,jl}}{6\sqrt{2}m^2_{\tilde{D}_L^j}m^2_{\tilde{D}_R^l}}\,,\hspace{2.5cm}C[{\cal V}_{ijk}^{\nu d\,R}](\mu_0)\approx\frac{g_1\lambda'_{ilj}M^2_{\tilde{D}_{LR}\,lk}}{3\sqrt{2}m^2_{\tilde{D}_L^l}m^2_{\tilde{D}_R^j}}\,,\\
&C[{\cal T}_{ijk}^{\nu d}](\mu_0)\approx\frac{g_1\lambda'_{ikl}}{3\sqrt{2}m^2_{\tilde{D}_R^j}}\Big[\delta_{lj}-\frac{M^2_{\tilde{D}_R\,jl}}{m^2_{\tilde{D}_R^l}}\Big]-\frac{g_1\lambda'_{ilj}}{6\sqrt{2}m^2_{\tilde{D}_L^k}}\Big[\delta_{lk}-\frac{M^2_{\tilde{D}_L\,lk}}{m^2_{\tilde{D}_L^l}}\Big]\,,\nonumber
\end{align}

\begin{align}
&C[{\cal S}_{ijk}^{eq\,LL}](\mu_0)\approx\frac{g_1}{\sqrt{2}}\Big[\tfrac{\lambda'_{lnj}V^{\text{CKM}\,*}_{kn}}{m^2_{\tilde{E}_L^i}}\Big(\delta_{il}
-\tfrac{M^2_{\tilde{E}_L\,li}}{m^2_{\tilde{E}_L^l}}\Big)+\tfrac{\lambda'_{ilj}V^{\text{CKM}\,*}_{ml}}{6m^2_{\tilde{U}_L^k}}\Big(\delta_{km}-
\tfrac{M^2_{\tilde{U}_L\,mk}}{m^2_{\tilde{U}_L^m}}\Big)\nonumber \\
&\null\hspace{2.7cm}+\tfrac{\lambda'_{ilm}V^{\text{CKM}\,*}_{kl}}{3m^2_{\tilde{D}_R^j}}\Big(\delta_{jm}-\tfrac{M^2_{\tilde{D}_R\,jm}}
{m^2_{\tilde{D}_R^m}}\Big)\Big]\,,\nonumber\\
&C[{\cal S}_{ijk}^{eq\,LR}](\mu_0)\approx0\,,\hspace{0.5cm}C[{\cal S}_{ijk}^{eq\,RL}](\mu_0)\approx\sqrt{2}g_1\lambda'_{lnj}V^{\text{CKM}\,*}
_{kn}\frac{M^2_{\tilde{E}_{LR}\,li}}{m^2_{\tilde{E}_L^l}m^2_{\tilde{E}_R^i}}\,,\hspace{0.5cm}C[{\cal S}_{ijk}^{eq\,RR}](\mu_0)\approx0\,,\nonumber\\
&C[{\cal V}_{ijk}^{eq\,LL}](\mu_0)\approx\frac{g_1\lambda'_{inm}V^{\text{CKM}\,*}_{kn}}{6\sqrt{2}}\frac{M^2_{\tilde{D}_{LR}\,jm}}{m^2_{\tilde{D}_L^j}m^2_{\tilde{D}_R^m}}\,,
\hspace{0.5cm}C[{\cal V}_{ijk}^{eq\,LR}](\mu_0)\approx\frac{\sqrt{2}g_1\lambda'_{inj}V^{\text{CKM}\,*}_{mn}}{3}\frac{M^2_{\tilde{U}_{LR}\,mk}}{m^2_{\tilde{U}_L^m}m^2_{\tilde{U}_R^k}}\nonumber\\
&C[{\cal V}_{ijk}^{eq\,RL}](\mu_0)\approx0\,,\hspace{2cm}C[{\cal V}_{ijk}^{eq\,RR}](\mu_0)\approx0\,,\hspace{2cm}C[{\cal T}_{ijk}^{eq\,R}](\mu_0)\approx0\,,\nonumber\\
&C[{\cal T}_{ijk}^{eq\,L}](\mu_0)\approx\frac{g_1}{\sqrt{2}}\Big[\tfrac{\lambda'_{ilj}V^{\text{CKM}\,*}_{mn}}{6m^2_{\tilde{U}_L^k}}\Big(\delta_{km}-\tfrac{M^2_{\tilde{U}_L\,mk}}{m^2_{\tilde{U}_L^m}}\Big)-\tfrac{\lambda'_{ilm}V^{\text{CKM}\,*}_{kl}}{3m^2_{\tilde{D}_R^j}}\Big(\delta_{jm}-\tfrac{M^2_{\tilde{D}_R\,jm}}{m^2_{\tilde{D}_R^m}}\Big)\Big]\,.
\end{align}

\subsection{Hadronic operators}
\begin{align}
&C[{\cal H}_{ijk}^{LL}](\mu_0)\approx-\frac{2\sqrt{2}g_1\lambda''_{njk}}{3m^2_{\tilde{U}_R^i}}\Big[\delta_{in}-\frac{M^2_{\tilde{U}_R\,in}}{m^2_{\tilde{U}_R^n}}\Big]\,,
\hspace{1.1cm}C[{\cal H}_{ijk}^{RL}](\mu_0)\approx-\frac{g_1\lambda''_{njk}}{3\sqrt{2}}\frac{M^2_{\tilde{U}_{LR}\,in}}{m^2_{\tilde{U}_L^i}m^2_{\tilde{U}_R^n}}\,,\\
&C[\widetilde{\cal H}_{ijk}^{LL}](\mu_0)\approx\frac{\sqrt{2}g_1\lambda''_{ink}}{3m^2_{\tilde{D}_R^j}}\Big[\delta_{jn}-\frac{M^2_{\tilde{D}_R\,jn}}{m^2_{\tilde{D}_R^n}}\Big]\,,
\hspace{1.2cm}C[\widetilde{\cal H}_{ijk}^{RL}](\mu_0)\approx-\frac{g_1\lambda''_{ink}}{3\sqrt{2}}\frac{M^2_{\tilde{D}_{LR}\,jn}}{m^2_{\tilde{D}_L^j}m^2_{\tilde{D}_R^n}}\,.\nonumber
\end{align}

\subsection{Electromagnetic dipole operators}
Neglecting any degree of mixing:
\begin{align}
C^{\text{MSSM}}[{\cal E}_i]\approx&\frac{g_1}{\sqrt{2}}\lambda_{ijj}m_{e_j}\Big[\frac{1}{m_{\tilde{E}_L^j}^2}{\cal I}^S_L\big(\tfrac{m^2_{\psi}}{m_{\tilde{E}_L^j}^2},\tfrac{m^2_{e_j}}{m_{\tilde{E}_L^j}^2}\big)+\frac{2}{m_{\tilde{E}_R^j}^2}{\cal I}^S_L\big(\tfrac{m^2_{\psi}}{m_{\tilde{E}_R^j}^2},\tfrac{m^2_{e_j}}{m_{\tilde{E}_R^j}^2}\big)\Big]\\
&-\frac{N_cg_1}{9\sqrt{2}}\lambda'_{ijj}m_{d_j}\Big[\frac{1}{m_{\tilde{D}_L^j}^2}{\cal I}^S_L\big(\tfrac{m^2_{\psi}}{m_{\tilde{D}_L^j}^2},\tfrac{m^2_{d_j}}{m_{\tilde{D}_L^j}^2}\big)-\frac{2}{m_{\tilde{D}_R^j}^2}{\cal I}^S_L\big(\tfrac{m^2_{\psi}}{m_{\tilde{D}_R^j}^2},\tfrac{m^2_{d_j}}{m_{\tilde{D}_R^j}^2}\big)\Big]\,.\nonumber
\end{align}

\section{Kinematic Integrals for the Three-body Leptonic Decays}\label{ap:lepkinintt}
The kinematic integrals of Eq.~(\ref{eq:lepkinint}) are defined by ($x_j+x_k<1$):
\begin{equation}
I_{\Omega,\Omega'}[x_j,x_k]=\int_{(x_j+x_k)^2}^{1}{\hspace{-0.7cm}ds\hspace{0.3cm}f_{\Omega,\Omega'}(s)(1-s)^2\Big(1-2\tfrac{x_j^2+x_k^2}{s}
+\tfrac{(x_j^2-x_k^2)^2}{s^2}\Big)^{1/2}}\,,
\end{equation}
where the functions $f_{\Omega,\Omega'}$ read (with $f_{\Omega',\Omega}=f_{\Omega,\Omega'}$):
\begin{align}
&f_{{\cal S}^L,{\cal S}^L}(s)=\tfrac{1}{2}(s-x_j^2-x_k^2)=f_{{\cal S}^R,{\cal S}^R}(s)\,, & & f_{{\cal S}^L,{\cal S}^R}(s)=-x_jx_k=f_{{\cal S}^R,{\cal S}^L}(s)\,,\nonumber\\
&f_{{\cal V}^L,{\cal S}^L}(s)=\tfrac{x_j}{2}[s^{-1}(x_j^2-x_k^2)-1]=f_{{\cal V}^R,{\cal S}^R}(s)\,, & & f_{{\cal V}^L,{\cal S}^R}(s)=\tfrac{x_k}{2}(s^{-1}[x_j^2-x_k^2]+1)=
f_{{\cal V}^R,{\cal S}^L}(s)\,,\nonumber \\
&f_{{\cal T}^L,{\cal S}^{L,R}}(s)=0\,, & & f_{{\cal V}^L,{\cal V}^R}(s)=2x_jx_k\,,\\
&f_{{\cal V}^L,{\cal V}^L}(s)=\tfrac{1}{3}\big\{1+2s+\tfrac{x_j^2+x_k^2}{s}(1-s)-\tfrac{(x_j^2-x_k^2)^2}{s^2}(2+s)\big\}=f_{{\cal V}^R,{\cal V}^R}(s)\,,\hspace{-4cm} & & \nonumber\\
&f_{{\cal T}^L,{\cal V}^L}(s)=\tfrac{3x_j}{2}(1+s^{-1}[x_k^2-x_j^2])\,, & & f_{{\cal T}^L,{\cal V}^R}(s)=\tfrac{3x_k}{2}(1+s^{-1}[x_j^2-x_k^2])\,,\nonumber\\
&f_{{\cal T}^L,{\cal T}^L}(s)=\tfrac{2+s}{6}\big\{1+\tfrac{x_j^2+x_k^2}{s}-2\tfrac{(x_j^2-x_k^2)^2}{s^2}\big\}\,. & & \nonumber
\end{align}
These can be written as linear combinations of the four base functions $f_{m}(s)=s^k$, $m\in\{-2,-1,0,1\}$; this decomposition can be used for the calculation of the integrals.
\begin{align}
	&\null\hspace{2cm}I_m[x_j,x_k]\equiv\int_{(x_j+x_k)^2}^{1}{\hspace{-0.7cm}ds\hspace{0.3cm}f_m(s)(1-s)^2\Big(1-2\tfrac{x_j^2+x_k^2}{s}+\tfrac{(x_j^2-x_k^2)^2}{s^2}\Big)^{1/2}}\,,\\
	&I_{-2}[x_j,x_k]=\left[\tfrac{x_j^2-x_k^2}{2}+\tfrac{x_j^2+x_k^2}{x_j^2-x_k^2}-\tfrac{x_j^2x_k^2}{(x_j^2-x_k^2)^3}\right]L_1+\left[1+\tfrac{x_j^2+x_k^2}{2}\right]L_2+\left[\tfrac{5}{2}+\tfrac{x_j^2+x_k^2}{2(x_j^2-x_k^2)^2}\right]S\,,\nonumber\\
	&I_{-1}[x_j,x_k]=-\left[x_j^2-x_k^2+\tfrac{x_j^2+x_k^2}{2(x_j^2-x_k^2)}\right]L_1-\left[\tfrac{1}{2}+x_j^2+x_k^2-x_j^2x_k^2\right]L_2-\tfrac{1}{2}[5+x_j^2+x_k^2]S\,,\nonumber\\
	&I_{0}[x_j,x_k]=\tfrac{x_j^2-x_k^2}{2}L_1+\tfrac{1}{2}\left[x_j^2+x_k^2+2x_j^2x_k^2(x_j^2+x_k^2-2)\right]L_2\nonumber \\
	&\null\hspace{2cm}+\tfrac{1}{6}\left[2+5(x_j^2+x_k^2)-x_j^4-x_k^4-10x_j^2x_k^2\right]S\,,\nonumber\\
	&I_{1}[x_j,x_k]=x_j^2x_k^2\left[1-3(x_j^2+x_k^2)+x_j^4+x_k^4+3x_j^2x_k^2\right]L_2\nonumber\\
	&\null\hspace{2.cm}+\tfrac{1}{12}\left[1-3(x_j^2+x_k^2)+3x_j^4+3x_k^4+32x_j^2x_k^2-x_j^6-x_k^6-29x_j^2x_k^2(x_j^2+x_k^2)\right]S\,,\nonumber\\
	&\null\hspace{1cm}L_1\equiv\ln\tfrac{1+X_1}{1-X_1}\hspace{0.1cm},\hspace{0.2cm}X_1\equiv\tfrac{x_j^2-x_k^2}{x_j^2+x_k^2-(x_j^2-x_k^2)^2}S\hspace{0.2cm};\hspace{0.3cm}L_2\equiv\ln\tfrac{1+X_2}{1-X_2}\hspace{0.1cm},\hspace{0.2cm}X_2\equiv-\tfrac{S}{1-x_j^2-x_k^2}\hspace{0.2cm};\hspace{0.3cm}\nonumber\\
	&\null\hspace{1cm}S\equiv\left[1-2(x_j^2+x_k^2)+(x_j^2-x_k^2)^2\right]^{1/2}\,.\nonumber
\end{align}
The functions $I_m$ are regular in the limits $x_j\to x_k$ and $x_j,x_k\to0$.

\section{Neutron-antineutron Oscillations}\label{ap:nnbarosc}
The chiral Lagrangian of Eq.~(\ref{eq:psiChPT}) implies neutralino-baryon mixing, which can mediate baryon-antibaryon oscillations. Here, we 
focus on the very constrained case of the neutron \cite{Super-Kamiokande:2011idx}. Eq.~(\ref{eq:psiChPT}) in particular contains the terms:
\begin{equation}\label{eq:neutronpsimix}
{\cal L}_{\psi}^{\chi PT}\ni-m_{\psi n}(\bar{\Psi}P_Ln^0)- m_{\psi n^c}^*(\bar{\Psi}P_Rn^0)+h.c.\ \ ;\ \ \begin{cases}
m_{\psi n}\equiv-\big(\tilde{\alpha}\,C[\widetilde{\cal H}_{111}^{RL}]^*+\tilde{\beta}\,C[\widetilde{\cal H}_{111}^{RR}]^*\big)(\mu)\\
m_{\psi n^c}\equiv-\big(\tilde{\alpha}\,C[\widetilde{\cal H}_{111}^{LR}]+\tilde{\beta}\,C[\widetilde{\cal H}_{111}^{LL}]\big)(\mu)
\end{cases}
\end{equation}
We can this in terms of a mass matrix  for the neutralino-neutron-antineutron system:
\begin{equation}\label{eq:neutronmass}
-\frac{1}{2}(\psi\,n_L\,n_R^c)\begin{bmatrix}
m_{\psi} & m_{\psi n} & m_{\psi n^c} \\ m_{\psi n} & m_{nn} & m_n \\ m_{\psi n^c} & m_n & m_{n^cn^c}
\end{bmatrix}\begin{pmatrix}
\psi\\n_L\\n_R^c
\end{pmatrix}+h.c.\,,
\end{equation}
where the entries $m_{nn}$ and $m_{n^cn^c}$ emerge from dimension 9 (six-fermion) operators and correspond to a direct high-energy 
contribution to the neutron-antineutron mixing. At tree-level, these are mediated by gluino + squark or heavy neutralino + squark diagrams 
--- see \textit{e.g.}~Ref.\,\cite{Calibbi:2016ukt} for a recent review. They entail a scale-suppression double to that of $m_{\psi n}$ and 
$m_{\psi n^c}$. Relevant lattice evaluations of the hadronic matrix elements for the dimension 9 operators can be found in Ref.\,\cite{Rinaldi:2019thf}.

Diagonalization of the mass matrix of Eq.~(\ref{eq:neutronmass}) generates a mass splitting between the neutron and antineutron. At 
leading order in $|m_{nn},m_{n^cn^c}|\ll|m_{\psi n},m_{\psi n^c}|\ll|m_n\pm m_{\psi}|$, the two mass eigenvalues for the hadron-dominated 
states read:
\begin{eqnarray}
m_{n^+}&\approx& m_n+\tfrac{1}{2}\text{Re}[m_{nn}+m_{n^cn^c}]+\tfrac{2m_n-m_{\psi}}{2(m_n-m_{\psi})^2}\text{Re}[m_{\psi n}+m_{\psi n^c}]^2 \nonumber\\
&&+\tfrac{2m_n+m_{\psi}}{2(m_n+m_{\psi})^2}\text{Im}[m_{\psi n}+m_{\psi n^c}]^2\,,\\
m_{n^-}&\approx& -m_n+\tfrac{1}{2}\text{Re}[m_{nn}+m_{n^cn^c}]-\tfrac{2m_n+m_{\psi}}{2(m_n+m_{\psi})^2}\text{Re}[m_{\psi n}-m_{\psi n^c}]^2\nonumber\\
&&-\tfrac{2m_n-m_{\psi}}{2(m_n-m_{\psi})^2}\text{Im}[m_{\psi n}-m_{\psi n^c}]^2\,.\nonumber
\end{eqnarray}
This results in a neutron-antineutron oscillation frequency of
\begin{equation}
\omega_{nn^c}\approx \text{Re}[m_{nn}+m_{n^cn^c}]+\tfrac{2m_n-m_{\psi}}{2(m_n-m_{\psi})^2}\text{Re}[(m_{\psi n}+m_{\psi n^c}^*)^2]- \tfrac{2m_n+m_{\psi}}{2(m_n+m_{\psi})^2}\text{Re}[(m_{\psi n}-m_{\psi n^c}^*)^2]\,,
\end{equation}
which can be compared to the experimental limit on the oscillation time, from Super Kamiokande \cite{Super-Kamiokande:2011idx}, $\tau_{nn^c}>2.7\cdot10^8$\,s. Nevertheless, we stress that this particular limit critically depends on the connection between $\tau_{nn^c}$ and the $O^{16}$ lifetime, which involves complex nuclear modeling. A comparable bound has been extracted from the deuteron lifetime \cite{SNO:2017pha} using EFT methods, but finding significant discrepancy with nuclear models \cite{Oosterhof:2019dlo}. Naturally, the existence of light new physics (in the presence of a light neutralino) may further modify corresponding analyses. More direct limits from neutron beam experiments are available in Ref.\cite{Baldo-Ceolin:1994hzw}, with further improvement expected at the European Spallation Source \cite{Theroine:2016chp}.

Let us focus on the contributions originating in neutron-neutralino mixing, \textit{i.e.}\ $m_{\psi n}$ and $m_{\psi n^c}$ of Eq.~(\ref{eq:neutronpsimix}). In the RpV MSSM at leading order, $C[\widetilde{\cal H}^{LR}_{111}]$ and $C[\widetilde{\cal H}^{RR}_{111}]$ vanish while non-zero $C[\widetilde{\cal H}^{LL}_{111}]$ and $C[\widetilde{\cal H}^{RL}_{111}]$ imply flavor mixing in the squark sector. This suggests a source of suppression from flavor-mixing in addition to that originating in the RpV violation. In the approximation of suppressed flavor mixing (see 
Appendix~\ref{ap:simpWilsonC}), we have:
\begin{equation}
C[\widetilde{\cal H}^{LL}_{111}](\mu_0)\approx\frac{\sqrt{2}g_1}{3}\lambda''_{1n1}\frac{M^2_{\tilde{D}_R1n}}{m^2_{\tilde{D}_R^1}m^2_{\tilde{D}_R^n}}\ ;\ \ \ \ C[\widetilde{\cal H}^{RL}_{111}](\mu_0)\approx\frac{g_1}{3\sqrt{2}}\lambda''_{1n1}\frac{M^2_{\tilde{D}_{LR}1n}}{m^2_{\tilde{D}_L^1}m^2_{\tilde{D}_R^n}}\,.
\end{equation}
Similarly, flavor-mixing is a necessary ingredient to generate non-trivial $m_{nn}$ and $m_{n^cn^c}$, so that the existence of neutron-antineutron oscillations in the RpV MSSM strongly depend on the chosen flavor template.

\section{Hadronic Decay Widths}\label{ap:haddecwi}
The mixing Lagrangian of Eq.~(\ref{eq:psiChPT}) generates mixing terms between the neutralino and the baryons, as well as trilinear couplings with mesons:
\begin{eqnarray}\label{eq:barlagr}
{\cal L}_{\chi}&\ni& \Omega^{B_k}_{L,R}\bar{\psi}P_{L,R}B_k+i\,C^{\psi M_iB_j}_{L,R}\bar{\psi}P_{L,R}B_jM_i\\[3.mm]
\Omega^{B_k}_{L}&\equiv&\!\!-(\tilde{\alpha} C^{LR\,*}_{mn}+\tilde{\beta} C^{RR\,*}_{mn})\text{Tr}[E_{mn}\lambda_k]\,;\ \ \Omega^{B_k}_{R}\equiv-(\tilde{\alpha} C^{RL\,*}_{mn}+\tilde{\beta} C^{LL\,*}_{mn})\text{Tr}[E_{mn}\lambda_k]\,,\\
C_L^{\psi M_iB_j}&\equiv&\!\!-\frac{\tilde{\alpha}}{f_{\chi}}C^{LR\,*}_{mn}\text{Tr}[E_{mn}(\lambda_i\lambda_j+\lambda_j\lambda_i)]-\frac{\tilde{\beta}}{f_{\chi}}C^{RR\,*}_{mn}\text{Tr}[E_{mn}(\lambda_i\lambda_j-\lambda_j\lambda_i)]\,,\\
C_R^{\psi M_iB_j}&\equiv&\!\!\frac{\tilde{\alpha}}{f_{\chi}}C^{RL\,*}_{mn}\text{Tr}[E_{mn}(\lambda_i\lambda_j+\lambda_j\lambda_i)]+\frac{\tilde{\beta}}{f_{\chi}}C^{LL\,*}_{mn}\text{Tr}[E_{mn}(\lambda_i\lambda_j-\lambda_j\lambda_i)]\,.
\end{eqnarray}
Here, $C_{mn}^{LL,RR,LR,RL}$ stand for the Wilson coefficients of the operators of ${\cal H}$- and ${\widetilde{\cal H}}$-types and can be straightforwardly identified by comparing with 
Eq.~(\ref{eq:psiChPT}). We have written the baryon and meson octets as $B\equiv\lambda_j B_j$, $M\equiv\lambda_i M_i$, where $\lambda_j$ are linear combinations of the Gell-Mann matrices, $B_j\in\{\Sigma^0,\Sigma^+,\Sigma^-,n^0,\Xi^0,p^+,\Xi^-,\Lambda^0\}$ and $M_i\in\{\pi^0,\pi^+,\pi^-,K^0,\bar{K}^0,K^+,K^-,\eta^0_8\}$.

Through the mixing terms of Eq.~(\ref{eq:barlagr}), the neutralino may decay into one baryon and one meson using the baryon-meson-baryon interaction
(in four-component spinor notation):
\begin{eqnarray}
{\cal L}_{\chi}&\ni& i\, C^{\bar{B}_kM_iB_j}\big(\bar{B}_k(-i\slashed{\partial}M_i)\gamma_5 B_j\big)+i\,\tilde{C}^{\bar{B}_kM_iB_j}M_i\big(\bar{B}_k\gamma_5 B_j\big)\,,\\
C^{\bar{B}_kM_iB_j}&\equiv&\frac{D-F}{f_{\chi}}\text{Tr}[\lambda^{\dagger}_k\lambda_j\lambda_i]+\frac{D+F}{f_{\chi}}\text{Tr}[\lambda^{\dagger}_k\lambda_i\lambda_j]\,,\\
\tilde{C}^{\bar{B}_kM_iB_j}&\equiv&-\frac{2b_1}{f_{\chi}}\text{Tr}[\lambda^{\dagger}_k(M_q\lambda_i+\lambda_i M_q)\lambda_j]-\frac{2b_2}{f_{\chi}}\text{Tr}[\lambda^{\dagger}_k\lambda_j(M_q\lambda_i+\lambda_i M_q)]\,,
\end{eqnarray}
where $M_q=\text{diag}[m_u,m_d,m_s]$ is the quark mass matrix, $D\approx0.8$, $F\approx 0.47$ while $b_{1,2}$ are ill-known (we will ignore them in practice). Further terms accounting for the weak interaction can be included: they are essential in the description of the hyperon decays --- see \textit{e.g.}\ Ref.\cite{Borasoy:1998ku}. We dispense with a detailed description here: they essentially matter in the immediate mass-vicinity of an hyperon or in situations where other sources of $s\to d$ transition are suppressed.

Then, evaluating the two topologies of Feynman diagrams contributing to the $\psi\to\bar{M}_i\bar{B}_j$ decay amplitude at tree-level, we obtain:
\begin{eqnarray}
{\cal A}_{\chi}[\psi&\to&\bar{M}_i\bar{B}_j]=-\bar{v}_{\psi}(p_{\psi})g_{L,R}^{\psi M_iB_j}P_{L,R}v_{B_j}(p_{B_j})\\
g_L^{\psi M_iB_j}&\equiv& C_L^{\psi M_iB_j}-C^{\bar{B}_kM_iB_j}\frac{\Omega^{B_k}_L(m_{\psi}^2+m_{B_k}m_{B_j})+\Omega^{B_k}_Rm_{\psi}
(m_{B_k}+m_{B_j})}{m_{\psi}^2-m_{B_k}^2+i m_{B_k}\Gamma_k}\nonumber\\
&&+\tilde{C}^{\bar{B}_kM_iB_j}\frac{\Omega^{B_k}_Lm_{B_k}+\Omega^{B_k}
_Rm_{\psi}}{m_{\psi}^2-m_{B_k}^2+i m_{B_k}\Gamma_k}\\
g_R^{\psi M_iB_j}&\equiv& C_R^{\psi M_iB_j}+C^{\bar{B}_kM_iB_j}\frac{\Omega^{B_k}_R(m_{\psi}^2+m_{B_k}m_{B_j})+\Omega^{B_k}_Lm_{\psi}
(m_{B_k}+m_{B_j})}{m_{\psi}^2-m_{B_k}^2+i m_{B_k}\Gamma_k}\nonumber\\
&&-\tilde{C}^{\bar{B}_kM_iB_j}\frac{\Omega^{B_k}_Rm_{B_k}+\Omega^{B_k}
_Lm_{\psi}}{m_{\psi}^2-m_{B_k}^2+i m_{B_k}\Gamma_k}
\end{eqnarray}

\section{Semi-leptonic Three-body Decays}\label{ap:semilepdec}
In the approximation where the lepton in the final state does not interact with the di-meson system, the decay width for an individual semi-leptonic process $\psi\to\bar{\ell}_iM_1M_2$ can be derived from the amplitude of Eq.~(\ref{eq:semilepamp}):
\begin{multline}
	\Gamma[\psi\to\bar{\ell}_iM_1M_2]=\frac{m_{\psi}^3B_0^2}{512(1+\delta_{M_1M_2})\pi^3f_{\chi}^2}\int_{(X_1+X_2)^2}^{(1-x_i)^2}{\hspace{-0.5cm}d\tilde{s}\,\Big[1-\tfrac{2}{\tilde{s}}(X_1^2+X_2^2)+\tfrac{1}{\tilde{s}^2}(X_1^2-X_2^2)^2\Big]^{1/2}}\\
	\times\Big[1-2(\tilde{s}+x_i^2)+(\tilde{s}-x_i^2)^2\Big]^{1/2}\sum_{\Omega,\Omega'}C[\Omega]C[\Omega']^*\,g^{\bar{\ell}_iM_1M_2}[\Omega,\Omega'](\tilde{s})\,,
\end{multline}
where $x_i\equiv m_{\ell_1}/m_{\psi}$, $X_{1,2}\equiv m_{M_{1,2}}/m_{\psi}$. This defines the kinematic integral $G^{M_1M_2}[\Omega,\Omega']$ of Eq.~(\ref{eq:SL3bdecwi}). After application of Eq.~(\ref{eq:semilepformfac}), $g^{\bar{\ell}_iM_1M_2}[\Omega,\Omega'](\tilde{s})$ can be related to the reduced form-factors ${\cal F}^{M_1M_2}_{\Omega}(\tilde{s})\equiv \sqrt{2}f_{\chi}\,F^{M_1M_2}_{\Omega}(m_{\psi}^2\tilde{s})/B_0$ for all relevant operator pairs $(\Omega,\Omega')$ (we omit the permutations that can be obtained from $g^{\bar{\ell}_iM_1M_2}[\Omega',\Omega]=g^{\bar{\ell}_iM_1M_2}[\Omega,\Omega']^*$):
\begin{eqnarray}\label{eq:SL3bint}
	g^{\bar{\ell}_iM_1M_2}[{\cal S}_{ijk}^{\ell q\,JK},{\cal S}_{i\tilde{\jmath}\tilde{k}}^{\ell q\,JK'}](\tilde{s})&\equiv&\tfrac{1}{8}(1-\tilde{s}+x_i^2){\cal F}^{M_1M_2}_{S\,jk}(\tilde{s})\,\big({\cal F}^{M_1M_2}_{S\,\tilde{\jmath}\tilde{k}}(\tilde{s})\big)^*\nonumber\\
	g^{\bar{\ell}_iM_1M_2}[{\cal S}_{ijk}^{\ell q\,JK},{\cal S}_{i\tilde{\jmath}\tilde{k}}^{\ell q\,J+1K'}](\tilde{s})&\equiv&\tfrac{x_i}{4}{\cal F}^{M_1M_2}_{S\,jk}(\tilde{s})\,\big({\cal F}^{M_1M_2}_{S\,\tilde{\jmath}\tilde{k}}(\tilde{s})\big)^*\\
	g^{\bar{\ell}_iM_1M_2}[{\cal V}_{ijk}^{\ell q\,JK},{\cal S}_{i\tilde{\jmath}\tilde{k}}^{\ell q\,JK'}](\tilde{s})&\equiv&-\tfrac{y_j-y_k}{8}\big[\tfrac{1}{\tilde{s}}(1-x_i^2)-1\big]{\cal F}^{M_1M_2}_{S\,jk}(\tilde{s})\,\big({\cal F}^{M_1M_2}_{S\,\tilde{\jmath}\tilde{k}}(\tilde{s})\big)^*\nonumber\\
	g^{\bar{\ell}_iM_1M_2}[{\cal V}_{ijk}^{\ell q\,JK},{\cal S}_{i\tilde{\jmath}\tilde{k}}^{\ell q\,J+1K'}](\tilde{s})&\equiv&-\tfrac{x_i}{8}(y_j-y_k)\big[\tfrac{1}{\tilde{s}}(1-x_i^2)+1\big]{\cal F}^{M_1M_2}_{S\,jk}(\tilde{s})\,\big({\cal F}^{M_1M_2}_{S\,\tilde{\jmath}\tilde{k}}(\tilde{s})\big)^*\nonumber\\
	g^{\bar{\ell}_iM_1M_2}[{\cal T}_{ijk}^{\ell q\,J},{\cal S}_{i\tilde{\jmath}\tilde{k}}^{\ell q\,J'K'}](\tilde{s})&\equiv&0\nonumber\\
	g^{\bar{\ell}_iM_1M_2}[{\cal V}_{ijk}^{\ell q\,JK},{\cal V}_{i\tilde{\jmath}\tilde{k}}^{\ell q\,JK'}](\tilde{s})&\equiv&\tfrac{1}{8\tilde{s}^2}(y_j-y_k)(y_{\tilde{\jmath}}-y_{\tilde{k}})\big[1-\tilde{s}-x_i^2(2+\tilde{s})+x_i^4\big]{\cal F}^{M_1M_2}_{S\,jk}(\tilde{s})\,\nonumber \\
	&&\times \big({\cal F}^{M_1M_2}_{S\,\tilde{\jmath}\tilde{k}}(\tilde{s})\big)^*
	+\tfrac{m^2_{\psi}}{24}\big[1-\tfrac{2}{\tilde{s}}(X_1^2+X_2^2)+\tfrac{1}{\tilde{s}^2}(X_1^2-X_2^2)^2\big]\nonumber\\
	&&\times \big[1+\tilde{s}-2\tilde{s}^2+x_i^2(\tilde{s}-2)+x_i^4\big]{\cal F}^{M_1M_2}_{V\,jk}(\tilde{s})\,\big({\cal F}^{M_1M_2}_{V\,\tilde{\jmath}\tilde{k}}(\tilde{s})\big)^*\nonumber\\
	g^{\bar{\ell}_iM_1M_2}[{\cal V}_{ijk}^{\ell q\,JK},{\cal V}_{i\tilde{\jmath}\tilde{k}}^{\ell q\,J+1K'}](\tilde{s})&\equiv&\tfrac{x_i}{4\tilde{s}}(y_j-y_k)(y_{\tilde{\jmath}}-y_{\tilde{k}}){\cal F}^{M_1M_2}_{S\,jk}(\tilde{s})\,\big({\cal F}^{M_1M_2}_{S\,\tilde{\jmath}\tilde{k}}(\tilde{s})\big)^*\nonumber\\
	&&-\tfrac{m_{\psi}^2}{4}x_i\,\tilde{s}\big[1-\tfrac{2}{\tilde{s}}(X_1^2+X_2^2)+\tfrac{1}{\tilde{s}^2}(X_1^2-X_2^2)^2\big]\nonumber \\
	&&{\cal F}^{M_1M_2}_{V\,jk}(\tilde{s})\,\big({\cal F}^{M_1M_2}_{V\,\tilde{\jmath}\tilde{k}}(\tilde{s})\big)^*\nonumber\\
	g^{\bar{\ell}_iM_1M_2}[{\cal T}_{ijk}^{\ell q\,J},{\cal V}_{i\tilde{\jmath}\tilde{k}}^{\ell q\,JK'}](\tilde{s})&\equiv&-\tfrac{m_{\psi}^3}{16}\tilde{s}\big(1-\tilde{s}-x_i^2\big)\big[1-\tfrac{2}{\tilde{s}}(X_1^2+X_2^2)+\tfrac{1}{\tilde{s}^2}(X_1^2-X_2^2)^2\big]\nonumber \\
	&&{\cal F}^{M_1M_2}_{T\,jk}(\tilde{s})\,\big({\cal F}^{M_1M_2}_{V\,\tilde{\jmath}\tilde{k}}(\tilde{s})\big)^*\nonumber\\
	g^{\bar{\ell}_iM_1M_2}[{\cal T}_{ijk}^{\ell q\,J},{\cal V}_{i\tilde{\jmath}\tilde{k}}^{\ell q\,J+1K'}](\tilde{s})&\equiv&\tfrac{m_{\psi}^3}{16}x_i\,\tilde{s}\big(1+\tilde{s}-x_i^2\big)\big[1-\tfrac{2}{\tilde{s}}(X_1^2+X_2^2)+\tfrac{1}{\tilde{s}^2}(X_1^2-X_2^2)^2\big]\nonumber\\
	&&{\cal F}^{M_1M_2}_{T\,jk}(\tilde{s})\,\big({\cal F}^{M_1M_2}_{V\,\tilde{\jmath}\tilde{k}}(\tilde{s})\big)^*\nonumber\\
	g^{\bar{\ell}_iM_1M_2}[{\cal T}_{ijk}^{\ell q\,J},{\cal T}_{i\tilde{\jmath}\tilde{k}}^{\ell q\,J}](\tilde{s})&\equiv&\tfrac{m_{\psi}^4}{96}\tilde{s}\big[2-\tilde{s}-\tilde{s}^2-x_i^2(4+\tilde{s})+2x_i^4\big]\nonumber\\
	&&\times\big[1-\tfrac{2}{\tilde{s}}(X_1^2+X_2^2)+\tfrac{1}{\tilde{s}^2}(X_1^2-X_2^2)^2\big]
	{\cal F}^{M_1M_2}_{T\,jk}(\tilde{s})\,\big({\cal F}^{M_1M_2}_{T\,\tilde{\jmath}\tilde{k}}(\tilde{s})\big)^*\nonumber\\
	g^{\bar{\ell}_iM_1M_2}[{\cal T}_{ijk}^{\ell q\,J},{\cal T}_{i\tilde{\jmath}\tilde{k}}^{\ell q\,J+1}](\tilde{s})&\equiv&-\tfrac{m_{\psi}^4}{16}x_i\,\tilde{s}^2\big[1-\tfrac{2}{\tilde{s}}(X_1^2+X_2^2)+\tfrac{1}{\tilde{s}^2}(X_1^2-X_2^2)^2\big]\nonumber \\
	&&{\cal F}^{M_1M_2}_{T\,jk}(\tilde{s})\,\big({\cal F}^{M_1M_2}_{T\,\tilde{\jmath}\tilde{k}}(\tilde{s})\big)^*\nonumber
\end{eqnarray}
with $y_{j,k,\tilde{\jmath},\tilde{k}}\equiv m_{q_{j,k,\tilde{\jmath},\tilde{k}}}/m_{\psi}$; the chiralities $J,J',K,K'$ take the values $L,R$, with the convention that $J+1\neq J$. Setting baryon, meson, quark and lepton masses to their numerical values, the kinematic integrals $G^{\bar{\ell}_iM_mM_n}[\Omega,\Omega']$ can be seen as functions of the sole mass $m_{\psi}$.

\section{Partonic Decay Widths}\label{ap:partdecwi}
For hadronic or semi-leptonic processes with a large phase space (\textit{i.e.}~far above thresholds), the partonic descriptions regains its legitimacy. We neglect quark and lepton masses below.

For the three hadronic decay channels $(i,j,k)\in\{(1,1,1),(1,1,2),(1,2,2)\}$ (and their CP-conjugates):
\begin{align}
	&\Gamma[\psi\to\bar{u}_i\bar{d}_j\bar{d}_k]=\frac{m_{\psi}^5}{1024(1+\delta_{jk})\pi^3}\bigg\{\big|C[{\cal H}^{J\,K}_{ijk}]\big|^2+\big|C[\widetilde{\cal H}^{J\,K}_{ijk}]\big|^2+\big|C[\widetilde{\cal H}^{J\,K}_{ikj}]\big|^2\\
	&\null\hspace{2.4cm}-\text{Re}\big[C[{\cal H}^{J\,J}_{ijk}]C[\widetilde{\cal H}^{J\,J}_{ijk}]^*\big]-\text{Re}\big[C[{\cal H}^{J\,J}_{ijk}]C[\widetilde{\cal H}^{J\,J}_{ikk}]^*\big]+\text{Re}\big[C[\widetilde{\cal H}^{J\,J}_{ijk}]C[\widetilde{\cal H}^{J\,J}_{ikj}]^*\big]\bigg\}\,.\nonumber
\end{align}

For the semi-leptonic decay channels (and their CP-conjugates):
\begin{align}
	&\Gamma[\psi\to\bar{\ell}_iq_j\bar{q}_k]=\frac{m_{\psi}^5}{2048\pi^3}\bigg\{\big|C[{\cal S}^{\ell q\,J\,K}_{ijk}]\big|^2+4\big|C[{\cal V}^{\ell q\,J\,K}_{ijk}]\big|^2+3\big|C[{\cal T}^{\ell q\,J}_{ijk}]\big|^2\bigg\}\,,\\[1.5mm]
	&\null\hspace{.5cm}(\ell_i,q_j,q_k)\in\{(\nu_i,u,u),(\nu_i,d,d),(\nu_i,s,s),(\nu_i,d,s),(\nu_i,s,d),(e_i,d,u),(e_i,s,u)\}\,.\nonumber
\end{align}

\section{Production from the Decays of SM Particles\label{ap:prod}}
\subsubsection*{Production in the decays of electroweak gauge bosons}
Single ($\chi_i^0\stackrel[]{!}{=}\psi$, $\chi_j^0\stackrel[]{!}{=}\nu_f,\bar{\nu}_f$) or pair production ($\chi_i^0\stackrel[]{!}{=}\psi\stackrel[]{!}{=}\chi_j^0$) can occur in decays of the $Z$-boson, relying on subleading neutralino components:
\begin{align}
&\Gamma[Z\to\chi_i^0\chi_j^0]=\frac{M_Z}{24(1+\delta_{ij})\pi}\bigg[1-2\tfrac{m_{\chi_i^0}^2+m_{\chi_j^0}^2}{M_Z^2}+\tfrac{(m_{\chi_i^0}^2+m_{\chi_j^0}^2)^2}{M_Z^4}\bigg]^{1/2}\times\\
&\null\hspace{0.8cm}\bigg\{\Big(|g_L^{Z\chi_i^0\chi_j^0}|^2+|g_R^{Z\chi_i^0\chi_j^0}|^2\Big)\Big[1-\tfrac{m_{\chi_i^0}^2+m_{\chi_j^0}^2}{2M_Z^2}-\tfrac{(m_{\chi_i^0}^2+m_{\chi_j^0}^2)^2}{2M_Z^4}\Big]+6\tfrac{m_{\chi_i^0}m_{\chi_j^0}}{M_Z^2}\text{Re}\Big[g_L^{Z\chi_i^0\chi_j^0}(g_R^{Z\chi_i^0\chi_j^0})^*\Big]\bigg\}\,.\nonumber
\end{align}
Single production in the decays of a $W$-boson:
\begin{align}
&\Gamma[W\to\psi e_i]=\frac{M_W}{24\pi}\bigg[1-2\tfrac{m_{\psi}^2+m_{e_i}^2}{M_W^2}+\tfrac{(m_{\psi}^2+m_{e_i}^2)^2}{M_W^4}\bigg]^{1/2}\times\\
&\null\hspace{1.4cm}\bigg\{\Big(|g_L^{W\psi e_i}|^2+|g_R^{W\psi e_i}|^2\Big)\Big[1-\tfrac{m_{\psi}^2+m_{e_i}^2}{2M_W^2}-\tfrac{(m_{\psi}^2+m_{e_i}^2)^2}{2M_W^4}\Big]+6\tfrac{m_{\psi}m_{e_i}}{M_W^2}\text{Re}\Big[g_L^{W\psi e_i\chi_j^0}(g_R^{W\psi e_i})^*\Big]\bigg\}\,.\nonumber
\end{align}

\subsubsection*{Production in the decays of a neutral scalar (Higgs)}
Single ($\chi_i^0\stackrel[]{!}{=}\psi$, $\chi_j^0\stackrel[]{!}{=}\nu_f,\bar{\nu}_f$) or pair production ($\chi_i^0\stackrel[]{!}{=}\psi\stackrel[]{!}{=}\chi_j^0$) can occur in decays of the Higgs boson, relying on subleading neutralino components or sneutrino-Higgs mixing:
\begin{align}
&\Gamma[S_p^0\to\chi_i^0\chi_j^0]=\frac{m_{S_p^0}}{16(1+\delta_{ij})\pi}\bigg[1-2\tfrac{m_{\chi_i^0}^2+m_{\chi_j^0}^2}{m_{S_p^0}^2}+\tfrac{(m_{\chi_i^0}^2+m_{\chi_j^0}^2)^2}{m_{S_p^0}^4}\bigg]^{1/2}\times\\
&\null\hspace{2.1cm}\bigg\{\big(|g_L^{S_p^0\chi_i^0\chi_j^0}|^2+|g_R^{S_p^0\chi_i^0\chi_j^0}|^2\big)\Big[1-\tfrac{m_{\chi_i^0}^2+m_{\chi_j^0}^2}{m_{S_p^0}}\Big]-4\tfrac{m_{\chi_i^0}m_{\chi_j^0}}{m_{S_p^0}^2}\text{Re}\Big[g_L^{S_p^0\chi_i^0\chi_j^0}(g_R^{S_p^0\chi_i^0\chi_j^0})^*\Big]\bigg\}\,.\nonumber
\end{align}

\subsubsection*{Production in leptonic transitions}
We collect the operators of Eq.~(\ref{eq:leptop}), last line, and Eq.~(\ref{eq:chipairop}), $f\stackrel[]{!}{=}e$, in a single basis, allowing to describe single and pair production in leptonic transitions simultaneously ($i\geq j$, $n\geq m$):
\begin{eqnarray}
{\cal S}_{ijmn}^{K\,J}&\equiv&(\bar{X}^0_iP_KX^0_j)(\bar{E}_mP_JE_n)\,,\qquad {\cal V}_{ijmn}^{K\,J}\equiv(\bar{X}^0_i\gamma^{\mu}P_KX^0_j)(\bar{E}_m\gamma_{\mu}P_JE_n)\,,
\nonumber \\
{\cal T}_{ijmn}^{K}&\equiv&\frac{1}{4}(\bar{X}^0_i\Sigma^{\mu\nu}P_KX^0_j)(\bar{E}_m\Sigma_{\mu\nu}P_KE_n)\,.
\end{eqnarray}
Then, the three-body decay of a charged lepton $e_n$ into the lighter $e_m$ and a light neutralino pair $\chi_i^0$, $\chi_j^0$ can be written as the following sum (with mass indices straightforwardly referring to the particle of corresponding index):
\begin{align}
&\Gamma[e_n\to e_m\chi_i^0\chi_j^0]=\frac{1+\delta_{ij}}{512\pi m_{n}}\int_{(m_i+m_j)^2}^{(m_n-m_{m})^2}{\hspace{-0.5cm}ds\,\Big[1-2\tfrac{m_i^2+m_j^2}{s}+\tfrac{(m_i^2-m_j^2)^2}{s^2}\Big]\Big[1-2\tfrac{s+m_m^2}{m_n^2}+\tfrac{(s-m_m^2)^2}{m_n^4}\Big]}\nonumber\\
&\null\hspace{3.3cm}\times\sum_{\Omega,\Omega'}C[\Omega]C[\Omega']^*\,g^{e_me_n\chi_i^0\chi_j^0}[\Omega,\Omega'](s)\,,
\end{align}
where
\begin{eqnarray} 
g^{e_me_n\chi_i^0\chi_j^0}[{\cal S}^{K\,J}_{ijmn},{\cal S}^{K\,J}_{ijmn}](s)&\equiv&(s-m_i^2-m_j^2)(m_n^2+m_m^2-s)\,,\nonumber\\
  g^{e_me_n\chi_i^0\chi_j^0}[{\cal S}^{K\,J}_{ijmn},{\cal S}^{K+1\,J}_{ijmn}](s)&\equiv&-2m_im_j(m_n^2+m_m^2-s)\,,\nonumber\\
  g^{e_me_n\chi_i^0\chi_j^0}[{\cal S}^{K\,J}_{ijmn},{\cal S}^{K\,J+1}_{ijmn}](s)&\equiv&2m_nm_m(s-m_i^2-m_j^2)\,,\nonumber\\
  g^{e_me_n\chi_i^0\chi_j^0}[{\cal S}^{K\,J}_{ijmn},{\cal S}^{K+1\,J+1}_{ijmn}](s)&\equiv&-4m_im_jm_nm_m\,,\nonumber\\
  g^{e_me_n\chi_i^0\chi_j^0}[{\cal V}^{K\,J}_{ijmn},{\cal S}^{K\,J}_{ijmn}](s)&\equiv&\tfrac{m_mm_i}{s}(s-m_i^2-m_j^2)(m_n^2-m_m^2+s)\,,\nonumber\\
  g^{e_me_n\chi_i^0\chi_j^0}[{\cal V}^{K\,J}_{ijmn},{\cal S}^{K+1\,J}_{ijmn}](s)&\equiv&-\tfrac{m_mm_j}{s}(s+m_i^2-m_j^2)(m_n^2-m_m^2+s)\,,\nonumber\\
  g^{e_me_n\chi_i^0\chi_j^0}[{\cal V}^{K\,J}_{ijmn},{\cal S}^{K\,J+1}_{ijmn}](s)&\equiv&\tfrac{m_nm_i}{s}(s-m_i^2+m_j^2)(m_n^2-m_m^2-s)\,,\nonumber\\
  g^{e_me_n\chi_i^0\chi_j^0}[{\cal V}^{K\,J}_{ijmn},{\cal S}^{K+1\,J+1}_{ijmn}](s)&\equiv&-\tfrac{m_nm_j}{s}(s+m_i^2-m_j^2)(m_n^2-m_m^2-s)\,,\nonumber\\
  g^{e_me_n\chi_i^0\chi_j^0}[{\cal T}^{K}_{ijmn},{\cal S}^{K'\,J}_{ijmn}](s)&\equiv&0\,,\nonumber\\
  g^{e_me_n\chi_i^0\chi_j^0}[{\cal V}^{K\,J}_{ijmn},{\cal V}^{K\,J}_{ijmn}](s)&\equiv&\big[(m_n^2-m_m^2)^2-s^2\big]\big[1-\tfrac{(m_i^2-m_j^2)^2}{s^2}\big]\nonumber\\
 && -\tfrac{1}{3}\big[(m_n^2-m_m^2)^2-2s(m_n^2+m_m^2)+s^2\big]\nonumber\\
 &&\big[1-2\tfrac{m_i^2+m_j^2}{s}+\tfrac{(m_i^2-m_j^2)^2}{s^2}\big]\,,\nonumber\\
  g^{e_me_n\chi_i^0\chi_j^0}[{\cal V}^{K\,J}_{ijmn},{\cal V}^{K+1\,J}_{ijmn}](s)&\equiv&4m_im_j(m_n^2+m_m^2-s)\,,\nonumber\\
  g^{e_me_n\chi_i^0\chi_j^0}[{\cal V}^{K\,J}_{ijmn},{\cal V}^{K\,J+1}_{ijmn}](s)&\equiv&-4m_nm_m(s-m_i^2-m_j^2)\,,\nonumber\\
  g^{e_me_n\chi_i^0\chi_j^0}[{\cal V}^{K\,J}_{ijmn},{\cal V}^{K+1\,J+1}_{ijmn}](s)&\equiv&-16m_im_jm_nm_m\,,\nonumber\\
 g^{e_me_n\chi_i^0\chi_j^0}[{\cal T}^{K}_{ijmn},{\cal V}^{K\,K}_{ijmn}](s)&\equiv&-3\tfrac{m_im_m}{s}(s-m_i^2+m_j^2)(m_n^2-m_m^2+s)\,,\nonumber\\
  g^{e_me_n\chi_i^0\chi_j^0}[{\cal T}^{K}_{ijmn},{\cal V}^{K+1\,K}_{ijmn}](s)&\equiv&-3\tfrac{m_jm_m}{s}(s+m_i^2-m_j^2)(m_n^2-m_m^2+s)\,,\nonumber\\
  g^{e_me_n\chi_i^0\chi_j^0}[{\cal T}^{K}_{ijmn},{\cal V}^{K\,K+1}_{ijmn}](s)&\equiv&3\tfrac{m_im_n}{s}(s-m_i^2+m_j^2)(m_n^2-m_m^2-s)\,,\nonumber\\
  g^{e_me_n\chi_i^0\chi_j^0}[{\cal T}^{K}_{ijmn},{\cal V}^{K+1\,K+1}_{ijmn}](s)&\equiv&3\tfrac{m_jm_n}{s}(s+m_i^2-m_j^2)(m_n^2-m_m^2-s)\,,\nonumber\\
  g^{e_me_n\chi_i^0\chi_j^0}[{\cal T}^{K}_{ijmn},{\cal T}^{K}_{ijmn}](s)&\equiv&\tfrac{1}{3}\big[2(m_n^2-m_m^2)^2-s(m_n^2+m_m^2)-s^2\big]\nonumber\\
  &&\big[1+\tfrac{m_i^2+m_j^2}{s}-2\tfrac{(m_i^2-m_j^2)^2}{s^2}\big]\,,\nonumber\\
  g^{e_me_n\chi_i^0\chi_j^0}[{\cal T}^{K}_{ijmn},{\cal T}^{K+1}_{ijmn}](s)&\equiv&-12m_im_jm_nm_m\,.\nonumber
\end{eqnarray}
\subsubsection*{Production in hadronic tau decays}
Two-body decays involving a pseudoscalar meson:
\begin{multline}
\Gamma[\tau\to \psi M]=\frac{m_{\tau}}{32\pi}\Big\{\big(1+\tfrac{m_{\psi}^2-m_{M}^2}{m_{\tau}^2}\big)\big(|g_L^{\psi\bar{M}\tau}|^2+|g_R^{\psi\bar{M}\tau}|^2\big)+4\tfrac{m_{\psi}}{m_{\tau}}\text{Re}\big[(g_L^{\psi\bar{M}\tau})^*g_R^{\psi\bar{M}\tau}\big]\Big\}\\
\times\Big[1-2\tfrac{m_{M}^2+m_{\psi}^2}{m_{\tau}^2}+\tfrac{(m_{M}^2-m_{\psi}^2)^2}{m_{\tau}^4}\Big]^{1/2}\,,
\end{multline}
with the couplings defined in Eq.~(\ref{eq:SL2bdecwi}).

For a light $\psi$, we expect the three-body decay channel, involving a meson pair, to be competitive:
\begin{multline}
	\Gamma[\tau\to\psi M_1M_2]=\frac{m_{\tau}^3B_0^2}{512\pi^3f_{\chi}^2}\int_{(X_1+X_2)^2}^{(1-m_{\psi}/m_{\tau})^2}{\hspace{-0.5cm}d\tilde{s}\,\Big[1-\tfrac{2}{\tilde{s}}(X_1^2+X_2^2)+\tfrac{1}{\tilde{s}^2}(X_1^2-X_2^2)^2\Big]^{1/2}}\\
	\times\Big[1-2(\tilde{s}+\tfrac{m_{\psi}^2}{m_{\tau}^2})+(\tilde{s}-\tfrac{m_{\psi}^2}{m_{\tau}^2})^2\Big]^{1/2}\sum_{\Omega,\Omega'}C[\Omega]C[\Omega']^*\,g^{\psi M_1M_2}[\Omega,\Omega'](\tilde{s})\,,
\end{multline}
where the relation $m_{\tau}^2g^{\psi M_1M_2}[\Omega,\Omega'](\tfrac{s}{m_{\tau}^2})=m_{\psi}^2g^{\bar{\ell}_3 M_1M_2}[\Omega,\Omega'](\tfrac{s}{m_{\psi}^2})$ determines the functions in the integrand in terms of their counterparts of Eq.~(\ref{eq:SL3bint}).

\subsubsection*{Production in baryon-number violating transitions}
A baryon $B_j$ may decay into a pseudoscalar meson $\bar{M}_i$ and a light neutralino according to:
\begin{eqnarray}
	\Gamma[B_j\to\psi\bar{M}_i]\!\!\!\!\!&=&\!\!\!\!\!\frac{m_{B_j}}{32\pi}\left\{\Big(1+\tfrac{m_{\psi}^2-m^2_{M_i}}{m_{B_j}^2}\Big)\Big[\big|g_{L}^{\psi M_i B_j}\big|^2+\big|g_{R}^{\psi M_i B_j}\big|^2\Big]+4\tfrac{m_{\psi}}{m_{B_j}}\text{Re}\Big[\big(g_{L}^{\psi M_i B_j}\big)^*g_{R}^{\psi M_i B_j}\Big]\right\}\nonumber\\
	&&\times\Big(1-2\tfrac{m_{\psi}^2+m^2_{M_i}}{m_{B_j}^2}+\tfrac{(m_{\psi}^2-m^2_{M_i})^2}{m_{B_j}^4}\Big)^{1/2}\,,
\end{eqnarray}
with the couplings of Eq.~(\ref{eq:haddecwi}). In the case of a nucleon, $B_j\stackrel[]{!}{=}p^+,n^0$, with a light neutralino, $m_{\psi}\lsim m_{\mu}$, the form factors computed on the lattice \cite{Aoki:2017puj} may offer a quantitatively more accurate description of these transitions. Higher orders in the chiral perturbation theory (\textit{e.g.}~encoding the electroweak charged interactions) can be added in order to account for \textit{e.g.}~the $s\to d$ transition.

For a heavy quark, \textit{i.e.}~with a large available phase space for the decay products, the partonic description may offer a good approximation of the decays of a $b$-flavored hadron ($\bar{B}$, $\Lambda_b^0$, etc.) into a light $\psi$ and light hadrons:
\begin{eqnarray}\label{eq:partdechad}
\Gamma[b\to\psi \bar{u} \bar{d}_j]&=&\frac{m_b^5}{1024\pi^3}\bigg\{\Big(\big|C[{\cal H}^{J\,K}_{1j3}]\big|^2+\big|C[\widetilde{\cal H}^{J\,K}_{1j3}]\big|^2+\big|C[\widetilde{\cal H}^{J\,K}_{13j}]\big|^2\nonumber\\
&&+\text{Re}\big[C[\widetilde{\cal H}^{J\,J}_{1j3}]C[\widetilde{\cal H}^{J\,J}_{13j}]^*-C[{\cal H}^{J\,J}_{1j3}]\big(C[\widetilde{\cal H}^{J\,J}_{1j3}]^*+C[\widetilde{\cal H}^{J\,J}_{13j}]^*\big)\big]\Big){\cal J}\big(\tfrac{m_{\psi}^2}{m_b^2}\big)\nonumber\\
&&+4\frac{m_{\psi}}{m_b}\text{Re}\Big[C[\widetilde{\cal H}^{J\,K}_{1j3}]C[\widetilde{\cal H}^{J+1\,K}_{13j}]^*-C[{\cal H}^{J\,J+1}_{1j3}]C[\widetilde{\cal H}^{J+1\,J}_{1j3}]^*\\
&&-C[{\cal H}^{J\,J}_{1j3}]C[\widetilde{\cal H}^{J+1\,J}_{13j}]^*+C[\widetilde{\cal H}^{J\,J}_{1j3}]C[\widetilde{\cal H}^{J+1\,J}_{13j}]^*\Big]{\cal K}\big(\tfrac{m_{\psi}^2}{m_b^2}\big)\bigg\}\,,\nonumber\\
{\cal J}(x)&\equiv&1-8x+8x^3-x^4-12x^2\ln{x}\ ,\hspace{0.5cm}{\cal K}(x)\equiv1+9x-9x^2-x^3+6x(1+x)\ln{x}\,.\nonumber
\end{eqnarray}

The same approach is evidently applicable to a top quark, provided higher-order operators do not compete with those of dimension $6$ 
that we consider here (\textit{i.e.}~provided the scalar mediators are much heavier than the top quark in the RpV MSSM):
\begin{eqnarray}
\Gamma[t\to\psi \bar{d}_j\bar{d}_k]&=&\frac{m_t^5}{1024\pi^3}\bigg\{\Big(\big|C[{\cal H}^{J\,K}_{3jk}]\big|^2+\big|C[\widetilde{\cal H}^{J\,K}_{3jk}]\big|^2+\big|C[\widetilde{\cal H}^{J\,K}_{3kj}]\big|^2\nonumber\\
&&+\text{Re}\big[C[\widetilde{\cal H}^{J\,J}_{3jk}]C[\widetilde{\cal H}^{J\,J}_{3kj}]^*-C[{\cal H}^{J\,J}_{3jk}]\big(C[\widetilde{\cal H}^{J\,J}_{3jk}]^*+C[\widetilde{\cal H}^{J\,J}_{3kj}]^*\big)\big]\Big){\cal J}\big(\tfrac{m_{\psi}^2}{m_b^2}\big)\nonumber\\
&&+4\frac{m_{\psi}}{m_t}\text{Re}\Big[C[{\cal H}^{J\,K}_{3jk}]C[{\cal H}^{J+1\,K}_{3jk}]^*-C[{\cal H}^{J\,J}_{3jk}]C[\widetilde{\cal H}^{J+1\,J}_{3jk}]^*\\
&&-C[{\cal H}^{J\,J}_{3jk}]C[\widetilde{\cal H}^{J+1\,J}_{3kj}]^*+C[\widetilde{\cal H}^{J\,J+1}_{3jk}]C[\widetilde{\cal H}^{J+1\,J}_{3kj}]^*\Big]{\cal K}\big(\tfrac{m_{\psi}^2}{m_b^2}\big)\bigg\}\nonumber\,.
\end{eqnarray}
A charm quark is likely too close to the threshold for baryon production for such an approximation to be meaningful in descriptions of the decays of $D$-mesons.

\subsubsection*{Production in leptonic and semi-leptonic meson decays}
Two-body decays of a pseudoscalar meson $\bar{M}$ into $\psi$ and a (anti)lepton $\bar{\ell}_i$:
\begin{eqnarray}
\Gamma[\bar{M}\to\psi \bar{\ell}_i]\!\!&=&\!\!\frac{m_{M}}{16\pi}\Big\{\big(1-\tfrac{m_{\ell_i}^2+m_{\psi}^2}{m_{M}^2}\big)\big(|g_L^{\psi\bar{M}\ell_i}|^2+|g_R^{\psi\bar{M}\ell_i}|^2\big)-4\tfrac{m_{\ell_i}m_{\psi}}{m_{M}^2}\text{Re}\big[(g_L^{\psi\bar{M}\ell_i})^*g_R^{\psi\bar{M}\ell_i}\big]\Big\}\nonumber\\
&&\times\sqrt{1-2\tfrac{m_{\ell_i}^2+m_{\psi}^2}{m_{M}^2}+\tfrac{(m_{\ell_i}^2-m_{\psi}^2)^2}{m_{M}^4}}\,,
\end{eqnarray}
with the couplings defined in Eq.~(\ref{eq:SL2bdecwi}) (and possibly extending the operator basis to include third-generation fermions). For a neutral meson, pair production of $\psi$ is obtained from a similar formula, up to a factor $2$ (for identical particles) and the Wilson coefficients of the operators of Eq.~(\ref{eq:chipairop}) replacing those of Eq.~(\ref{eq:seleptop}) in the definition of the couplings: $C[{\cal S}^{\ell q\,L\,J}_{ijk}]\to C[\dot{\cal S}^{q\,J}_{jk}]$, $C[{\cal S}^{\ell q\,R\,J}_{ijk}]\to C[\dot{\cal S}^{q\,J+1}_{kj}]^*$, $C[{\cal V}^{\ell q\,L\,J}_{ijk}]\to C[\dot{\cal V}^{q\,J}_{jk}]$, $C[{\cal V}^{\ell q\,R\,J}_{ijk}]\to -C[\dot{\cal V}^{q\,J}_{kj}]^*$, $C[{\cal T}^{\ell q\,J}_{ijk}]\to 0$.

Similar decays are also accessible to a vector meson $M^*$ with the couplings defined in Eq.~(\ref{eq:SL2bVdecwi}):
\begin{align}
	&\Gamma[M^*\to \psi\bar{\ell}_i]=\frac{m_{M^*}}{48\pi} \Big[1-2\tfrac{m_{\psi}^2+m_{\ell_i}^2}{m_{M^*}^2}+\tfrac{(m_{\psi}^2-m_{\ell_i}^2)^2}{m_{M^*}^4}\Big]^{1/2}\\
	&\null\hspace{1cm}\times\Bigg\{\Big(|g_{V\,L}^{M^*\psi\ell_i}|^2+|g_{V\,R}^{M^*\psi\ell_i}|^2\Big)\Big[2-\tfrac{m^2_{\psi}+m^2_{\ell_i}}{m^2_{M^*}}-\tfrac{(m_{\psi}^2-m^2_{\ell_i})^2}{m^4_{M^*}}\Big]+6\frac{m_{\ell_i}m_{\psi}}{m^2_{M^*}}\text{Re}\big[g_{V\,L}^{M^*\psi\ell_i\,*}g_{V\,R}^{M^*\psi\ell_i}\big]\nonumber\\
	&\null\hspace{1.cm}+\Big[1-2\tfrac{m_{\psi}^2+m_{\ell_i}^2}{m_{M^*}^2}+\tfrac{(m_{\psi}^2-m_{\ell_i}^2)^2}{m_{M^*}^4}\Big]\Big[m^2_{M^*}\big(|g_{S\,L}^{M^*\psi\ell_i}|^2+|g_{S\,R}^{M^*\psi\ell_i}|^2\big)\big(1+\tfrac{m^2_{\psi}-m^2_{\ell_i}}{m^2_{M^*}}\big)\nonumber\\
	&\null\hspace{2cm}-4\,m_{\ell_i}m_{\psi}\,\text{Re}\big[g_{S\,L}^{M^*\psi\ell_i\,*}g_{S\,R}^{M^*\psi\ell_i}\big]+2m_{\ell_i}\text{Re}\big[g_{V\,L}^{M^*\psi\ell_i\,*}g_{S\,L}^{M^*\psi\ell_i}+g_{V\,R}^{M^*\psi\ell_i\,*}g_{S\,R}^{M^*\psi\ell_i}\big]\nonumber\\
	&\null\hspace{6cm}+2m_{\psi}\text{Re}\big[g_{V\,L}^{M^*\psi\ell_i\,*}g_{S\,R}^{M^*\psi\ell_i}+g_{V\,R}^{M^*\psi\ell_i\,*}g_{S\,L}^{M^*\psi\ell_i}\big]\Big]\Bigg\}\nonumber\,.
\end{align}
This expression is again straightforwardly extended to pair production through the substitution of the Wilson coefficients and the application of a factor $2$.

As soon as enough phase space is accessible, semi-leptonic decays become competitive. We restrict ourselves to a three-body decay involving pseudoscalar mesons in the initial and final states. Then, in analogy to Eq.~(\ref{eq:semilepformfac}), one can introduce the form-factors for a $M_1\to M_2$ transition ($s\equiv(p_1-p_2)^2$):
\begin{align}
&\left<M_2(p_2)\right|\bar{Q}_jQ_k\left|M_1(p_1)\right>\equiv \overline{F}^{M_1M_2}_{S\,jk}(s)\\
&\left<M_2(p_2)\right|\bar{Q}_j\gamma^{\mu}Q_k\left|M_1(p_1)\right>\equiv -(p_1-p_2)^{\mu}\tfrac{m_j-m_k}{s}\overline{F}^{M_1M_2}_{S\,jk}(s)\nonumber\\
&\null\hspace{4.95cm}-\big[p_1^{\mu}+p_2^{\mu}-\tfrac{m_1^2-m_2^2}{s}(p_1-p_2)^{\mu}\big]\overline{F}^{M_1M_2}_{V\,jk}(s)\nonumber\\
&\left<M_2(p_2)\right|\bar{Q}_j\Sigma^{\mu\nu}Q_k\left|M_1(p_1)\right>\equiv -i(p_1^{\mu}p_2^{\nu}-p_2^{\mu}p_1^{\nu})\overline{F}^{M_1M_2}_{T\,jk}(s)\nonumber
\end{align}
The decay width for $M_1\to M_2\psi\bar{\ell}_i$ then reads ($m_{1,2}\equiv m_{M_{1,2}}$):
\begin{align}
&\Gamma[M_1\to M_2\psi\bar{\ell}_i]=\frac{1}{256\pi^3m_1}\int_{(m_{\psi}+m_{\ell_i})^2}^{(m_1-m_2)^2}{\hspace{-0.5cm}ds\,\Big[1-2\tfrac{m_{\psi}^2+m_{\ell_i}^2}{s}+\tfrac{(m_1^2-m_2^2)^2}{s^2}\Big]^{1/2}}\nonumber \\
&\null\hspace{3.3cm}\times\Big[1-2(\tfrac{s+m_2^2}{m_1^2})+\tfrac{(s-m_2^2)^2}{m_1^4}\Big]^{1/2}
\sum_{\Omega,\Omega'}C[\Omega]C[\Omega']^*\,g^{M_2\psi \bar{\ell}_i}[\Omega,\Omega'](s)
\end{align}
where
\begin{eqnarray}
g^{M_2\psi \bar{\ell}_i}[{\cal S}_{ijk}^{\ell q\,JK},{\cal S}_{i\tilde{\jmath}\tilde{k}}^{\ell q\,JK'}](s)&\equiv&\tfrac{1}{4}(s-m_{\psi}^2-m_{\ell_i}^2)\overline{F}^{M_1M_2}_{S\,jk}(s)\,\big(\overline{F}^{M_1M_2}_{S\,\tilde{\jmath}\tilde{k}}(s)\big)^*\nonumber\,,\\
g^{M_2\psi \bar{\ell}_i}[{\cal S}_{ijk}^{\ell q\,JK},{\cal S}_{i\tilde{\jmath}\tilde{k}}^{\ell q\,J+1K'}](s)&\equiv&-\tfrac{m_{\psi}m_{\ell_i}}{2}\overline{F}^{M_1M_2}_{S\,jk}(s)\,\big(\overline{F}^{M_1M_2}_{S\,\tilde{\jmath}\tilde{k}}(s)\big)^*\nonumber\,,\\
g^{M_2\psi \bar{\ell}_i}[{\cal V}_{ijk}^{\ell q\,JK},{\cal S}_{i\tilde{\jmath}\tilde{k}}^{\ell q\,JK'}](s)&\equiv&-\tfrac{m_j-m_k}{4}\big[1-\tfrac{m^2_{\psi}-m_{\ell_i}^2}{s}\big]\overline{F}^{M_1M_2}_{S\,jk}(s)\,\big(\overline{F}^{M_1M_2}_{S\,\tilde{\jmath}\tilde{k}}(s)\big)^*\nonumber\,,\\
g^{M_2\psi \bar{\ell}_i}[{\cal V}_{ijk}^{\ell q\,JK},{\cal S}_{i\tilde{\jmath}\tilde{k}}^{\ell q\,J+1K'}](s)&\equiv&\tfrac{m_{\ell_i}}{4}(m_j-m_k)\big[1+\tfrac{m^2_{\psi}-m_{\ell_i}^2}{s}\big]\overline{F}^{M_1M_2}_{S\,jk}(s)\,\big(\overline{F}^{M_1M_2}_{S\,\tilde{\jmath}\tilde{k}}(s)\big)^*\nonumber\,,\\
g^{M_2\psi \bar{\ell}_i}[{\cal T}_{ijk}^{\ell q\,J},{\cal S}_{i\tilde{\jmath}\tilde{k}}^{\ell q\,J'K'}](s)&\equiv&0\nonumber\,,\\
g^{M_2\psi \bar{\ell}_i}[{\cal V}_{ijk}^{\ell q\,JK},{\cal V}_{i\tilde{\jmath}\tilde{k}}^{\ell q\,JK'}](s)&\equiv&\tfrac{1}{4}(m_j-m_k)(m_{\tilde{\jmath}}-m_{\tilde{k}})\big[\tfrac{m^2_{\psi}+m_{\ell_i}^2}{s}-\tfrac{(m_{\psi}^2-m_{\ell_i}^2)^2}{s^2}\big]\nonumber\\
&&\big(\overline{F}^{M_1M_2}_{S\,\tilde{\jmath}\tilde{k}}(s)\big)^*\overline{F}^{M_1M_2}_{S\,jk}(s)\,\nonumber\\
&&+\tfrac{m^4_1}{12}\big[1-2\tfrac{s+m^2_2}{m^2_1}+\tfrac{(s-m^2_2)^2}{m_1^4}\big]\big[2-\tfrac{m^2_{\psi}+m_{\ell_i}^2}{s}-\tfrac{(m_{\psi}^2-m_{\ell_i}^2)^2}{s^2}\big]\,\nonumber\\
&&\overline{F}^{M_1M_2}_{V\,jk}(s)\big(\overline{F}^{M_1M_2}_{V\,\tilde{\jmath}\tilde{k}}(s)\big)^*\nonumber\,,\\
g^{M_2\psi \bar{\ell}_i}[{\cal V}_{ijk}^{\ell q\,JK},{\cal V}_{i\tilde{\jmath}\tilde{k}}^{\ell q\,J+1K'}](s)&\equiv&-\tfrac{m_{\ell_i}m_{\psi}}{2s}(m_j-m_k)(m_{\tilde{\jmath}}-m_{\tilde{k}})\overline{F}^{M_1M_2}_{S\,jk}(s)\,\big(\overline{F}^{M_1M_2}_{S\,\tilde{\jmath}\tilde{k}}(s)\big)^*\nonumber\\
&&+\tfrac{m_{\ell_i}m_{\psi}}{2s}m_1^4\big[1-2\tfrac{s+m^2_2}{m^2_1}+\tfrac{(s-m^2_2)^2}{m_1^4}\big]\overline{F}^{M_1M_2}_{V\,jk}(s)\,\big(\overline{F}^{M_1M_2}_{V\,\tilde{\jmath}\tilde{k}}(s)\big)^*\nonumber\,,\\
g^{M_2\psi \bar{\ell}_i}[{\cal T}_{ijk}^{\ell q\,J},{\cal V}_{i\tilde{\jmath}\tilde{k}}^{\ell q\,JK'}](s)&\equiv&-\tfrac{m_{\psi}}{8}m_1^4\big(1-\tfrac{m^2_{\psi}-m_{\ell_i}^2}{s}\big)\big[1-\tfrac{s+2m_2^2}{m_1^2}+\tfrac{m_2^2(3s-m^2_2)}{m_1^4}\big]\nonumber\\
&&\overline{F}^{M_1M_2}_{T\,jk}(s)\,\big(\overline{F}^{M_1M_2}_{V\,\tilde{\jmath}\tilde{k}}(s)\big)^*\nonumber\,,\\
g^{M_2\psi \bar{\ell}_i}[{\cal T}_{ijk}^{\ell q\,J},{\cal V}_{i\tilde{\jmath}\tilde{k}}^{\ell q\,J+1K'}](s)&\equiv&-\tfrac{m_{\ell_i}}{8}m_1^4\big(1+\tfrac{m_{\psi}^2-m_{\ell_i}^2}{s}\big)\big[1-\tfrac{s+2m_2^2}{m_1^2}+\tfrac{m_2^2(3s-m_2^2)}{m_1^4}\big]\nonumber\\
&&\overline{F}^{M_1M_2}_{T\,jk}(s)\,\big(\overline{F}^{M_1M_2}_{V\,\tilde{\jmath}\tilde{k}}(s)\big)^*\,,\nonumber\\
g^{M_2\psi \bar{\ell}_i}[{\cal T}_{ijk}^{\ell q\,J},{\cal T}_{i\tilde{\jmath}\tilde{k}}^{\ell q\,J}](s)&\equiv&\tfrac{m_1^4}{48}s\big[1-2\tfrac{s+m_2^2}{m_1^2}+\tfrac{(s-m_2^2)^2}{m_1^4}\big]\big[1+\tfrac{m^2_{\psi}+m_{\ell_i}^2}{s}-2\tfrac{(m^2_{\psi}-m_{\ell_i}^2)^2}{s^2}\big]\nonumber\\
&&\overline{F}^{M_1M_2}_{T\,jk}(s)\,\big(\overline{F}^{M_1M_2}_{T\,\tilde{\jmath}\tilde{k}}(s)\big)^*\,,\nonumber\\
g^{M_2\psi \bar{\ell}_i}[{\cal T}_{ijk}^{\ell q\,J},{\cal T}_{i\tilde{\jmath}\tilde{k}}^{\ell q\,J+1}](s)&\equiv&\tfrac{m_{\ell_i}m_{\psi}}{8}m_1^4\big[1-2\tfrac{s+m_2^2}{m_1^2}+\tfrac{(s-m_2^2)^2}{m_1^4}\big]\overline{F}^{M_1M_2}_{T\,jk}(s)\,\big(\overline{F}^{M_1M_2}_{T\,\tilde{\jmath}\tilde{k}}(s)\big)^*\,,\nonumber
\end{eqnarray}
where the sum runs over all semi-leptonic operators. As before, the case of $\psi$-pair production can be addressed by substituting $\ell_i$ by $\psi$ in the expression above, hence employing the Wilson coefficients of the operators of Eq.~(\ref{eq:chipairop}), and introducing a factor $2$ for identical particles in the final state.

When substantial phase space is accessible in the final state, \textit{i.e.}~for hadrons involving a heavy quark, the partonic description may again prove a useful approximation. Neglecting the masses of the light leptons and quarks, we can write the relevant decay widths for a bottom quark:
\begin{align}
	&\Gamma[b\to\psi \bar{\ell}_i q_j]=\frac{m_b^5}{2048\pi^3}\bigg\{\Big(\big|C[{\cal S}^{\ell q\,J\,K}_{ij3}]\big|^2+4\big|C[{\cal V}^{\ell q\,J\,K}_{ij3}]\big|^2+3\big|C[{\cal T}^{\ell q\,J}_{ij3}]\big|^2\Big){\cal J}\big(\tfrac{m_{\psi}^2}{m_b^2}\big)\\
	&\null\hspace{2.5cm}+8\frac{m_{\psi}}{m_b}\text{Re}\Big[C[{\cal V}^{\ell q\,J\,K}_{ij3}]C[{\cal S}^{\ell q\,J\,K+1}_{i3j}]^*+3C[{\cal T}^{\ell q\,J}_{ij3}]C[{\cal V}^{\ell q\,J\,J+1}_{ij3}]^*\Big]{\cal K}\big(\tfrac{m_{\psi}^2}{m_b^2}\big)\bigg\}\nonumber\,,\\
	&\Gamma[\bar{b}\to\psi \bar{\ell}_i \bar{q}_j]=\frac{m_b^5}{2048\pi^3}\bigg\{\Big(\big|C[{\cal S}^{\ell q\,J\,K}_{i3j}]\big|^2+4\big|C[{\cal V}^{\ell q\,J\,K}_{i3j}]\big|^2+3\big|C[{\cal T}^{\ell q\,J}_{i3j}]\big|^2\Big){\cal J}\big(\tfrac{m_{\psi}^2}{m_b^2}\big)\nonumber\\
	&\null\hspace{2.6cm}-8\frac{m_{\psi}}{m_b}\text{Re}\Big[C[{\cal V}^{\ell q\,J\,K}_{i3j}]C[{\cal S}^{\ell q\,J\,K}_{i3j}]^*+3C[{\cal T}^{\ell q\,J}_{i3j}]C[{\cal V}^{\ell q\,J\,J}_{i3j}]^*\Big]{\cal K}\big(\tfrac{m_{\psi}^2}{m_b^2}\big)\bigg\}\nonumber\,,
\end{align}
with ${\cal J}$ and ${\cal K}$ the functions defined in Eq.~(\ref{eq:partdechad}) and the considered `$\ell q$' operators extend the base of Eq.~(\ref{eq:seleptop}) to bottom quarks (the index `3' referring to $b$). The decays of a top quark would take exactly the same form (but the operators now include $t$ instead of $b$). The decay widths of a $b$ quark producing a pair of light neutralinos and a light quark $q_j=d,s$ (of negligible mass) read:
\begin{align}
	&\Gamma[b\to\psi \psi q_j]=\frac{m_b^5}{1024\pi^3}\bigg\{\Big(\big|C[\dot{\cal S}^{d\,J}_{j3}]\big|^2+\big|C[\dot{\cal S}^{d\,J}_{3j}]\big|^2+4\big|C[\dot{\cal V}^{d\,J}_{j3}]\big|^2+4\big|C[\dot{\cal V}^{d\,J}_{3j}]\big|^2\Big)\widetilde{\cal J}\big(\tfrac{m_{\psi}^2}{m_b^2}\big)\\
	&\null\hspace{2.6cm}-8\frac{m_{\psi}^2}{m_b^2}\text{Re}\Big[C[\dot{\cal S}^{d\,J}_{j3}]C[\dot{\cal S}^{d\,J+1}_{3j}]+4C[\dot{\cal V}^{d\,J}_{j3}]C[\dot{\cal V}^{d\,J}_{3j}]\Big]\widetilde{\cal K}\big(\tfrac{m_{\psi}^2}{m_b^2}\big)\nonumber\\
	&\null\hspace{2.6cm}+8\frac{m_{\psi}}{m_b}\text{Re}\Big[C[\dot{\cal V}^{d\,J}_{j3}]C[\dot{\cal S}^{d\,J+1}_{j3}]^*-C[\dot{\cal V}^{d\,J}_{3j}]C[\dot{\cal S}^{d\,J}_{3j}]^*
	\nonumber \\
	&\null\hspace{2.6cm}-C[\dot{\cal V}^{d\,J}_{j3}]C[\dot{\cal S}^{d\,J+1}_{3j}]+C[\dot{\cal V}^{d\,J}_{3j}]C[\dot{\cal S}^{d\,J}_{j3}]\Big]\widetilde{\cal K}
	\big(\tfrac{m_{\psi}^2}{m_b^2}\big)\bigg\}\nonumber\,,\\
	&\null\hspace{1cm}\widetilde{\cal J}(x)\equiv\sqrt{1-4x}(1-14x-2x^2-12x^3)-48x^2(1-x^2)\ln\tfrac{1+\sqrt{1-4x}}{2\sqrt{x}}\,,\nonumber\\
	&\null\hspace{1cm}\widetilde{\cal K}(x)\equiv\sqrt{1-4x}(1+5x-6x^2)-12x(1-2x+x^2)\ln\tfrac{1+\sqrt{1-4x}}{2\sqrt{x}}\,.\nonumber
\end{align}
Again, this formula can be applied to the decays of a top quark (and $q_j\stackrel[]{!}{=}u,c$) up to the replacement of $\dot{\cal S},\dot{\cal V}^{d\,J}_{jk}$ 
by $\dot{\cal S},\dot{\cal V}^{u\,J}_{jk}$.

\end{document}